\documentclass[
 aps,
 prd,
 reprint,
 nofootinbib
 ]{revtex4-2}
\usepackage[utf8]{inputenc} 
\usepackage{hyperref} 
\usepackage{amsmath} 
\usepackage{amsfonts}
\usepackage{amsthm}
\usepackage{amssymb}
\usepackage{float} 
\usepackage{graphicx} 
\usepackage{lmodern} 
\usepackage{braket} 
\usepackage{mathrsfs} 
\usepackage{xcolor} 
\usepackage[normalem]{ulem} 
\usepackage{cancel}
\usepackage{simpler-wick}

\usepackage{tikz}  
\DeclareMathOperator*{\sumint}{%
\mathchoice%
  {\ooalign{$\displaystyle\sum$\cr\hidewidth$\displaystyle\int$\hidewidth\cr}}
  {\ooalign{\raisebox{.14\height}{\scalebox{.7}{$\textstyle\sum$}}\cr\hidewidth$\textstyle\int$\hidewidth\cr}}
  {\ooalign{\raisebox{.2\height}{\scalebox{.6}{$\scriptstyle\sum$}}\cr$\scriptstyle\int$\cr}}
  {\ooalign{\raisebox{.2\height}{\scalebox{.6}{$\scriptstyle\sum$}}\cr$\scriptstyle\int$\cr}}
}

\definecolor{Persianturquoise}{rgb}{0.4, 0.67, 0.63}

\graphicspath{
    {./}
    {./figs/}
}

\newcommand{\Nottingham}{\affiliation{
		School of Physics and Astronomy,
		University Park,
		University of Nottingham,
		Nottingham NG7 2RD,
		United Kingdom
}}

\begin{document}

\title{Effective field theory for strong phase transitions}
\date{\today}

\author{Oliver Gould}
\email{oliver.gould@nottingham.ac.uk}
\Nottingham

\author{Farbod-Sayyed Rassouli}
\email{sayyed.rassouli@nottingham.ac.uk}
\Nottingham

\author{Paul M. Saffin}
\email{paul.saffin@nottingham.ac.uk}
\Nottingham

\begin{abstract}
We show how to construct perturbative effective field theories for strong phase transitions, where the standard tools of high-temperature dimensional reduction fail.
By identifying the relevant power counting relations for the couplings, masses and background fields, we construct a controlled perturbative expansion adapted to this regime.
We find that the order parameter field remains light at the transition, so that
there is a controlled derivative expansion, even though large external fields need to be resummed.
The resulting effective field theories for strong phase transitions live in three dimensions and have nonpolynomial field dependence.
Applying our framework to
a two-scalar model, we verify renormalisation scale cancellation through
next-to-next-to-leading
order and find decreasing corrections to the latent heat as the perturbative order is increased.
Extending to include fermions and gauge fields, we
identify a broad class of models in which strong-transition thermodynamics are systematically improvable.
\end{abstract}

\maketitle

\section{Introduction}

Gravitational waves from cosmological phase transitions are one of the few possible signals that can reach us today from the early universe.
In recent years there has been much progress in understanding such signals, and how their features relate to underlying microphysics.
In this paper, we outline a promising class of models that can yield very strong signals, and describe how they can be systematically analysed using effective field theory (EFT).

To make predictions for a given microphysical model, the first step is to understand the phase diagram of the model, and how this was traversed as the universe cooled after the Hot Big Bang. Determining this reliably raises significant theoretical challenges, such as the necessity to resum various infinite classes of diagrams~\cite{Kapusta:1979fh}, and the strong coupling of infrared modes at high temperatures~\cite{Linde:1980ts}.
Further challenges arise for making predictions of the resulting gravitational wave signal, which typically suffer from relative errors of several orders of magnitude~\cite{Carena:2019une, Croon:2020cgk, Gould:2021oba}.
These are much larger than projected experimental uncertainties~\cite{Lewicki:2024xan}, so should be solved before the next generation of gravitational wave detectors are up and running.

EFT is perhaps our most powerful theoretical tool to tackle these challenges.
In particular, \textit{high-temperature dimensional reduction}~\cite{Appelquist:1981vg, Nadkarni:1982kb, Farakos:1994kx, Kajantie:1995dw, Braaten:1995cm, Braaten:1995jr} is the EFT framework that describes the long-wavelength modes of perturbative quantum field theories (QFTs) at high temperatures.
This was a crucial tool for determining the phase history of the electroweak sector in the Standard Model~\cite{Kajantie:1996mn, DOnofrio:2015gop}.
The approach has also yielded the most reliable and accurate predictions of the gravitational wave signals of cosmological phase transitions~\cite{Croon:2020cgk, Lewicki:2024xan, Ramsey-Musolf:2024ykk, Gould:2024jjt}.

Yet dimensional reduction appears to fail for the strongest phase transitions, which give the largest signals.
In such cases, the standard assumptions underlying dimensional reduction break down, and higher dimensional operators become important~\cite{Chala:2024xll, Gould:2024jjt, Bernardo:2025vkz, Chala:2025oul}.

The early studies that developed dimensional reduction methods for the Standard Model~\cite{Farakos:1994kx, Kajantie:1995dw, Kajantie:1995kf} assumed the Higgs and magnetic gauge fields to be very light.
In terms of the temperature $T$ and a perturbative gauge coupling $g/\pi \ll 1$, the assumption is that their masses are typically of order $O(g^2 T)$,
leading to a strongly coupled EFT, which can be studied on the lattice.
The transitions may be crossovers, second-order transitions or weakly first-order transitions, with a typical latent heat of order $L=O(g^4T^4)$.
Much stronger transitions will push the magnetic gauge field masses to break the underlying assumptions of the EFT.
We discuss these EFTs, and their \textit{weak} transitions, further in Sec.~\ref{sec:ultrasoft_transitions}.

The methods of high-temperature dimensional reduction have also been extended to stronger transitions.
These arise in two distinct model classes.
The first class occurs for Higgs-like phase transitions, when the Higgs self-coupling is anomalously small $\lambda= O(g^3)$~\cite{Arnold:1992rz, Ekstedt:2022zro, Gould:2023ovu}.
The second class occurs for $\lambda= O(g^2)$ when a potential barrier appears at tree-level within the dimensional reduced theory, and requires either cubic scalar interactions or multiple scalar fields~\cite{Niemi:2020hto, Niemi:2021qvp, Gould:2023jbz}.
In both cases, the transitions are then more strongly first order, with larger latent heat $L=O(g^2T^4)$, though the EFT still breaks down if the transition becomes too strong.
These EFTs, which admit \textit{intermediate} strength transitions, are discussed further in Sec.~\ref{sec:soft_supersoft_transitions}.


For stronger transitions still, it has been assumed that EFT methods break down altogether.
In this absence, we have no reliable methods to study very strong transitions, as it is unclear which infinite classes of diagrams to resum.
We fill this gap, showing how EFT methods can be systematically applied to strong phase transitions.
The resulting EFTs describe fields in three dimensions, so that their construction can be described as dimensional reduction, though it goes beyond the high-temperature approximation.

The strong transitions that we focus on will typically have a latent heat of order $L=O(T^4)$.
Their existence relies on a hierarchy of couplings, whereby the self-coupling of the nucleating field is much smaller than its strongest coupling to other fields $\lambda = O(g^4)$.
In Sec.~\ref{sec:strong_transitions} we show how to construct such EFTs, and demonstrate explicitly that our construction leads to a controlled perturbative expansion, within which the renormalisation scale cancels order-by-order.

The models that we find lead to strong transitions involve scalar fields with very weak self-interactions, and can involve couplings to other scalars, fermions or gauge fields.
Some of the results of this paper were foreshadowed in Ref.~\cite{Lofgren:2023sep}, in particular as \textit{Hierarchy IV}, see also \textit{RSB} and \textit{RSB~6} in Ref.~\cite{Camargo-Molina:2024sde}.

Our setup is similar to that of classically conformal models~\cite{Espinosa:2008kw, Jinno:2016knw, Prokopec:2018tnq}, which allow for a large degree of supercooling.
However, we note that it does not rely on classical conformality, but rather on a hierarchy of couplings, which in turn yields a hierarchy of scales.
Note that we do not compute the bubble nucleation rate, and hence do not consider the completion nor degree of supercooling of the phase transition.
Nevertheless, our EFT-based approach does allow for calculations of the bubble nucleation rate within a controlled derivative expansion~\cite{Gould:2021ccf}.

In this paper, we start in Sec.~\ref{sec:scale_hierarchies} by motivating the use of EFT to study phase transitions.
We then review the existing EFT-based approaches to weaker transitions, in Secs.~\ref{sec:ultrasoft_transitions} and \ref{sec:soft_supersoft_transitions}, highlighting where they break down as the transition strength increases.
Building on this, we outline how EFT methods can be applied to strong transitions in Sec.~\ref{sec:strong_transitions}, and demonstrate that the EFT is under perturbative control.
In Sec.~\ref{sec:other_models}, we generalise this construction to fermionic and gauge models, and discuss its application to extensions of the Standard Model.
Finally, we take stock and outline some future directions in Sec.~\ref{sec:discussion}.

\section{Why EFT?}
\label{sec:why_eft}

\subsection{What's wrong with the loop-expansion?} \label{sec:whats_wrong_with_the_loop_expansion}

A common approach to studying the cosmological phase transitions of a given model is first to construct the loop-expanded effective potential~\cite{Quiros:1999jp, Wainwright:2011kj, Caprini:2019egz, Athron:2023mer},
\begin{equation}
    V_\text{loop}(\phi; T) = V_{\text{tree}}(\phi) + V_{\text{1-loop}}(\phi; T) + \dots ,
\end{equation}
truncated to some loop order.
For instance, the 1-loop term can be expressed as
\begin{equation}
    V_{\text{1-loop}}(\phi; T) = V_{\text{ct}}(\phi) + V_{\text{CW}}(\phi) + V_{T}(\phi; T) ,
\end{equation}
in terms of the counterterm, Coleman-Weinberg~\cite{Coleman:1973jx}, and thermal~\cite{Dolan:1973qd} contributions respectively.
These terms can be constructed simply and algorithmically, based only on the perturbative mass spectrum of the model.

To determine the phase history for a given parameter point, one then studies the minima of $V_\text{loop}$ with respect to the background fields $\phi$.
With a little more work, one can also use $V_\text{loop}$ to determine the thermodynamics and other properties of any phase transitions.

However, there is a tension between two assumptions that this approach is based upon:
\begin{itemize}
    \item[(i)] thermal effects significantly modify the potential;
    \item[(ii)] higher loop orders are suppressed.
\end{itemize}
Assumption (i) typically holds when there are phase transitions, as thermal corrections modify the shape of the tree-level potential.
Assumption (ii) holds when the only small parameters are couplings,
so that the loop expansion and the coupling expansion agree.
The tension arises because thermal corrections to the potential appear from loops, and (i) assumes these are large, while (ii) assumes these are small.

As a consequence of this tension, the loop expansion is generically not applicable to the study of thermal phase transitions.
The 2-loop potential does no better at resolving this tension than the 1-loop potential.
The loop expansion itself is at fault, not the order of the expansion.

One can try to fix the problem with $V_\text{loop}$, by resumming the effects of thermal loops on the propagators used to compute the loop diagrams.
At leading order, and for static observables such as the effective potential, this is just a shift to the masses.%
\footnote{Leading order thermal effects also yield tadpoles, and induce mixing, through shifts to off-diagonal terms in the mass matrix, but we will neglect this effect for simplicity here.}
At leading order, this resummation leads to
\begin{equation} \label{eq:one-loop-potential-resummed}
    V_{\text{1-loop}}(\phi; T) \to V_{\text{1-loop}}(\phi; T) + V_{\text{daisy}}(\phi; T) ,
\end{equation}
where the daisy term takes the form
\begin{equation} \label{eq:daisy}
    V_{\text{daisy}}(\phi; T)
    = \sum_{a}\left[\Delta(M^2_a+\Pi_{a,T}) - \Delta(M^2_a) \right]
\end{equation}
and $a$ runs over the mass eigenstates to be resummed.
Here $\Delta$ is the part of the one-loop potential that is resummed.
It is contained within the 1-loop potential, $\Delta(M^2_a) \subseteq V_{\text{1-loop}}$, so that the last term in Eq.~\eqref{eq:daisy} cancels in the sum of Eq.~\eqref{eq:one-loop-potential-resummed}, avoiding double counting of modes.
The result of the resummation is to replace $M^2_a \to M^2_a+\Pi_{a,T}$ inside $\Delta(\cdot)$.

Historically, there has been some disagreement on the precise form of this resummation, with one proposal that $\Delta=V_{\text{1-loop}}$~\cite{Parwani:1991gq} and another that $\Delta$ be only the part of the 1-loop potential the \textit{needs} resummation~\cite{Arnold:1992rz}.

In either approach, daisy resummation goes some way to alleviate the tension between the assumptions (i) and (ii) that underlie the loop expansion.
However, it provides only a temporary sticking plaster, as new tensions arise depending on the nature of any phase transitions and generically at higher orders.

On a practical level this approach often fails to correctly predict the order of phase transitions, plus it suffers from strong dependence on the gauge-fixing parameter and renormalisation scale, and imaginary contributions to real physical observables~\cite{Croon:2020cgk}. This is partly because of the slow convergence of perturbation theory at high temperatures. Perhaps more importantly though, the ad-hoc nature of the resummation obscures its physical origin, and so typically leads to inconsistencies in subsequent calculations, such as inconsistent derivative expansions and double counting modes in computations of bubble nucleation.

To resolve these problems inevitably requires further resummations, which can in principle be tackled directly at the diagram level by identifying infinite classes of diagrams that are enhanced by thermal effects, being careful to avoid double-counting as well as inconsistent derivative expansions for spatially varying background fields. However, rather than wrestle with the labyrinthine by-ways that this route offers, we would repeat the wisdom of the old joke about asking for directions: \textit{If I were you, I wouldn't start from here.} For a systematic solution, we need to investigate more closely \textit{why} resummation is necessary.

\subsection{Scale hierarchies in phase transitions}
\label{sec:scale_hierarchies}

To find a way out of the catch-22 outlined in Sec.~\ref{sec:whats_wrong_with_the_loop_expansion}, it is helpful to formulate the key assumptions mathematically. Within a weakly-coupled theory, assumption (i), that thermal effects significantly modify the potential, can be formulated as 
\begin{equation} \label{eq:potential_changes_O1}
    \text{(i)} \implies \frac{V_T}{V_\text{tree}} = O(1) ,
\end{equation}
which must hold over a range of values of $\phi$, for the potential to really change shape with temperature.
Our assumption that the theory is weakly coupled implies that the tree-level potential is a good approximation of the effective potential at zero temperature, and that the leading thermal corrections arise at 1-loop.

The terms in Eq.~\eqref{eq:potential_changes_O1} are constrained in form by the structure of quantum field theory. To gain some insight, consider a simple scalar field theory with tree-level potential
\begin{equation}
    V_\text{tree}(\phi) = -\frac{1}{2}\mu_\phi^2 \phi^2 + \frac{1}{4!}\lambda_\phi \phi^4 .
\end{equation}
In the vicinity of the broken phase minimum $\phi=\pm\phi_\text{min}$, where the quadratic and quartic terms balance, the field and potential are of order 
\begin{align}
    \phi^2 = O\left(\frac{m_\phi^2}{\lambda_\phi}\right)  , \qquad
    V_\text{tree} = O\left(\frac{m_\phi^4}{\lambda_\phi}\right)  ,
\end{align}
here written in terms of the scalar mass $m_\phi = \sqrt{2}\mu_\phi$. At zero temperature, higher loops are suppressed by powers of the coupling
\begin{equation}
    V_{n\text{-loop}} = O(\lambda_\phi^n V_\text{tree}) = O\left(\lambda_\phi^{n-1}m_\phi^4\right),
\end{equation}
So that the ratio,
\begin{equation}
    \frac{V_\text{1-loop}}{V_\text{tree}}=O(\lambda_\phi),
\end{equation}
cannot become $O(1)$ without the theory being strongly coupled, noting also that including angular factors, this ratio is really $O(\lambda_\phi/(4\pi)^2)$. Thus, loop corrections to the potential are small at weak coupling.

However, at nonzero temperature loop corrections can become large. To understand why, it is helpful to consider the specific form of the thermal corrections (see Eq.~\eqref{eq:app_Jthermal})
\begin{equation} \label{eq:potential_ratio_equality_1}
    \frac{V_T}{V_\text{tree}} = O\left( \lambda_\phi \frac{T^4}{m_\phi^4}  \mathcal{J}\left(\frac{m_\phi}{T}\right)\right),
\end{equation}
where $\mathcal{J}(x)$ is a dimensionless function of its argument $x$ that decreases in magnitude monotonically from an $O(1)$ constant at $x=0$, vanishing exponentially $\mathcal{J}(x)\propto e^{-x}$ as $x\to \infty$, so that
\begin{equation} \label{eq:potential_ratio_inequality_1}
    \mathcal{J}\left(\frac{m_\phi}{T}\right) \leq O(1)
    .
\end{equation}
Combining this with Eq.~\eqref{eq:potential_changes_O1} and the assumption of weak coupling, we can immediately conclude that $T\gg m_\phi$.
However, the leading contribution of $\mathcal{J}(x)$ at high temperatures
(small $x$) is simply a field-independent constant, followed by a quadratic
$x^2$ term.
Considering instead a change in the potential, the constant term cancels leaving
\begin{equation}
    \frac{V_T(\phi) - V_T(0)}{V_\text{tree}(\phi) - V_\text{tree}(0)} \leq O\left( \lambda_\phi \frac{T^2}{m_\phi^2}\right),
\end{equation}
where $\phi=O(\phi_\text{min})$, leading us to
\begin{equation} \label{eq:hierarchy_of_scales}
    \text{(i) and }\lambda_\phi\ll (4\pi)^2\implies  \frac{m_\phi^2}{T^2} = O(\lambda_\phi) .
\end{equation}
In words, the shape of the effective potential can only change significantly in this weakly coupled theory if the temperature is parametrically large compared with the scalar mass.
Thus, phase transitions happen at high temperatures.

For more complicated models with multiple fields, the same structures appear for
a joint expansion in all the couplings.
If $\phi$ couples to a second field $\xi$ with larger thermal contributions,
Eq.~\eqref{eq:potential_ratio_equality_1} generalises to
\begin{equation} \label{eq:potential_ratio_equality_2}
    \frac{V_T}{V_\text{tree}} = O\left( \lambda_\phi \frac{T^4}{m_\phi^4}  \mathcal{J}\left(\frac{m_\xi}{T}\right)\right).
\end{equation}
The fact that different masses enter the tree-level and thermal contributions
means that the argument of the thermal function $\mathcal{J}$ is not necessarily
small. If $m_\xi$ depends on the background field $\phi$ at leading order and
$m_\xi=O(T)$, then $\mathcal{J}$ has no constant term at leading order, and
Eq.~\eqref{eq:hierarchy_of_scales} is replaced by
\begin{equation} \label{eq:hierarchy_of_scales_2}
    \text{(i), }\lambda_\phi\ll (4\pi)^2 \text{ and }\frac{m_\xi}{T}=O(1) \implies  \frac{m_\phi^4}{T^4} = O(\lambda_\phi) .
\end{equation}
Thus, the scalar field $\phi$ undergoing the transition must still be light
compared with the temperature, but the hierarchy is somewhat relaxed. A more detailed argument will be presented in
a forthcoming work~\cite{Gould:2026forthcoming}.

Once there is a hierarchy of scales, as in Eq.~\eqref{eq:hierarchy_of_scales} or
\eqref{eq:hierarchy_of_scales_2}, then the weak-coupling expansion should be
modified to resum enhancements by large ratios of scales, and for this EFT
methods are the natural tools. In an EFT framework, degrees of freedom are
integrated out one scale at a time, starting from the highest energy modes and
working down. After an energy scale is integrated out, its effects on lower
energy scales are resummed. This allows us to escape the catch-22: we can
construct a controlled weak-coupling expansion, while the hierarchy of scales
allows for significant thermal corrections to the potential.

When either Eq.~\eqref{eq:hierarchy_of_scales} or \eqref{eq:hierarchy_of_scales_2} is satisfied, the scale $\pi T$ is a high energy scale compared to the mass of the $\phi$ field, and is referred to as the \textit{hard scale}. Integrating out the hard scale will lead to large modifications of the effective mass of the $\phi$ field, it being a parameter of positive mass dimension. The requirement that the hard scale gives $O(1)$ corrections to the leading order potential of the EFT is
\begin{equation} \label{eq:effective_hierarchy_of_scales}
    \frac{m_{\phi, \text{eff}}^4}{T^4} = O(\lambda_{\phi,\text{eff}})  \quad\text{or}\quad \frac{m_{\phi, \text{eff}}^2}{T^2} = O(\lambda_{\phi,\text{eff}}),
\end{equation}
with the former applying if there is a field with mass $O(T)$ that depends on
$\phi$ at leading order,
and the latter applying if there are no such fields.
Here $m_{\phi, \text{eff}}$ is the effective mass of the $\phi$ field, i.e.\
the second field derivative of the EFT potential, and
$\lambda_{\phi,\text{eff}}$ is the effective self-coupling. The difference between
Eq.~\eqref{eq:effective_hierarchy_of_scales} and \eqref{eq:hierarchy_of_scales}
or \eqref{eq:hierarchy_of_scales_2}
can become important at a phase transition, where it is possible that $m_{\phi,
\text{eff}} \ll m_{\phi}$.

\section{EFTs for weaker transitions} \label{sec:weak_transitions}

EFT has played a key role for studying the thermodynamics of quantum field theories. This was spurred in particular by questions about the order of the electroweak phase transition, and the nature of the quark-gluon plasma. In this section, we review some of the key assumptions made in this historical development, for comparison to our later EFT construction for strong transitions. Those primarily interested in the latter may skip this section.

\subsection{Weak transitions at the ultrasoft scale ($\frac{g^2T}{\pi}$)} \label{sec:ultrasoft_transitions}

In this subsection, we take as our example the $\mathrm{SU}(2)$ Higgs model, essentially a simplified electroweak model, for which the Euclidean action reads
\begin{align} \label{eq:electroweak_lagrangian}
    S_\text{E} &= \int\mathrm{d}\tau \mathrm{d}^3x\bigg[
    (D_\mu H)^\dagger D_\mu H + V(H)
    +\frac{1}{4}F_{\mu\nu}^a F_{\mu\nu}^a
    \bigg], \\
    V(H) &= - \mu_H^2 H^\dagger H + \lambda (H^\dagger H)^2,
\end{align}
where $H$ is the Higgs field, $D_\mu$
the gauge-covariant derivative,
and $F^a_{\mu\nu}$ the field strength tensor. The gauge field is $A_\mu^a$, though sometimes we will suppress the colour index $a$.
To avoid distracting from the narrative, in this section we do not include gauge fixing explicitly.

The nature of the phase transition in this model
is known to depend sensitively on the
ratio between the Higgs self-coupling and the gauge coupling, $\lambda/g^2$.
The early works
made the relatively natural assumption that $\lambda/g^2=O(1)$~\cite{Farakos:1994kx, Kajantie:1995dw, Kajantie:1995kf, Kajantie:1996mn}. They are therefore applicable within this subset of the full parameter space of the model.

The weak-coupling expansion with $\lambda/g^2=O(1)$ can be formalised by introducing a power-counting parameter $\alpha$, such that
\begin{equation} \label{eq:gsq_lambda_ultrasoft_power_counting}
    \{g^2, \lambda\} \to \{\alpha^2 g^2, \alpha^2\lambda\} .
\end{equation}
Both couplings then approach zero in the same way as $\alpha\to 0_+$. For
application to a specific parameter point, one sets $\alpha=1$ at the end, so
that $g^2$ and $\lambda$ actually determine convergence, while $\alpha$ is just used
to organise the expansion.
In this section, we follow the literature and use $g$ to represent power-counting magnitudes, with $g^2/\pi^2\sim \alpha^2$ the natural loop-expansion parameter.

As we argued in Sec.~\ref{sec:why_eft},
at the transition, thermal effects should significantly modify the potential; Eq.~\eqref{eq:potential_changes_O1}. Following identical reasoning as led to Eq.~\eqref{eq:hierarchy_of_scales}, and assuming there are no fields with $m_\text{eff}=O(T)$, one finds that the Higgs mass vacuum parameter should be of order $\mu_H^2=O(g^2 T^2)$, or equivalently
\begin{equation} \label{eq:higgs_mass_parameter_scaling}
    \{\mu_H^2, T\} \to \{\alpha^2 \mu_H^2, T\} ,
\end{equation}
implying that the temperature is large compared to the vacuum mass.

The scale hierarchy between the temperature and the Higgs mass appears within
a series of scale hierarchies,
\begin{align}
    \frac{g^2T}{\pi} \ll gT \ll \pi T .
\end{align}
These scales, called \textit{ultrasoft}, \textit{soft} and \textit{hard} respectively, reflect the inverse correlation lengths of different field modes.
As we explain below, the temperature comes pared with a factor of $\pi$, and as is usual in perturbative quantum field theory, $g$ comes together with $1/\pi$, so that e.g.\ $gT=(g/\pi)\times \pi T$~\cite{Gould:2023ovu}

The origin of the hard scale, $\pi T$, is clearest in the imaginary time formalism,
formulated on $\mathbb{R}^3\times S^1$, with $\beta\equiv 1/T$ the circumference of the circle~\cite{Laine:2016hma}.
Expanding the fields in a Fourier series around the circle leads to a tower of modes, called Matsubara modes. For example, for the gauge field, this reads
\begin{align} \label{eq:matsubara}
    A_\mu(\tau, \mathbf{x}) = \sum_{n=-\infty}^{\infty} A^{(n)}_\mu(\mathbf{x}) e^{i 2\pi n T \tau} ,
\end{align}
where $\tau \in [0, \beta)$ parameterises the circle. Inserting this mode expansion back into the Lagrangian, one finds that the mode $A^{(n)}_\mu(\mathbf{x})$ acquires a mass $2\pi n T$. Thus, the nonzero Matsubara modes, with $n\neq 0$, have a mass at the hard scale.

By contrast, the zero Matsubara modes are comparatively light: At tree level, the gauge field appears massless and the Higgs mass is $\sqrt{2}\mu_H = O(gT)$; see Eq.~\eqref{eq:higgs_mass_parameter_scaling}.
To tackle this hierarchy of scales (which can enhance loop integrals) within a weak-coupling expansion, the hard nonzero Matsubara modes should then be integrated out.
In essence, the fields are split as
\begin{align}
    A_\mu = A_\mu^\text{IR} + A_\mu^\text{UV} , \qquad H = H^\text{IR} + H^\text{UV} ,
\end{align}
where the infrared (IR) parts contain only the parts of the zero Matsubara modes with momentum $p < \Lambda \ll \pi T$, and the ultraviolet (UV) parts contain the
rest.
This split can be made explicitly with a hard cut off at the matching scale $\Lambda$~\cite{Greiner:1996dx}. However, it is more efficient to make the split implicitly using dimensional regularisation and an expansion by regions~\cite{Braaten:1995cm, Braaten:1995jr, Jantzen:2011nz, Hirvonen:2022jba}.

Integrating out the hard scale is referred to as \textit{high temperature dimensional reduction}~\cite{Appelquist:1981vg, Nadkarni:1982kb, Farakos:1994kx, Kajantie:1995dw, Braaten:1995cm, Braaten:1995jr}, because the resulting EFT describes fields living in $\mathbb{R}^3$. For our example model, the high-temperature EFT is
\begin{align} \label{eq:electroweak_3d_lagrangian_soft}
    S_\text{soft} =& \beta \int \mathrm{d}^3 x \bigg[(D_i H)^\dagger D_i H + (D_i A_{0})^a (D_i A_{0})^a
    \nonumber \\
    &+ V_\text{soft}(H, A_0) + \frac{1}{4}F_{ij}^a F_{ij}^a \bigg], \\
    V_\text{soft}(H) =& m_\text{soft}^2 H^\dagger H + \lambda (H^\dagger H)^2
    \nonumber \\
    &
    + \frac{1}{2}m_\text{D}^2 A_0^aA_0^a 
    + \frac{1}{4}g^2 A_0^aA_0^a H^\dagger H,
    \label{eq:electroweak_3d_potential_soft}
\end{align}
where we have dropped the superscript IR on the fields in the EFT, to avoid overburdening the notation. All the Lagrangian parameters are temperature-dependent Wilson coefficients, but for simplicity, we have approximated them at leading order in the weak-coupling expansion, in which case the effective masses are
\begin{align}
    m_\text{D}^2 &= \frac{5}{6}g^2 T^2 , \label{eq:debye_mass_soft}\\
    m_3^2 &= -\mu_H^2 + \left(\frac{3}{16}g^2 + \frac{1}{2}\lambda\right)T^2, \label{eq:higgs_mass_soft}
\end{align}
and the couplings are unchanged from tree-level.%
\footnote{Note that we have not canonically normalised the EFT fields, which would require absorbing the overall factor of $\beta$~\cite{Farakos:1994kx, Kajantie:1995dw, Braaten:1995cm, Braaten:1995jr}.}
They are known to higher orders~\cite{Kajantie:1995dw, Chala:2026wjp}.

At sufficiently high temperatures, both Eqs.~\eqref{eq:debye_mass_soft} and \eqref{eq:higgs_mass_soft} are  $O(g^2T^2)$ and hence lie at the soft scale.
The thermal loop corrections in Eq.~\eqref{eq:higgs_mass_soft} are boosted to leading order because of the scale hierarchy between the soft and hard scales.

As the temperature is decreased, Eq.~\eqref{eq:higgs_mass_soft} shows that the
Higgs effective mass goes through zero at
\begin{align} \label{eq:T0}
    T_{0}^2 = \frac{\mu_H^2}{\frac{3}{16}g^2 + \frac{1}{2}\lambda},
\end{align}
below which the symmetric phase becomes absolutely unstable.
However, on the approach to $T_{0}$ first the Higgs becomes much lighter than the $A_0^a$ field,
\begin{align} \label{eq:second_hierarchy_weak}
    m_3 \ll m_\text{D} .
\end{align}
To construct the coupling expansion in the presence of this hierarchy of scales, the temporal components of the gauge field, $A_0$, should be integrated out.

The hierarchy of Eq.~\eqref{eq:second_hierarchy_weak} does not fix the power-counting of the effective Higgs mass.
However, a new scale does dynamically arise: the scale on which the $A_i$ field is screened, the ultrasoft scale $g^2 T/\pi$~\cite{Linde:1980ts, Kajantie:1995dw, Braaten:1995jr}.
The early works assumed that the Higgs also lies at the ultrasoft scale~\cite{Farakos:1994kx, Kajantie:1995dw, Kajantie:1995kf, Kajantie:1996mn}, as summarised in Table~\ref{tab:hierarchy_electroweak}.

\begin{table}[t]
    \centering
    \begin{tabular}{c|c|c}
        & $m_{\rm eff}$ & Fields \\
        \hline
        Hard & $\pi T$         & $A_\mu^\text{UV}$, $H^\text{UV}$ \\
        Soft & $g T$           & $A_0^\text{IR}$ \\
        Ultrasoft & $g^{2} T/\pi$       &  $A_i^\text{IR}$, $H^\text{IR}$ 
    \end{tabular}
    \caption{Scale hierarchies for the $\mathrm{SU}(2)$ Higgs model, when the IR modes of the Higgs are ultrasoft.}
    \label{tab:hierarchy_electroweak}
\end{table}

Integrating out the $A_0$ field under the assumption that the Higgs field lies at the ultrasoft scale leads to the following EFT,
\begin{align} \label{eq:electroweak_3d_lagrangian_ultrasoft}
    &S_\text{ultra.} = \beta \int \mathrm{d}^3 x \bigg[(D_{i} H)^\dagger D_{i} H + V_\text{ultra.}(H)+ \frac{1}{4}F_{ij}^{a} F_{ij}^{a} \bigg] ,
\end{align}
with scalar potential
\begin{align}
    V_\text{ultra.}(H) = m^2_\text{ultra.} H^\dagger H + \lambda (H^\dagger H)^{2} ,
\end{align}
where by assumption $m_\text{ultra.}^2=O(g^4T^2)$.
The effect of the $A_0$ field appears only through modifications of the
effective parameters.
Due to the leading-order cancellation, the dominant effect is on the Higgs
mass $m^2_\text{ultra.}$,
giving additional $O(g^3T^2)$ and $O(g^4T^2)$ terms on top of
Eq.~\eqref{eq:higgs_mass_soft}. One must also go back and compute subleading
corrections in dimensional reduction to reach $O(g^4T^2)$ accuracy.

The ultrasoft EFT has the following tree-level scalar potential, with $H=(0,v/\sqrt{2})$,
\begin{align} \label{eq:potential_ultrasoft}
    V_\text{ultra.}(v) = \frac{1}{2} m_\text{ultra.}^2 v^2 + \frac{1}{4}\lambda v^4 .
\end{align}
By assumption there are no further hierarchies of scale, both $H$ and the IR modes of $A_i$ living at the ultrasoft scale, and no hierarchies of couplings. The mean-field analysis of the phase transition can therefore be applied to Eq.~\eqref{eq:potential_ultrasoft}. Requiring detailed balance, i.e.\ that the two terms in the potential are the same order of magnitude at the transition, gives
\begin{align} \label{eq:ultrasoft_vev_midpoint}
    v = O\left(\frac{m_\text{ultra.}}{\sqrt{\lambda}}\right)
    = O\left(\frac{g T}{\pi}\right),
\end{align}
where we have used the ultrasoft assumption in the vicinity of $T_\text{c}$,
$m_\text{ultra.}\sim g^2 T/\pi$, and the assumption that $\lambda/g^2=O(1)$.
This order-of-magnitude estimate for the field expectation value implies one for the latent heat, which can be power counted as
\begin{align} \label{eq:latent_heat_for_counting}
    L = \frac{d\Delta V_{\rm eff}(T)}{d\log T} 
 \bigg|_{T_\text{c}}= O\left(\frac{d m_\text{eff}^2}{d\log T} v^2 \right),
\end{align}
where we define the free energy density difference between the two phases as
\begin{align}
    \Delta V_{\rm eff}(T)
    \equiv V_{\rm eff}\left(v;T\right)
    -V_{\rm eff}\left(0;T\right).
    \label{eq:two_scalar_free_energy_difference}
\end{align}
Here $v$ denotes the broken-phase minimum at temperature $T$, and the critical temperature is determined by $\Delta V_{\rm eff}(T_c)=0$.
Next, we note that
\begin{align} \label{eq:eta_y}
    \frac{d m_\text{ultra.}^2}{d\log T}=O(g^2 T^2) ,
\end{align}
which is implied by Eq.~\eqref{eq:higgs_mass_soft}. This also holds in much more general models~\cite{Kajantie:1995dw}.
Note that $\frac{d m_\text{ultra.}^2}{d\log T} \ll m_\text{ultra.}^2$, because of the partial cancellation between vacuum and thermal contributions near $T_0$.
Putting this together with Eq.~\eqref{eq:ultrasoft_vev_midpoint}, we find that
\begin{align} \label{eq:latent_heat_ultrasoft}
    L_\text{ultra.} =O\left(\frac{g^4 T^4}{\pi^2}\right).
\end{align}
This follows from the set of assumptions made in constructing the ultrasoft EFT for the Higgs. 

It is possible to relax some of the power-counting assumptions without the EFT of Eq.~\eqref{eq:electroweak_3d_lagrangian_ultrasoft} breaking down.
In particular, one can relax the strict assumption on the magnitude of the Higgs effective mass, maintaining only that $m_\text{ultra.}^2 \ll m_\text{D}^2$.
Doing so relaxes Eq.~\eqref{eq:latent_heat_ultrasoft}, replacing it with a bound
that follows from the requirement that the expansion in powers
of the Higgs field can be truncated.
Higher powers of $H^\dagger H$, assumed suppressed,
would enter the effective Lagrangian as
\begin{align}\label{eq:Lultra_cn}
    \mathcal{L}_\text{ultra.}\supset c_n (H^\dagger H)^n, \quad \text{with} \quad c_n = O\left(\frac{g^{2n} T}{m_\text{D}^{2n-3}}\right) .
\end{align}
The parametric magnitude of $c_n$ can be derived by considering the background field-expansion of the one-loop vacuum diagram for $A_0^a$ with constant Higgs field $H^\dagger H=\tfrac{1}{2}v^2$.
Noting from Eq.~\eqref{eq:electroweak_3d_potential_soft} that the effective mass-squared for $A_0^a$ is $m_{\rm D}^2+\frac{1}{4}g^2 v^2$, the one-loop contribution to the effective potential is then
\begin{align}
&T\int_p\ln(p^2+m_{\rm D}^2+\frac{1}{2}g^2v^2)\\
&= \text{const.}+T\int_p\ln\left(1+\frac{\frac{1}{2}g^2v^2}{p^2+m_{\rm D}^2}\right)  .
\end{align}
Expanding the logarithm and performing the $p$-integral gives \eqref{eq:Lultra_cn}.
This expansion can be truncated at finite $n$ only if
\begin{align} \label{eq:v3_ultrasoft_break_down}
    v \ll O\left( \frac{m_\text{D}}{g} \right) = O\left( T \right) .
\end{align}
Combining Eq.~\eqref{eq:v3_ultrasoft_break_down} with Eqs.~\eqref{eq:latent_heat_for_counting} and \eqref{eq:eta_y}, implies the following upper bound on the latent heat
\begin{align} \label{eq:latent_heat_weak}
    0 \leq L_\text{weak} \ll g^2 T^4
    .
\end{align}
This bound is the key conclusion of this section. In what follows, we will see how to go beyond it.

While we have focused on the $\mathrm{SU}(2)$ Higgs model, many of the conclusions carry over to other models. The key constraint is that the background-field dependence of masses remains much smaller than the soft scale, i.e.\ Eq.~\eqref{eq:v3_ultrasoft_break_down}. Given this, the upper bound in Eq.~\eqref{eq:latent_heat_weak} follows.

\subsection{Derivative and field expansions}
\label{sec:derivative_and_field_expansions}

The ultrasoft-scale EFT of Sec.~\ref{sec:ultrasoft_transitions} is constructed
as a dual expansion in derivatives and powers of fields.
As
the magnitude of the background field $v$ grows, eventually 
Eq.~\eqref{eq:v3_ultrasoft_break_down} no longer holds,
leading to a breakdown in the expansion of powers of fields as in Eq.~\eqref{eq:Lultra_cn}. However, as we argue below, this does not prevent us from constructing a new EFT that resums the expansion in powers of fields, but retains the derivative expansion.

The Euclidean effective Lagrangian for a general perturbative EFT of a scalar field $\phi$ can be written in the form~\cite{Burgess:2020tbq}
\begin{align} \label{eq:general_eft_lagrangian}
    \mathcal{L}_\text{EFT} = \frac{f^4}{M^2 u^2}\bigg[&(\partial_\mu\phi)^2 + m^2\phi^2
    \nonumber\\
    &+ \sum_n \frac{\hat{c}_n}{M^{d_n}u^{f_n}} \mathcal{O}_n(\phi)\bigg] ,
\end{align}
where $\hat{c}_n$ is dimensionless, and each operator $\mathcal{O}_n$ contains $d_n$ derivatives and $f_n$ powers of the field, e.g.\ $\mathcal{O}_n=\phi^2\partial^6\phi^2$ has $d_{n}=6$ and $f_{n}=4$.
It is assumed that the mass $m$ and momentum $p \sim m$ of the field $\phi$ are small compared to the UV scales
$m, p\ll f, M, u$.
An $E$-point, $L$-loop, amplitude for the scattering of $\phi$-particles
with $V_n$ vertices of each operator $\mathcal{O}_n$
is then of order~\cite{Manohar:1983md,
Burgess:2003jk, Gavela:2016bzc, Burgess:2020tbq}
\begin{align} \label{eq:general_eft_amplitude}
    \mathcal{A}_E(p) \sim
    f^4
    \left(\frac{1}{u}\right)^E
    \left(\frac{M p}{4\pi f^2}\right)^{2 L}
    \left(\frac{p}{M}\right)^{2 + \sum_n(d_n-2)V_n}
    .
\end{align}
The three scales $f$, $M$ and $u$ give distinct contributions to the amplitude. This justifies their introduction as logically separate scales in Eq.~\eqref{eq:general_eft_lagrangian}, and implies that the expansion in powers of derivatives is distinct from the expansion in powers of fields.

The EFT is local iff the expansion in $\partial/M$ can be truncated. Similarly,
the EFT has a polynomial potential iff the expansion in $\phi/u$ can be
truncated. Both expansions could be truncated for the weak transition EFT in
Sec.~\ref{sec:ultrasoft_transitions}. But, as one considers stronger
transitions, at some point the expansion in $\phi/u$ fails as in Eq.~\eqref{eq:v3_ultrasoft_break_down}, while the expansion
in $\partial/M$ continues to hold. Similar
situations arise in Chiral EFT~\cite{Weinberg:1968de,
Weinberg:1978kz}, the Higgs EFT~\cite{Brivio:2025yrr}, and
inflationary EFTs~\cite{Weinberg:2008hq, Burgess:2017ytm}.

\subsection{Intermediate strength transitions at the soft ($gT$) and supersoft ($g^{\frac{3}{2}}T/\sqrt{\pi}$) scales} \label{sec:soft_supersoft_transitions}
We now consider transitions for which the characteristic background is $v=O(T)$, the point at which EFTs for weak transitions break down, as in Eq.~\eqref{eq:v3_ultrasoft_break_down}.
At the phase transition, thermal corrections must still modify the potential at
leading order, as encapsulated in Eq.~\eqref{eq:potential_changes_O1}. The barrier
between phases can already be present at tree level in a soft-scale EFT, or can
be generated by soft fluctuations when the transitioning field is supersoft. As
discussed in Sec.~\ref{sec:derivative_and_field_expansions}, the latter
construction will require retaining the full background-field dependence of
soft-scale loop integrals while expanding in derivatives.
In what follows, we illustrate these different EFTs for intermediate strength transitions through concrete examples.
We keep the field
normalisation of Sec.~\ref{sec:ultrasoft_transitions}: the EFT action carries an
overall factor of $\beta$, and scalar fields have mass dimension one.
\\

\paragraph{Tree-level barriers at the soft scale.}
In this case, the soft-scale EFT, constructed through high-temperature dimensional reduction, has (at least) two phases separated by a barrier already at tree-level. This typically occurs for transitions which do not involve symmetry breaking.

Consider, for example, a real scalar $\phi$ coupled via a Yukawa interaction to a Dirac fermion $\psi$, with
background-dependent fermion mass
$m_\psi=\mu_{\psi}+y\phi$~\cite{Gould:2021dzl, Gould:2023jbz}.
Fermions have antiperiodic boundary conditions on the thermal imaginary time circle, so that they have only odd Matsubara modes $(2n+1)\pi T$ and no zero mode~\cite{Laine:2016hma}. Therefore, all fermion modes
are integrated out together with $\phi_{\rm
UV}$.
The validity of high-temperature dimensional reduction requires $m_\psi \ll \pi T$.
The resulting soft-scale dimensionally reduced EFT has potential
\begin{align}\label{eq:bridge_soft_potential}
    V_{\rm soft}(\phi_{\rm IR})=\sigma_3 \phi_{\rm IR}
    +\frac{1}{2}m^2_3\phi_{\rm IR}^2
    +\frac{\kappa}{3!}\phi_{\rm IR}^3 +\frac{\lambda}{4!}\phi_{\rm IR}^4,
\end{align}
where at leading order
\begin{align} \label{eq:sigma3_yukawa}
    \sigma_3&=
    \left(\frac{\kappa}{24} +\frac{y\mu_{\psi}}{6}\right)T^2,\\
    m_3^2&=\mu_{\phi}^2+\left(\frac{\lambda}{24} +\frac{y^2}{6}\right)T^2.\label{eq:msq3_yukawa}
\end{align}
The vacuum quadratic term is $\mu_{\phi}^2\phi^2/2$, and we have shifted the scalar field to remove a possible vacuum linear term~\cite{Gould:2021dzl}. The cubic and quartic scalar couplings are unchanged by thermal screening at this order.

We adopt the default assumption that different interactions are of comparable strength in vacuum, and hence $\lambda=O(y^2)$ and $\kappa^2/\mu_\phi^2=O(y^2)$.
To fix power-counting relations in terms of the temperature, we demand that the thermal corrections to the potential are of the same order as the vacuum tree-level potential, Eq.~\eqref{eq:potential_changes_O1}, finding
\begin{align}
    \mu_{\phi}^2=O(y^2T^2), \qquad \mu_{\psi}=O(yT).
\end{align}
The former agrees with Eq.~\eqref{eq:hierarchy_of_scales}, as expected. Inserting these relations into the soft-scale potential, and barring parametric cancellations, we find
\begin{align}
    \frac{\sigma_3}{T^3},\; \frac{m_{3}^2}{T^2},\; \frac{\kappa}{T},\; \lambda = O(y^2),
\end{align}
so that all four terms in the potential are balanced when $\phi_{\rm IR}=O(T)$.
The two minima and barrier between are already present in the soft scale EFT at leading order. Soft scalar loops give subleading corrections to the potential, suppressed by $O(y)$~\cite{Gould:2021dzl}. The hierarchy is summarised in Table \ref{tab:hierarchy_tree_level}. Note that $y^2$ here plays the role that $g^2$ did in the $\mathrm{SU}(2)$ Higgs model, as the (only and hence largest) coupling between the order parameter and another field.
\begin{table}[H]
    \centering
    \begin{tabular}{c|c}
         $m_{\rm eff}$ & Fields \\
        \hline
         $\pi T$         & $\phi_{\rm UV}$, $\psi$  \\
         $y T$           & $\phi_\text{IR}$ \\
    \end{tabular}
    \caption{Scale hierarchy for a tree level barrier in the Yukawa model. The scalar zero mode is soft, while all fermionic Matsubara modes are hard.}
    \label{tab:hierarchy_tree_level}
\end{table}

\paragraph{2-step model at the soft scale.}
A tree level barrier can also arise between phases aligned along different scalar field directions. Consider two real scalars $\phi$ and $\xi$, each with a $\mathrm{Z}_2$ symmetry, and coupled with a positive portal coupling $g^2$.
After integrating out the hard modes, through dimensional reduction, the leading-order potential is 
\begin{equation}
\begin{aligned}
    V_{\rm soft}(\phi_{\rm IR},\xi_{\rm IR})=&\frac{1}{2}m_{\phi,3}^2\phi_{\rm IR}^2 
    +\frac{1}{2} m_{\xi,3}^2\xi_{\rm IR}^2\\
    &+\frac{\lambda_{\phi}}{4!}\phi^4_{\rm IR}
    +\frac{\lambda_{\xi}}{4!}\xi^4_{\rm IR}
    +\frac{1}{4}g^2\phi^4_{\rm IR}\xi^4_{\rm IR},
\end{aligned}
\end{equation}
where we have assumed that all couplings are comparable: $\lambda_{\phi}$, $\lambda_\xi=O(g^2)$.
The leading-order thermal masses are
\begin{align}
    m_{\phi,3}^2&=
    -\mu_{\phi}^2
    +\frac{\lambda_{\phi }+g^2}{24}T^2,\\
    m_{\xi,3}^2&=
    +\mu_{\xi}^2
    +\frac{\lambda_{\xi }+g^2}{24}T^2 ,
\end{align}
here the vacuum quadratic terms are $-\mu_{\phi }^2\phi^2/2$ and $+\mu_{\xi }^2\xi^2/2$.

To determine appropriate power-counting relations, we require
that thermal effects compete with vacuum masses, Eq.~\eqref{eq:potential_changes_O1}, giving 
\begin{align}
    \mu_{\phi}^2,\; \mu_{\xi}^2=O(g^2T^2).
\end{align}
Nontrivial extrema of the potential occur when the quadratic and quartic terms balance in magnitude, at which point we find
\begin{align}
    \quad \phi_{\rm IR},\; \xi_{\rm IR}=O(T).
\end{align}

For suitable parameters, cooling can first take the system from the origin to a minimum along the $\xi_{\rm IR}$ direction, and then to a minimum along the $\phi_{\rm IR}$ direction. We focus on the second step, where two broken minima compete. For negative quadratic coefficients, the nonzero field values satisfy 
\begin{align}
    (\phi_{\rm IR}^{\rm min})^2=-\frac{6m_{\phi,3}^2}{\lambda_{\phi}},
    \qquad (\xi_{\rm IR}^{\rm min })^2=-\frac{6m_{\xi,3}^2}{\lambda_{\xi}}
\end{align}
with the other field set to zero in each case. At tree level they are locally stable in both scalar directions and degenerate (at $T_\text{c}$) when 
\begin{align}
    \frac{(m_{\phi,3}^2)^2}{\lambda_{\phi}}=\frac{(m_{\xi,3}^2)^2}{\lambda_{\xi}}.
\end{align}
The portal interaction then produces a barrier between the two different broken minima.
Both scalar masses are soft, $O(g^2T^2)$, and hence must be retained, there
being no parametric hierarchy that would justify integrating one of them out.
Again, IR scalar loops are subdominant. This setup is summarised in Table \ref{tab:hierarchy_2_step}.
\begin{table}[H]
    \centering
    \begin{tabular}{c|c}
         $m_{\rm eff}$ & Fields \\
        \hline
         $\pi T$         & $\phi_{\rm UV}$, $\xi_{\rm UV}$  \\
         $g T$           & $\phi_\text{IR}$,  $\xi_{\rm IR}$ \\
    \end{tabular}
    \caption{Power counting hierarchy for the tree level second step model. Both scalar zero modes remain in the soft scale EFT.}
    \label{tab:hierarchy_2_step}
\end{table}

\paragraph{Radiative barriers at the supersoft scale.}

We now return to the SU(2) Higgs model of
Sec.~\ref{sec:ultrasoft_transitions}.
Given that
integrating out the temporal gauge field $A_0$
failed for background
fields as large as $v=O(T)$, we must
return to the first soft-scale EFT~\eqref{eq:electroweak_3d_lagrangian_soft}, still
containing the field $A_0$. In this case, the scalar potential
reads
\begin{align} \label{eq:soft_tree_level_potential}
    V_{\rm soft}(v)=\frac{1}{2}m_{\rm 3}^2v^2+\frac{\lambda}{4!}v^4.
\end{align}
Under the assumption of an intermediate-strength transition, $v=O(T)$, the
Higgs mechanism induces a soft-scale mass $\sim gT$ for all four components of the gauge
field.

At high enough temperatures, the Higgs is also at the soft scale, and hence
there are no heavier fields to significantly modify its potential. Then, as the
temperature decreases, so does the Higgs mass until $T_0$ defined in
Eq.~\eqref{eq:T0}, where the Higgs mass goes through zero. On the approach to
this temperature, the Higgs field becomes much lighter than the gauge field $m_3\ll gT$.
Such a scale hierarchy will give large corrections to the Higgs effective action.

For $v=O(T)$, all
components of the gauge field have background-field-dependent masses of $O(gT)$
and hence
should be integrated out to describe the parametrically lighter Higgs.
In so doing, it is not possible to expand the result in powers of $v$ because the
$v$-dependence of the gauge field masses is of leading order.

The derivative expansion for this light Higgs EFT does still hold, because we are interested in describing Higgs-field
configurations with momenta $p \sim m_3 \ll g T$, in line with Eq.~\eqref{eq:general_eft_amplitude}.
Therefore, the leading contribution from the gauge fields to the Higgs
effective action is through the potential. At one-loop this comes through terms
of the form
\begin{align}
    \frac{T}{2}\int_p\ln[p^2+m_{A}^2(v)]=-\frac{T}{12\pi}\left[m_A^2(v)\right]^{\frac{3}{2}},
\end{align}
where $m_A(v)$ is a background-field dependent mass of one component of the gauge field. In
total, the resulting EFT potential is
\begin{align}
    V_{\rm super.}(v)&=
    \frac{1}{2}m_{\rm 3}^2v^2
    + \frac{\lambda}{4!}v^4 \nonumber \\
    &- \frac{T}{2\pi}\left(\frac{g^2 v^2}{4}\right)^{\frac{3}{2}}
    - \frac{T}{4\pi}\left(m_{\rm D}^2+\frac{g^2 v^2}{4}\right)^{\frac{3}{2}},
\end{align}
where the terms on the second line are due to $A_i$ and $A_0$ fluctuations respectively.
For further discussion of this EFT, see
Refs.~\cite{Arnold:1992rz, Farakos:1994kx, Ekstedt:2020abj, Ekstedt:2022zro, Gould:2023ovu}.

This potential shows coexisting minima, and a first-order phase transition. At the
transition, the different terms in the potential must be of the same order,
which fixes the Higgs-mass scale,
\begin{align}
    \frac{\frac{T}{12\pi}\left[m_A^2(v)\right]^{\frac32}}{m_{3}^2v^2}=O(1)\,\,\Rightarrow\,\, m^2_{3}=O(g^3T^2).
\end{align}
Further, balancing the quadratic and quartic terms at $v=O(T)$ then fixes the
ratio of couplings,
\begin{align}
    \frac{m_{3}^2 v^2}{\lambda v^4}=O(1) \quad \Rightarrow \quad \lambda=O(g^3).
\end{align}
The Higgs lives at the \textit{supersoft scale}, with mass of order $m_{3}=O(g^{3/2}T)$.
As in the weak-transition case, there are two hierarchies of scales, and
correspondingly two layers of EFTs.
Here, however, the spatial gauge fields are also integrated out.
The truncation in powers of fields is controlled by $v/T$, which is not
small and therefore must be resummed, while
the derivative expansion for momenta $p\sim m_3$ is controlled by
$p^2/m_A^2(v)=O(g/\pi)$, and therefore remains under control.

The hierarchies are summarised in Table \ref{tab:hierarchy_intermediate}.
\begin{table}[H]
    \centering
    \begin{tabular}{c|c}
         $m_{\rm eff}$ & Fields \\
        \hline
         $\pi T$     & $H_{\rm UV}$, $A_{0}^{\rm UV},A_{i}^{\rm UV}$  \\
         $g T$       & $A_{0}^\text{IR},A_{i}^\text{IR}$ \\
         $g^{\frac{3}{2}}T/\sqrt{\pi}$        & $H_{\rm IR}$
    \end{tabular}
    \caption{Scale hierarchies for the radiative barrier case in the $SU(2)$ Higgs model. The Higgs is supersoft, while the gauge fields are soft.}
    \label{tab:hierarchy_intermediate}
\end{table}

\paragraph{Transition strength and the limit of the expansion.}
The latent-heat power counting follows from
Eq.~\eqref{eq:latent_heat_for_counting}, as in Sec.~\ref{sec:ultrasoft_transitions}.
The temperature-dependent mass of the order parameter generically satisfies, for
all the different models,
\begin{align} \label{eq:dmeff_dlogT}
    \frac{dm_{\rm eff}^2}{d\log T}=O(g^2T^2),
\end{align}
or with $y$ switched for $g$ in the Yukawa model.
So, for the characteristic intermediate-strength background field $v=O(T)$, the
latent heat is
\begin{align}
    L_{\rm Intermediate}=O(g^2T^4).
\end{align}
This applies both to a soft-scale tree-level barrier and to the supersoft
radiative barrier cases above. 

All the intermediate-strength EFTs rely on the validity of high-temperature dimensional reduction.
Thus, all the background-field dependent masses of all relevant fields must remain below
the hard scale.
On the other hand, if the background-field dependent masses are too small, then
the $A_0$ field gives important thermal corrections to the $A_i$ field, which
are missed when all components are integrated out together.
Parametrically, these considerations require the background field to remain between 
\begin{align}\label{eq:v_bound_intermediate}
   O\left(gT\right) \ll v \ll O\left(\frac{T}{g}\right).
\end{align}
Towards the lower end, the transition is weak and
the EFT of Sec.~\ref{sec:ultrasoft_transitions} becomes more
appropriate.
Towards the upper end, $m_\xi(v)\sim gv$ approaches the hard scale $\pi T$.
The transition becomes too strong for the validity of the high-temperature expansion.
We show how to overcome this in Sec.~\ref{sec:strong_transitions}, constructing EFTs
for such strong transitions.

Combining Eqs.~\eqref{eq:latent_heat_for_counting} and \eqref{eq:eta_y}, this corresponds to the broad range
\begin{equation} \label{eq:latent_heat_intermediate}
    O(g^4T^4) \ll L_\text{intermediate} \ll O( T^4) ,
\end{equation}
with $L_\text{intermediate}=O(g^2T^4)$
at the natural midpoint of the power counting.

\subsection{Mapping EFTs onto phase diagrams}
\label{sec:phase_diagrams}

Any given UV model will have a phase diagram as a function of its parameters, and the temperature. Different regions of the phase diagram may have first-order, second-order or crossover transitions.

\begin{figure}[t]
    \centering\hspace{-0.5 cm}
    \includegraphics[width=0.95\linewidth]{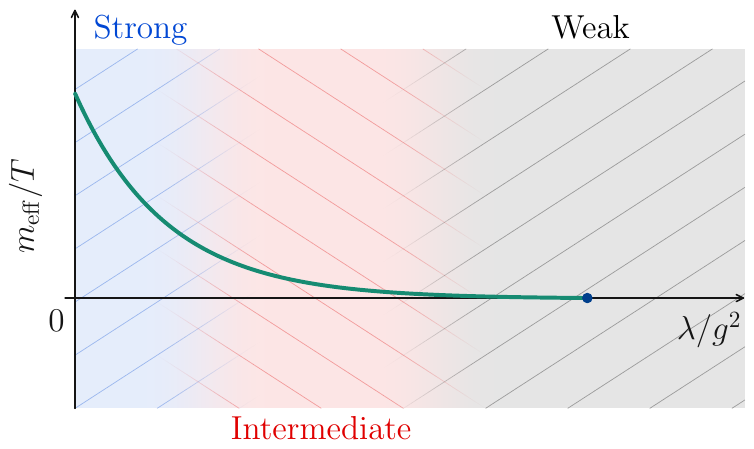}
    \caption{
    A cartoon of the phase diagram of the electroweak model, showing a line of first-order phase transitions ending at a critical point~\cite{Kajantie:1996mn, Laine:1998jb, Gould:2022ran}. The horizontal axis $\lambda/g^2$ is the ratio of the Higgs self-coupling to the gauge coupling, while the vertical axis is the ratio between the thermally corrected Higgs effective mass and the temperature. The coloured regions labelled `Weak', `Intermediate' and `Strong' show the overlapping regions of validity of the corresponding EFTs given in Secs.~\ref{sec:ultrasoft_transitions}, \ref{sec:soft_supersoft_transitions} and \ref{sec:strong_transitions} respectively.}
    \label{fig:phase_diagram}
\end{figure}

As a concrete example, consider the phase diagram of the electroweak model, shown in Fig.~\ref{fig:phase_diagram}. This shows a line of first-order phase transitions ending in a critical point, where there is a second-order phase transition, and for larger $\lambda/g^2$ there are only crossovers. At the critical point the effective Higgs mass goes to zero.%
\footnote{Where, in fact, another EFT is valid, describing the macroscopic critical modes in the Ising universality class~\cite{Rummukainen:1998as}.}
Nearby, the Higgs effective mass is not heavier than the ultrasoft gauge bosons, so that weak phase transition EFT of Sec.~\ref{sec:ultrasoft_transitions} is the correct description. Moving to smaller $\lambda/g^2$, the Higgs field becomes heavier at the transition, moving from the ultrasoft to the supersoft scale, described by the intermediate strength phase transition EFT of Sec.~\ref{sec:soft_supersoft_transitions}. Finally, at smaller $\lambda/g^2$ still, the Higgs effective mass at the transition moves from the supersoft to the soft scale, described by the strong phase transition EFT of Sec.~\ref{sec:strong_transitions}.

Importantly, neighbouring EFTs on the phase diagram have overlapping regimes of validity, as summarised in Eqs.~\eqref{eq:latent_heat_weak}, \eqref{eq:latent_heat_intermediate} and \eqref{eq:latent_heat_strong}. This allows predictions to be cross checked. In the overlap region, there will be a change in the dominant balance of different terms in the effective potential. Fixing the EFT and moving towards the edge of its regime of validity, higher dimensional operators become increasingly important~\cite{Kajantie:1995dw, Andersen:2017ika, Gorda:2018hvi, Chala:2024xll, Bernardo:2025vkz}. By choosing appropriate power-counting prescriptions, it is possible to define additional EFTs that are targeted to a given overlap region~\cite{Gould:2023ovu, Camargo-Molina:2024sde}.

\section{EFTs for strong transitions}\label{sec:strong_transitions}

We now turn to stronger transitions, for which
the scalar background becomes parametrically larger than the temperature, 
$v\sim \pi T/g$, pushing another field's mass $m\sim gv$ to the hard scale.
Together with the temperature dependence of the effective mass, Eq.~\eqref{eq:dmeff_dlogT}, this enhances the latent heat to $L=O(T^4)$. More generally, the full range of validity of
the strong transition EFT is
\begin{equation}\label{eq:latent_heat_strong}
    O(g^2T^4) \ll L_{\rm strong} \leq O(T^4) .
\end{equation}
The range overlaps with the intermediate-strength EFT, but at $O(g^2T^4)$ the
strong phase transition EFT breaks down because missing thermal
resummations become necessary.

We demonstrate how to construct a perturbative expansion for this regime using a simple two-scalar theory as a worked example.
We construct the hard-scale effective action
through LO and NLO, and include the first infrared correction at N$^2$LO. We then test the expansion through UV and IR finiteness, renormalisation-scale independence, order-by-order convergence of the latent heat, and verify the hierarchy of masses at the transition. 

\subsection{Two scalar model}
We begin constructing and testing the strong transition power counting in a simple theory of two real scalar fields, $\phi$ and $\xi$, with Euclidean Lagrangian
\begin{equation}
\begin{aligned}
\mathcal{L}_\text{E}
=&
\frac12(\partial_\mu\phi)^2
+ \frac12(\partial_\mu\xi)^2
- \frac{1}{2}\mu_\phi^2\phi^2
+ \frac12\mu_\xi^2\xi^2\\
&
+ \frac{\lambda_\phi}{4!}\phi^4
+ \frac{\lambda_\xi}{4!}\xi^4+\frac{1}{4} g^2 \phi^2 \xi^2, 
\end{aligned}
\end{equation}
here the field $\phi$ will play the role of the order parameter for the phase
transition, while $\xi$ provides the fluctuations that generate the leading
thermal barrier. The background field dependent masses are
\begin{equation}
\begin{aligned}
    \label{eq:strong_field_dependent_effective_masses}
m_\phi^{2}(\phi)&=-\mu_\phi^{2}+\frac12 \lambda_\phi \phi^2,\\
m_\xi^{2}(\phi)&=+\mu_\xi^{2}+\frac12 g^2 \phi^2.
\end{aligned}
\end{equation}

To discover the appropriate power-counting, we require that thermal effects
significantly modify the potential, and that different potential terms balance at
the transition, just as for the EFTs describing weaker transitions. These are combined with the specific strong-transition condition that the scalar background field pushes another field to the hard scale.

In this model, strong phase transitions occur when the background field pushes $m_\xi(\phi)$
to become of order the temperature, i.e.\
\begin{align} \label{eq:m_xi_order_T}
    g^2 \phi^2 = O\left(T^2\right).
\end{align}
When Eq.~\eqref{eq:m_xi_order_T} holds, the requirement that thermal
corrections modify the potential at leading order becomes
\begin{align} \label{eq:hierarchy_of_scales_2_repeat}
    \frac{m_\phi^4}{T^4} = O\left( \lambda_\phi \right),
\end{align}
repeating Eq.~\eqref{eq:hierarchy_of_scales_2} of
Sec.~\ref{sec:scale_hierarchies}.

Further, at the transition, there should be a balance between different
terms in the leading-order potential,
\begin{align} \label{eq:potential_balance_strong}
    \mu_\phi^2\phi^2 \sim \lambda_\phi \phi^4
    \quad \implies \quad
    \phi^2\sim \frac{\mu _\phi ^2}{\lambda_\phi}.
\end{align}
Together Eqs.~\eqref{eq:m_xi_order_T}, \eqref{eq:hierarchy_of_scales_2_repeat} and \eqref{eq:potential_balance_strong} imply
\begin{align} \label{eq:strong_coupling_hierarchy}
    \lambda_\phi &= O(g^4).
\end{align}
This is an important conclusion. Strong phase transitions require a large hierarchy of
couplings. The beta function of
$\lambda_\phi$ contains a term $O(g^4)$, so that this is the largest hierarchy of
couplings that can be maintained under renormalisation group running.

Starting from the assumption of a strong phase transition, we derived
Eq.~\eqref{eq:strong_coupling_hierarchy}, a condition on the parameter space of the model. Let's now reverse the argument,
and focus on the following region of parameter space
\begin{equation}
\begin{aligned}
     \label{eq:strong_two_scalar_power_counting}
    \{\lambda_\phi,\lambda_\xi, g^2\} 
    &\rightarrow 
    \{\alpha^4\lambda_\phi, \alpha^2\lambda_\xi, \alpha^2g^2\},\\
    \{\mu_\phi^2, \mu_\xi^2, T\} 
    &\rightarrow 
    \{\alpha^2\mu_\phi^2,\mu^2_\xi, T\},
\end{aligned}
\end{equation}
where $\alpha\to 0_+$ is a power-counting parameter.

The scalings with $\alpha$ of $\lambda_\phi$ and $\mu_\phi$ are key for
there to be a strong transition.
The other two parameters are less constrained: the construction that
follows is largely unmodified for the range $\mu_\xi^2=O(\alpha^a T^2$) with $a\geq0$ and $\lambda_\xi=O(\alpha^b)$
with $b\geq 2$. For $a>0$, the mass of the $\xi$ field shifts parametrically
between phases, being lighter in the symmetric phase.
However, its dominant effects arise at larger field values where it is
heavier, with $m_\xi(\phi)=O(T)$~\cite{Gould:2021ccf}.

The choice of power-counting is motivated physically, in this case by our
interest in strong phase transitions. However, once it is chosen, the
perturbative expansion is uniquely fixed~\cite{Bender:1999box, Lofgren:2023sep}.
The power counting, rather than the number of loops, determines which
contributions enter at each order.

Starting from Eq.~\eqref{eq:strong_two_scalar_power_counting} and repeating the arguments of
Sec.~\ref{sec:scale_hierarchies}, one finds that they imply
Eqs.~\eqref{eq:m_xi_order_T}, \eqref{eq:hierarchy_of_scales_2_repeat}
and \eqref{eq:potential_balance_strong} at the transition.
The small self-coupling $\lambda_\phi$ forces the minimum out to large field values,
\begin{align}
    \phi = O\left(\frac{T}{\alpha}\right).
\end{align}

This is what drives the whole construction, because the two fields respond very
differently to such a background.
The field-dependent masses, evaluated at the minimum, are of order
\begin{align} \label{eq:strong_masses_order}
    m_{\phi}^2=O(\alpha^2 T^2), \qquad m_\xi^2=O(\alpha^0 T^2) .
\end{align}
The mass of the order parameter remains soft, as it must, while the mass of
$\xi$ is hard, of order the temperature. The large background
field compensates for the small coupling $g$. For our choice of power-counting relations,
Eq.~\eqref{eq:strong_two_scalar_power_counting}, the additive terms in
the field-dependent masses~\eqref{eq:strong_field_dependent_effective_masses} are of the same order,
but this can be relaxed as long as Eq.~\eqref{eq:strong_masses_order} holds.
Table \ref{tab:hierarchy2scalar} summarises the strong transition scale
hierarchy,
contrasted with that of the intermediate strength and weak regimes of Sec.~\ref{sec:weak_transitions}. 

Given that the $\phi$ field is light compared with the temperature, we can
separate its modes by energy scale,
\begin{align}
    \phi= \phi_{\rm IR}+\phi_{\rm UV},
\end{align}
where $\phi_{\rm IR}$ denotes the slowly varying, long-wavelength part of the field, which is the background field retained in the EFT. Whereas $\phi_{\rm UV}$ denotes the hard fluctuations, whose momenta are of order $\pi T$ and hence integrated out. 
These include the nonzero Matsubara modes~\eqref{eq:matsubara} and the high-momentum part of the zero Matsubara mode.

In terms of power counting, the shape of the potential pushes the IR field to large field values, $\phi_\text{IR}=O(T/\alpha)$, while the UV field describes local fluctuations which are not so enhanced, $\phi_\text{UV},\xi=O(T)$. The latter follows because the action for a typical fluctuation is $O(1)$, but it can also be verified post-hoc, by noting that the free energy of UV fluctuations is $O(T^4)$.

\begin{table}[t]
    \centering
    \begin{tabular}{c|c|c|c}
        $m_{\rm eff}$ & \textbf{Strong} & \textbf{Intermediate} & \textbf{Weak} \\
        \hline
        $\pi T$         & $\xi, \phi_{\rm UV}$ & $\xi_{\rm UV}, \phi_{UV} $ & $\xi_{\rm UV}, \phi_{UV} $ \\
        $g T$           &$\phi_{\rm IR}$  & $\xi_{\rm IR}$ & $\xi_{\rm IR}$ \\
        $g^{\frac{3}{2}}T/\sqrt{\pi}$       &  &$ \phi_{\rm IR}$ & \\
        $g^{2} T/\pi$       &  & & $\phi_{\rm IR}$  
    \end{tabular}
    \caption{Scale hierarchies for the strong, intermediate and weak strength
    \label{tab:hierarchy2scalar}
    $\mathrm{Z}_2$ symmetry-breaking transitions in the two-scalar model.
    }
\end{table}
With the strong-transition power counting, Eq.~\eqref{eq:latent_heat_for_counting} gives for the latent heat
\begin{align}
L = O\left(
\frac{d m_{\phi,\rm eff}^2}{d\log T}\,
\phi_{\rm IR}^2
\right)
= O(T^4).
\end{align}
This corresponds parametrically to the strongest transitions accessible
within this mechanism.
Increasing the background field further also increases the mass of the field inducing the transition, $m_\xi \sim g\phi_{\rm IR}$. Once $m_\xi \gg \pi T$, its thermal effects become exponentially suppressed and can no longer produce an $O(1)$ modification of the potential as required by Eq.~\eqref{eq:potential_changes_O1}. Thus, without additional parametric fine tuning, the strong-transition regime is bounded from above at $L = O(T^4)$.

In the strong transition case the whole of $\xi$ lives at the hard scale and is integrated out in one step, together with $\phi_{\rm UV}$. It is useful to make this separation explicit directly at the level of the path integral, and so to construct integrate out the UV scale in the manner of Ref.~\cite{Hirvonen:2022jba}.
The EFT is
constructed as
\begin{align}
Z&=\int\mathcal D\phi_{\rm IR}\,e^{-S_{\rm EFT}[\phi_{\rm IR}]},\\
e^{-S_{\rm EFT}[\phi_{\rm IR}]}
&=e^{-S_{\rm IR}[\phi_{\rm IR}]}
\int\mathcal D\phi_{\rm UV}\,\mathcal D\xi\,
 e^{-S_{\rm UV}[\phi,\xi;\alpha]}. \label{eq:SEFT_definition}
\end{align}
We then organise $S_\text{UV}$ based on powers of $\alpha$.
Factoring out the powers implied by the strong-transition counting, we write
\begin{align}
S_{\rm UV}
=S_{\rm UV}^{(0)}+\alpha S_{\rm UV}^{(1)}
+\alpha^2S_{\rm UV}^{(2)}+\cdots .
\end{align}
Here $S_{\rm UV}^{(n)}$ denotes the coefficient of the corresponding explicit power of $\alpha$. Using dimensional regularisation and working strictly within this power-counting scheme ensures that only UV contributions to loop integrals are included~\cite{Braaten:1995cm, Hirvonen:2022jba}. This enacts an expansion by regions for loop integrals which include IR scales~\cite{Jantzen:2011nz}.

At zeroth order in $\alpha$, the Lagrangian for the hard fields is
\begin{align}\label{eq:UVLagrangian}
\mathcal{L}_{\rm UV}^{(0)}=\frac12(\partial_\mu\phi_{\rm UV})^2
+\frac12(\partial_\mu\xi)^2
+\frac12m_\xi^2(\phi_{\rm IR})\xi^2,
\end{align}
noting that $\partial_\mu \phi_\text{UV} \sim \pi T \phi_\text{UV}$, and similarly for $\xi$.
This is taken to be the free Lagrangian for UV modes.
Thus, the $\phi_{\rm UV}$ propagator is massless at leading order, while the full background-dependent $\xi$ mass is retained exactly,
\begin{align}
\wick{\c\xi\,\c\xi}
&=\frac{1}{P^2+m_\xi^2},\qquad
\wick{\c\phi\vphantom{\phi}_{\rm UV}\,\c\phi\vphantom{\phi}_{\rm UV}}
=\frac{1}{P^2}.
\end{align}
For the subleading terms it is more transparent to display the corresponding Lagrangian densities,
\begin{align} \label{eq:two_scalar_lagrangian_O1}
\mathcal{L}_{\rm UV}^{(1)}={}&
-\phi_{\rm UV}(\partial^2+\mu_\phi^2)\phi_{\rm IR} \nonumber\\
&+\phi_{\rm UV}\left(
\frac{\lambda_\phi}{3!}\phi_{\rm IR}^3
+\frac{g^2}{2}\phi_{\rm IR}\xi^2\right),\\
\mathcal{L}_{\rm UV}^{(2)}={}&
\frac12m_\phi^2(\phi_{\rm IR})\phi_{\rm UV}^2
+\frac{g^2}{4}\phi_{\rm UV}^2\xi^2
+\frac{\lambda_\xi}{4!}\xi^4,
\\
\mathcal{L}_{\rm UV}^{(3)}={}& \frac{\lambda_\phi}{3!}\phi_{\rm IR}\phi^3_{\rm UV},\\
\mathcal{L}_{\rm UV}^{(4)}={}& \frac{\lambda_\phi}{4!}\phi^4_{\rm UV},
\end{align}
where for simplicity we have not shown the counterterms explicitly.
The terms in $\mathcal{L}_{\rm UV}^{(1)}$ are odd in $\phi_{\rm UV}$, so a single insertion vanishes under Gaussian integration. Two insertions, however, contribute at NLO. 
The higher-order interactions $\mathcal{L}_{\rm UV}^{(3)}$ and $\mathcal{L}_{\rm
UV}^{(4)}$ contribute only beyond NLO and therefore are not needed here.
This
illustrates a key difference from the loop expansion: perturbative orders
are determined by powers of $\alpha$, not by the number of loops.
As we
show below, LO already contains one-loop diagrams, while NLO combines
one-loop insertions with a subset of two-loop diagrams. Other diagrams at
the same loop order are parametrically higher order in $\alpha$.

Writing $Z_{\rm UV}^{(0)}$ for the Gaussian path integral defined by $S_{\rm UV}^{(0)}$, and $\langle\cdots\rangle_0$ for the corresponding normalised Gaussian average, the expansion via correlation functions makes the power counting clear,
\begin{widetext}
\begin{align}
S_{\rm EFT}[\phi_{\rm IR}]=S_{\rm IR}[\phi_{\rm IR}]-\log Z_{\rm UV}^{(0)}[\phi_{\rm IR}]
+\alpha\underbrace{\langle S_{\rm UV}^{(1)}\rangle_0}_{\equiv0}+\alpha^2\left[\langle S_{\rm UV}^{(2)}\rangle_0
-\frac12\langle S_{\rm UV}^{(1)}S_{\rm UV}^{(1)}\rangle_{0
}\right]
+O(\alpha^4).
\end{align}
\end{widetext}
The above expansion determines the diagrams contributing at each order. In the following, orange lines denote massless $\phi_{\rm UV}$ propagators, crosses denote mass insertions of $m_\phi^2$, and circled crosses denote counterterm insertions.

Given the hierarchy of scales, the construction of the effective action can be carried out in a derivative expansion. Let us focus first on the potential, and return to derivative corrections later.
This is an expansion in integer powers of $\alpha^2$:
\begin{equation}
        V_{\rm EFT}=V_{\rm LO}+V_{\rm NLO} + O(\alpha^4 T^4),
\end{equation}
where the first two terms take the form:
\begin{equation}\label{eq:feyscalarLO}
    \begin{gathered}
\tikzset{every picture/.style={line width=0.75pt}} 
\begin{tikzpicture}[x=0.75pt,y=0.75pt,yscale=-1,xscale=1]

\draw  [color={rgb, 255:red, 245; green, 166; blue, 35 }  ,draw opacity=1 ] (140,30) .. controls (140,18.95) and (148.95,10) .. (160,10) .. controls (171.05,10) and (180,18.95) .. (180,30) .. controls (180,41.05) and (171.05,50) .. (160,50) .. controls (148.95,50) and (140,41.05) .. (140,30) -- cycle ;
\draw  [dash pattern={on 4.5pt off 4.5pt}] (200,30) .. controls (200,18.95) and (208.95,10) .. (220,10) .. controls (231.05,10) and (240,18.95) .. (240,30) .. controls (240,41.05) and (231.05,50) .. (220,50) .. controls (208.95,50) and (200,41.05) .. (200,30) -- cycle ;

\draw (69,29) node    {$V_{\text{LO}} =V_{\text{tree}}+V_{\rm LO}^{\rm ct}$};
\draw (189.68,29.2) node    {$+$};
\draw (130,29.2) node    {$+$};
\end{tikzpicture}
    \end{gathered}
\end{equation}
\begin{equation}\label{eq:feyscalarNLO}
\begin{gathered}
\tikzset{every picture/.style={line width=0.75pt}} 
\begin{tikzpicture}[x=0.75pt,y=0.75pt,yscale=-1,xscale=1]
\draw  [color={rgb, 255:red, 245; green, 166; blue, 35 }  ,draw opacity=1 ] (213.82,30) .. controls (213.82,18.95) and (222.77,10) .. (233.82,10) .. controls (244.87,10) and (253.82,18.95) .. (253.82,30) .. controls (253.82,41.05) and (244.87,50) .. (233.82,50) .. controls (222.77,50) and (213.82,41.05) .. (213.82,30) -- cycle ;
\draw  [dash pattern={on 4.5pt off 4.5pt}] (213.82,70) .. controls (213.82,58.95) and (222.77,50) .. (233.82,50) .. controls (244.87,50) and (253.82,58.95) .. (253.82,70) .. controls (253.82,81.05) and (244.87,90) .. (233.82,90) .. controls (222.77,90) and (213.82,81.05) .. (213.82,70) -- cycle ;
\draw  [dash pattern={on 4.5pt off 4.5pt}] (273.82,30) .. controls (273.82,18.95) and (282.77,10) .. (293.82,10) .. controls (304.87,10) and (313.82,18.95) .. (313.82,30) .. controls (313.82,41.05) and (304.87,50) .. (293.82,50) .. controls (282.77,50) and (273.82,41.05) .. (273.82,30) -- cycle ;
\draw  [dash pattern={on 4.5pt off 4.5pt}] (273.82,70) .. controls (273.82,58.95) and (282.77,50) .. (293.82,50) .. controls (304.87,50) and (313.82,58.95) .. (313.82,70) .. controls (313.82,81.05) and (304.87,90) .. (293.82,90) .. controls (282.77,90) and (273.82,81.05) .. (273.82,70) -- cycle ;
\draw  [color={rgb, 255:red, 245; green, 166; blue, 35 }  ,draw opacity=1 ] (156.82,50) .. controls (156.82,38.95) and (165.77,30) .. (176.82,30) .. controls (187.87,30) and (196.82,38.95) .. (196.82,50) .. controls (196.82,61.05) and (187.87,70) .. (176.82,70) .. controls (165.77,70) and (156.82,61.05) .. (156.82,50) -- cycle ;
\draw    (171.82,25) -- (181.82,35) ;
\draw    (171.82,35) -- (181.82,25) ;
\draw  [color={rgb, 255:red, 245; green, 166; blue, 35 }  ,draw opacity=1 ] (186.82,120) .. controls (186.82,108.95) and (195.77,100) .. (206.82,100) .. controls (217.87,100) and (226.82,108.95) .. (226.82,120) .. controls (226.82,131.05) and (217.87,140) .. (206.82,140) .. controls (195.77,140) and (186.82,131.05) .. (186.82,120) -- cycle ;
\draw  [fill={rgb, 255:red, 255; green, 255; blue, 255 }  ,fill opacity=1 ] (203.54,96.23) .. controls (205.62,94.41) and (208.78,94.64) .. (210.59,96.72) .. controls (212.41,98.8) and (212.18,101.96) .. (210.1,103.77) .. controls (208.02,105.58) and (204.86,105.36) .. (203.05,103.28) .. controls (201.24,101.2) and (201.46,98.04) .. (203.54,96.23) -- cycle ;
\draw [fill={rgb, 255:red, 255; green, 255; blue, 255 }  ,fill opacity=1 ]   (203.28,96.46) -- (210.36,103.54) ;
\draw [fill={rgb, 255:red, 255; green, 255; blue, 255 }  ,fill opacity=1 ]   (210.36,96.46) -- (203.28,103.54) ;

\draw  [dash pattern={on 4.5pt off 4.5pt}] (246.82,120) .. controls (246.82,108.95) and (255.77,100) .. (266.82,100) .. controls (277.87,100) and (286.82,108.95) .. (286.82,120) .. controls (286.82,131.05) and (277.87,140) .. (266.82,140) .. controls (255.77,140) and (246.82,131.05) .. (246.82,120) -- cycle ;
\draw  [fill={rgb, 255:red, 255; green, 255; blue, 255 }  ,fill opacity=1 ] (263.54,96.23) .. controls (265.62,94.41) and (268.78,94.64) .. (270.59,96.72) .. controls (272.41,98.8) and (272.18,101.96) .. (270.1,103.77) .. controls (268.02,105.59) and (264.86,105.36) .. (263.05,103.28) .. controls (261.24,101.2) and (261.46,98.04) .. (263.54,96.23) -- cycle ;
\draw [fill={rgb, 255:red, 255; green, 255; blue, 255 }  ,fill opacity=1 ]   (263.28,96.46) -- (270.36,103.54) ;
\draw [fill={rgb, 255:red, 255; green, 255; blue, 255 }  ,fill opacity=1 ]   (270.36,96.46) -- (263.28,103.54) ;

\draw  [dash pattern={on 4.5pt off 4.5pt}] (126.82,120) .. controls (126.82,108.95) and (135.77,100) .. (146.82,100) .. controls (157.87,100) and (166.82,108.95) .. (166.82,120) .. controls (166.82,131.05) and (157.87,140) .. (146.82,140) .. controls (135.77,140) and (126.82,131.05) .. (126.82,120) -- cycle ;
\draw [color={rgb, 255:red, 245; green, 166; blue, 35 }  ,draw opacity=1 ]   (126.82,120) -- (166.82,120) ;

\draw (207.5,49.2) node    {$+$};
\draw (263.82,48.5) node    {$+$};
\draw (117.5,119.2) node    {$+$};
\draw (236.82,118.5) node    {$+$};
\draw (177.5,119.2) node    {$+$};
\draw (105,49) node    {$V_{\text{NLO}} =V_{\text{NLO}}^{\text{ct}} +$};

\end{tikzpicture}
    \end{gathered}
\end{equation}
Here we have collected the $\alpha^0$ counterterms in $V_{\rm LO}^{\rm ct}$ and
$\alpha^2$ counterterms in $V_{\rm NLO}^{\rm ct}$, these counterterms are listed
in Appendix \ref{app:renormalisation}. Given that we are focusing on the potential at this point, loop diagrams are evaluated for constant background fields.

The term $\log Z_{\rm UV}^{(0)}$ generates the two LO determinants. At NLO,
$\alpha^2\langle S_{\rm UV}^{(2)}\rangle_0$ generates the $\phi_{\rm UV}$ mass
insertion, the mixed and $\xi$ figure-8 diagrams, and the counterterm
insertions. The remaining sunset arises from $\alpha^2\langle S_{\rm
UV}^{(1)}S_{\rm UV}^{(1)}\rangle_{0,c}$, through two insertions of the
$g^2\phi_{\rm IR}\phi_{\rm UV}\xi^2$ vertex. Thus the $\alpha$ power counting
naturally places one-loop insertions and genuine two-loop diagrams at the same
order.

Note that here the $\xi$ field does not have thermal corrections in its propagator. These are $O(\alpha^2T^2)$ and hence subleading, so should not be
resummed. The small effects of thermal screening on the $\xi$ field instead appear order-by-order in the $\alpha$ expansion.
This also highlights the limit of validity of the strong-transition EFT, which breaks down when thermal screening effects compete at the same order as the background-field dependence of $m_\xi$.
This happens when $v = O(T)$, at which point the correct framework is instead the intermediate-strength EFT discussed in Sec.~\ref{sec:soft_supersoft_transitions}.

Having identified which diagrams contribute at each order, the path integral construction
enforces that the propagators of light fields are expanded within the sum-integrand, i.e.\ the sum-integrals are automatically expanded by regions.
In the usual $\hbar$ expansion one first evaluates the one-loop determinant with the full $m_\phi$ dependence, and only afterwards expands the result at high temperature,
\begin{equation}
    \begin{gathered}
\tikzset{every picture/.style={line width=0.75pt}} 
\begin{tikzpicture}[x=0.75pt,y=0.75pt,yscale=-1,xscale=1]

\draw  [color={rgb, 255:red, 0; green, 0; blue, 0 }  ,draw opacity=1 ] (80,30) .. controls (80,18.95) and (88.95,10) .. (100,10) .. controls (111.05,10) and (120,18.95) .. (120,30) .. controls (120,41.05) and (111.05,50) .. (100,50) .. controls (88.95,50) and (80,41.05) .. (80,30) -- cycle ;

\draw (190,30) node    {$\displaystyle =\frac{1}{2}\sumint \log\left( P^{2} +m_{\phi }^{2}\right)$};
\draw (236,68.5) node   {$\displaystyle = -\frac{\pi^2 T^4}{90}
+\frac{m_\phi^2 T^2}{24}
-\frac{m_\phi^3 T}{12\pi}
+O\!\left(m_\phi^4\right),$};
\end{tikzpicture}
    \end{gathered}
\end{equation}
In the $\alpha$ expansion, or more generally for constructing an EFT, the order of operations is
reversed.
Since we are integrating out the hard $\phi$ modes using the expansion by
regions~\cite{Braaten:1995cm, Jantzen:2011nz, Hirvonen:2022jba}, the hard
$\phi_{\rm UV}$ propagator is taken as massless.
The massless determinant and subsequent mass insertions therefore appear separately,
\begin{equation}\hspace{+0.7cm}
    \begin{gathered}
\tikzset{every picture/.style={line width=0.75pt}} 
\begin{tikzpicture}[x=0.75pt,y=0.75pt,yscale=-1,xscale=1]

\draw  [color={rgb, 255:red, 245; green, 166; blue, 35 }  ,draw opacity=1 ] (80,30) .. controls (80,18.95) and (88.95,10) .. (100,10) .. controls (111.05,10) and (120,18.95) .. (120,30) .. controls (120,41.05) and (111.05,50) .. (100,50) .. controls (88.95,50) and (80,41.05) .. (80,30) -- cycle ;
\draw  [color={rgb, 255:red, 245; green, 166; blue, 35 }  ,draw opacity=1 ] (140,30) .. controls (140,18.95) and (148.95,10) .. (160,10) .. controls (171.05,10) and (180,18.95) .. (180,30) .. controls (180,41.05) and (171.05,50) .. (160,50) .. controls (148.95,50) and (140,41.05) .. (140,30) -- cycle ;
\draw    (155,5) -- (165,15) ;
\draw    (155,15) -- (165,5) ;

\draw (190.5,77) node    {$\displaystyle=\frac{1}{2}\sumint \log P^{2} +\frac{\alpha^2 }{2}\sumint \frac{m_{\phi }^{2}}{P^{2}} +O\left( \alpha ^{4} T^4\right)$};

\draw (164,120) node    {$\displaystyle=-\frac{\pi^2T^4}{90}+\frac{m_\phi^2 T^2}{24}+O(\alpha^4 T^4).$};
\draw (129.68,29.2) node    {$+$};
\draw (203.27,29.2) node    {$+\cdots $};
\end{tikzpicture}
    \end{gathered}
\end{equation}
Notably, only even powers of $m_\phi^2$ appear, and non-analytic terms
like $(m_\phi^2)^{3/2}$ are absent in this expansion.
To include them here would miss necessary resummations of the IR propagator, which are correctly accounted for only after constructing the EFT; see Eq.~\eqref{eq:VN2LO_two_scalar}.

For the hard field $\xi$, by contrast, the mass is of leading order, $m_\xi^2(\phi_{\rm IR})=O(\alpha^0T^2)$, and is therefore contained in the leading-order action $S_{\rm UV}^{(0)}$. Its propagator is hence not expanded, and the full background-field dependent determinant is retained:
\begin{equation}
    \begin{gathered}
\tikzset{every picture/.style={line width=0.75pt}} 
\begin{tikzpicture}[x=0.75pt,y=0.75pt,yscale=-1,xscale=1]
 
\draw  [color={rgb, 255:red, 0; green, 0; blue, 0 }  ,draw opacity=1 ][dash pattern={on 4.5pt off 4.5pt}] (80,30) .. controls (80,18.95) and (88.95,10) .. (100,10) .. controls (111.05,10) and (120,18.95) .. (120,30) .. controls (120,41.05) and (111.05,50) .. (100,50) .. controls (88.95,50) and (80,41.05) .. (80,30) -- cycle ;

\draw (190,30) node    {$\displaystyle =\frac{1}{2}\sumint \log\left( P^{2} +m_{\xi }^{2}\right)$};
\draw (242,68.5) node   {$\displaystyle = \frac{m_\xi^4}{64\pi^2\epsilon}+\frac{m^4_\xi}{64\pi^2}\left[ \ln\left(\frac{m_\xi^2}{\bar\mu^2}\right)-\frac{3}{2} \right]
    +\frac{T^4}{2\pi^2} \mathcal{J}\left(\frac{m_\xi}{T}\right)$};
\end{tikzpicture}
    \end{gathered}
\end{equation}
where $\bar\mu$ is the $\overline{\mathrm{MS}}$ renormalisation scale and $\mathcal J(x)$ is the bosonic thermal function defined in Eq.~\eqref{eq:app_Jthermal}.
Through $m_\xi(\phi_{\rm IR })$, the logarithm and the last term are non-polynomial functions of the background field, and it is the last term that generates a temperature-dependent phase transition.
Crucially, it does so at leading order: since $V_{\rm tree}=O(\alpha^0 T^4)$ and $T^4 \mathcal{J}=O(\alpha^0  T^4)$. 
That $\mathcal{J}$ is not expanded in $m_\xi$ is a key structural difference from the weaker-transition EFTs reviewed in Sec.~\ref{sec:weak_transitions}.

Once we have collected the terms needed up to NLO we can give the explicit potential of the EFT,
\begin{align}\hspace{-0.7 cm}
    V_{\rm LO}&=V_{\rm tree}+\frac12 J^0(0)+\frac12J^0(m_\xi),\label{eq:VLO2scalar}\\
    V_{\rm NLO}&=\frac12 m_\phi^2A^0(0)\nonumber\\
    &+\frac{g^2}{4}\left[ A^0(0)A^0(m_\xi )+A^1(0)A^{-1}(m_\xi)+A^1(m_\xi)A^{-1}(0)\right]\nonumber\\
    &+\frac{\lambda_\xi}{8}\left[ A^0(m_\xi)^2+2A^1(m_\xi)A^{-1}(m_\xi) \right]\nonumber \\
    &-\frac{g^4\phi_{\rm IR}^2}{4}H^0(m_\xi,m_\xi,0)\nonumber\\
    &+\frac{1}{2}\delta m_\phi^2\epsilon A^1(0)+\frac{1}{2}\delta m_\xi^2\epsilon A^1(m_\xi),\label{eq:VNLO2scalar}
\end{align}
where the sum-integrals $J^n$, $A^n$ and $H^n$ are defined in the Appendix \ref{app:sumintegrals}, and the diagram-by-diagram results, including their divergent parts, are given in Appendix \ref{app:2scalar}.
Every term in Eq.~\eqref{eq:VNLO2scalar} is suppressed by exactly $\alpha^2$ relative to \eqref{eq:VLO2scalar}.

All $1/\epsilon$ poles at both orders cancel exactly, including those with nontrivial temperature-dependent coefficients. This cancellation does not happen if one simply applies thermal resummation to all modes~\cite{Parwani:1991gq, Laine:2017hdk}.

An important consequence of this construction is that the hard-matched potential
is real order by order within the assumed scale hierarchy, provided
$m_\xi^2(\phi_{\rm IR})>0$ throughout the field range of interest. The
background-field dependent $\xi$ mass is retained in the propagator, whereas $m_\phi^2(\phi_{\rm IR})$
enters only through insertions into massless hard propagators. Hence $m_\phi^2(\phi_{\rm IR})$ only appears in positive integer powers. A negative tree-level curvature in the $\phi$
direction, which always occurs at the barrier between phases, therefore does not generate an imaginary part in the construction of
the EFT. No ad hoc prescription such as replacing $m_\phi^2$ by $|m_\phi^2|$, or taking
the real part of the potential, is required.
\\\\
A decisive test of the consistency of the $\alpha$ expansion is order-by-order renormalisation scale invariance, through the Callan--Symanzik equation.
Writing $\partial_t\equiv \partial/\partial \ln \bar\mu$ at fixed couplings, 
this reads
\begin{align}\label{eq:cancellationRG}
    \left(\partial_t +\sum_{\lambda_i}\beta_{\lambda_i}\frac{\partial}{\partial \lambda_i} + \gamma_\phi \phi_{\rm IR}\frac{\partial}{\partial\phi_{\rm IR}}\right)V_{\rm eff} &=0.
\end{align}
The beta functions are collected in Appendix \ref{app:renormalisation}.
The anomalous dimension $\gamma_\phi$ in this model is $O(\alpha^4)$, so does not contribute through NLO.
The beta functions themselves contain contributions at different orders in the strong-transition power counting. Consequently, expanding Eq.~\eqref{eq:cancellationRG} order by order in $\alpha$ relates the potential at different orders.
At leading order the explicit scale dependence may be found from Eq.~\eqref{eq:V_LO} to be
\begin{align}
    \partial_t V_{\rm LO}= -\frac{m_\xi^4}{2(4\pi)^2},
\end{align}
which indeed cancels the leading order running in
\begin{align}
    \alpha^0\!:\qquad
     0=\partial_t V_{\rm LO} + \frac{\partial V_{\rm LO}}{\partial \mu_\phi^2}\beta_{\mu_\phi^2}^{\rm LO} + \frac{\partial V_{\rm LO}}{\partial \lambda_\phi}\beta_{\lambda_\phi}^{\rm LO}  .
\end{align}
At next-to-leading order we have from Eq.~\eqref{eq:full2loopcompled}
\begin{align}
\partial_t V_{\rm NLO}=&
+\frac{g^4\phi_{\rm IR}^2m_\xi^2}{(4\pi)^4}
-\frac{g^4\phi_{\rm IR}^2T^2}{24 (4\pi)^2}
-\frac{g^2T^2m_\xi^2}{24(4\pi)^2}\nonumber\\
&-A^0(m_\xi)\left(
\frac{\lambda_\xi m_\xi^2}{2(4\pi)^2}
+\frac{g^4\phi_{\rm IR}^2}{(4\pi)^2}
\right), \label{eq:NLO_rg_cancellation_first}
\end{align}
this also cancels the leading order running in 
\begin{align}
\alpha^2\!:\qquad
0={}&\partial_t V_{\rm NLO}
+\frac{\partial V_{\rm LO}}{\partial\mu_\phi^2}
 \beta_{\mu_\phi^2}^{\rm NLO}
+\frac{\partial V_{\rm NLO}}{\partial\mu_\phi^2}
 \beta_{\mu_\phi^2}^{\rm LO}
\nonumber\\
&+\frac{\partial V_{\rm LO}}{\partial\lambda_\phi}
 \beta_{\lambda_\phi}^{\rm NLO}
+\frac{\partial V_{\rm NLO}}{\partial\lambda_\phi}
 \beta_{\lambda_\phi}^{\rm LO}
\nonumber\\
&+\frac{\partial V_{\rm LO}}{\partial\mu_\xi^2}
 \beta_{\mu_\xi^2}^{\rm LO}
+\frac{\partial V_{\rm LO}}{\partial g^2}
 \beta_{g^2}^{\rm LO}
 +\frac{\partial V_{\rm LO}}{\partial \lambda_\xi}
 \beta_{\lambda_\xi}^{\rm LO}.
 \label{eq:NLO_rg_cancellation_last}
\end{align}
The explicit $\bar\mu$ dependence therefore cancels against parameter running through NLO, providing a nontrivial consistency check of the $\alpha$ expansion.
\\\\
The corrections considered so far arise from integrating out the hard modes, with $\phi_{\rm IR}$ treated as a background field. The next correction in powers of $\alpha$ has a different origin. After the hard modes have been integrated out, $\phi_{\rm IR}$ remains as a dynamical light degree of freedom, and its fluctuations generate loops within the EFT.

In this context, we distinguish two different effective potentials:
the EFT potential $V_{\rm EFT}$, which makes up the non-derivative part of $S_\text{EFT}$ in Eq.~\eqref{eq:SEFT_definition} and hence only includes loops from UV fluctuations,
and the full effective potential $V_{\rm eff}$, which is the free energy density for given constant background field, and hence includes loops of both UV and IR fluctuations. The relation between the two potentials is shown by~\cite{Laine:2016hma}
\begin{align}
    Z &= \int \mathrm{d}\bar{\phi}_\text{IR}\int \mathcal{D}\phi'_\text{IR}\ e^{-S_{\rm EFT}[\phi_\text{IR}]} ,\\
    &= \int \mathrm{d}\bar{\phi}_\text{IR}\ e^{-V_{\rm eff}(\bar{\phi}_\text{IR})\mathcal{V}} ,
\end{align}
where $\bar{\phi}_\text{IR}$ is the constant field mode, $\phi'_\text{IR}$ are the nonconstant modes, and $\mathcal{V}$ is the spatial volume.

The effective mass of $\phi_\text{IR}$ is
\begin{align}
    m_{\rm eff}^2 =V''_{\rm EFT}\simeq\alpha^2m^2_{\phi ,\rm eff}|_{\rm LO}+\alpha^4 m^2_{\phi, \rm eff}|_{\rm NLO}+O(\alpha^6).
\end{align}
The leading loop contribution to the effective potential from $\phi_{\rm IR}$ is its (three-dimensional) one-loop determinant,
\begin{align}
    V_{\rm eff}&\supset \frac{T}{2} \int_{p} \log \left(p^2+m_{\phi,\mathrm{eff}}^2\right) =\\
    &=-\frac{T}{12\pi}(m^2_{\phi, \rm eff})^{\frac32}\simeq -\frac{T}{12\pi}(m^2_{\phi, \rm eff}|_{\rm LO})^{\frac32}+O(\alpha^5).
\end{align}
Since $m_{\phi,\rm eff}^2|_{\rm LO}= O(\alpha^2T^2)$, this contribution is of order $\alpha^3T^4$, and therefore enters $V_{\rm eff}$ at N$^{2}$LO,
\begin{align} \label{eq:VN2LO_two_scalar}
    V_{\text{N$^2$LO}}= -\frac{T}{12\pi}(m^2_{\phi,\rm eff}|_{\rm LO})^{\frac32},
\end{align}
Thus, while LO and NLO are generated by hard-scale contributions, N$^{2}$LO is
the first correction arising from infrared fluctuations within the EFT.

More generally the expansion takes the form
\begin{align}
    V_{\rm eff} \supset\cdots + \alpha^3 V_{ \text{N$^2$LO}}+\alpha^4V_{ \text{N$^3$LO}}+\alpha^5V_{ \text{N$^4$LO}}+\cdots 
\end{align}
The odd powers of $\alpha$ involve infrared contributions from loops within the EFT, analogous to the odd powers of the coupling familiar from the high-temperature expansion.
The expansion of $V_\text{EFT}$ involves only even powers of $\alpha$.
Note also that the $\alpha^3$ term does not introduce any new renormalisation-scale dependence, being constructed from $m^2_{\rm eff}|_{\rm LO}$, whose scale dependence already cancels by Eq.~\eqref{eq:cancellationRG}. Running of the effective action within the EFT appears first through a vacuum sunset diagram~\cite{Farakos:1994kx}. The corresponding counterterm is
\begin{equation}
    V_\text{EFT,ct}(\phi_\text{IR}) = \frac{T^2\left(V_\text{EFT}^{'''}(\phi_\text{IR})\right)^2}{48(4\pi)^2 \epsilon} ,
\end{equation}
and is N${}^5$LO or $O(\alpha^6 T^4)$, noting that $V_\text{EFT}^{'''}=O(\alpha^3 T)$.
\\\\
The EFT construction is valid beyond just the effective potential, and allows to
construct higher derivative terms in
the EFT action.
For a static, slowly varying $\phi_{\rm IR}$, the leading derivative correction stems from expanding the hard $\xi$ one-loop determinant. Using Eq.~(22) of Ref.~\cite{Moss:1985ve} and integrating by parts gives
\begin{equation}
\label{eq:xi_derivative_expansion}
\begin{aligned}
&\frac12\operatorname{Tr}_{\beta}
\log\!\left(-\partial_\tau^2-\boldsymbol{\nabla}^2
+m_\xi^2\right)
\\
&\quad=\beta\int_x\,\biggl[V_{\rm LO}^{\xi}
+\frac{g^4\phi_{\rm IR}^2}{12}A_3^0(m_\xi)
(\boldsymbol{\nabla}\phi_{\rm IR})^2
+ O(\boldsymbol{\nabla}^4)\biggr].
\end{aligned}
\end{equation}
Here $\operatorname{Tr}_{\beta}$ includes all bosonic Matsubara modes and $V_{\rm LO}^{\xi}=\tfrac12J(m_\xi)$ is the potential correction from $\xi$. 
Together with the tree-level kinetic term, this gives the $\phi_\text{IR}$ field normalisation
\begin{equation}
    z=1+\frac{g^4\phi_{\rm IR}^2}{6} A_3^0(m_\xi).
\end{equation}
We remind the reader that $m_\xi$ depends on the background field $\phi_\text{IR}$.
The correction is $O(\alpha^2)$, and therefore enters at NLO; terms with additional derivatives are suppressed by further powers of the separation between the soft and hard scales. The effective action through this order is
\begin{align}
S_{\rm EFT}[\phi_{\rm IR}]\!=\!
\beta\!\int_x \left[ \frac{1}{2}z(\phi_{\rm IR})(\boldsymbol{\nabla} \phi_{\rm IR})^2 + V_{\rm EFT}+ \cdots \right],
\end{align}
where the integral is over three-dimensional space.
The derivative corrections are essential for computing e.g.\ the nucleation rate
or bubble wall speed consistently to NLO (see for instance Refs.~\cite{Ekstedt:2021kyx, Hirvonen:2021zej, Kierkla:2025qyz}), but in what follows we focus on the
latent heat, for which only the potential is needed. Note that the field
$\phi_{\rm IR}$ lives in three dimensions, so that the construction of this EFT
can still be called dimensional reduction, just not high-temperature dimensional reduction.
\\\\
For homogeneous equilibrium field configurations, the derivative terms do not contribute, so the thermodynamics is determined entirely by the effective potential. We use the latent heat normalised to $T^4_c$, as a simple measure of the transition strength and as a numerical test of the convergence of the EFT expansion. 
The latent heat is as in Eq.~\eqref{eq:latent_heat_for_counting},
\begin{align}\label{eq:two_scalar_latent_heat}
    L=\frac{d\Delta V_{\rm EFT}(T)}{d\log T}\Bigg|_{T_c},
\end{align}
where $\Delta V_{\rm EFT}$ is the broken-phase free energy density minus the symmetric phase free energy density, each evaluated at its minimum $\phi_{\rm IR}^{\rm phase}$, and $T_c$ is determined by solving $\Delta V_{\rm EFT}(T_c)=0$. 
To solve for $\phi_\text{IR}^\text{phase}$, we directly minimise the numerical potential, rather than using a strict expansion~\cite{Gould:2023ovu}. 
The derivative in Eq.~\eqref{eq:two_scalar_latent_heat} is then computed as a finite difference with some small $\delta T$ and for fixed $\phi_{\rm IR}^{\rm phase}$, valid because the field value is an extremum. Note that for all temperatures the potential is real in and near both phases.

\begin{figure}[t]
    \centering
    \includegraphics[width=\linewidth]{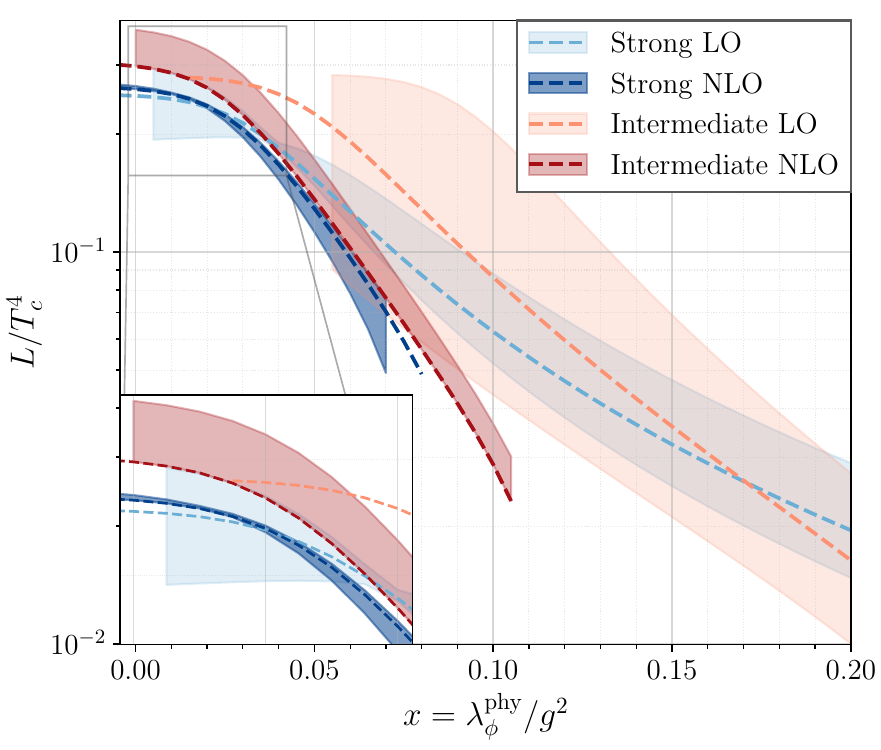}
    \caption{The latent heat obtained in the strong and intermediate transition EFTs. In the region of small $x$ where the strong transition hierarchy is realised, the NLO strong-EFT predictions become less sensitive to the renormalisation scale, while the intermediate expansion begins to deteriorate. Conversely, as $x$ is increased and the hierarchy becomes less pronounced, the intermediate EFT becomes the more appropriate description. The dashed curves show the predictions at the central scale choice $\bar\mu=\pi T$. The light bands show the residual renormalisation scale variation at LO, while the darker bands show the corresponding NLO variation, obtained by varying $0.1\pi T\leq\bar\mu\leq10\pi T$. The bands end where the computation fails for some renormalisation scale.
    }
    \label{fig:Lathentheat2scalar}
\end{figure}

We study the benchmark point
\begin{equation}
\begin{aligned}
    \mu_\phi^2&=0.044784\;\text{GeV}^2, &
    \mu_\xi^2&=0.005\;\text{GeV}^2,\\
    \lambda_\xi&=0.2, &
    \lambda_\phi&=0.003,\\
    g^2&=1.5,
\end{aligned}
\label{eq:two_scalar_benchmark}
\end{equation}
with these $\overline{\text{MS}}$ parameters specified at the input scale $\bar\mu_0=1\;\text{GeV}$. The scan shown in Fig.~\ref{fig:Lathentheat2scalar} compares the strong and intermediate strength EFTs as a function of
\begin{equation}
    x=\frac{\lambda_\phi^{\rm phy}}{g^2},\qquad
    \lambda_\phi^{\rm phy}
    =\frac{\partial^4 V_{\rm LO}}{\partial\phi_{\rm IR}^4}
    \Bigg|_{T=0,\,\phi_{\rm IR}^{\rm phase}}.
    \label{eq:two_scalar_physical_quartic}
\end{equation}
Here we use the physical quartic $\lambda_\phi^{\rm phy}$, rather than the $\overline{\rm MS}$ parameter, so that the horizontal axis is itself free from renormalisation scale ambiguity at leading order. Small $x$ corresponds to stronger transitions, including the $\lambda_\phi\sim g^4$ hierarchy that underlies the strong power counting, while larger $x$ moves towards the intermediate strength and weak regimes. Over this scan, the zero-temperature mass $m_\xi(\phi_{\rm IR})|_{T=0}$ decreases from approximately $2.7\;\text{GeV}$ at $x=0$ to $0.8\;\text{GeV}$ at $x=0.2$, so that $m_\xi(\phi_{\rm IR})|_{T=0}\sim \mu_0$. The bands in Fig.~\ref{fig:Lathentheat2scalar} show the residual dependence on the renormalisation scale, obtained by varying $\bar\mu/(\pi T)\in[0.1,10]$.

Two features of Fig.~\ref{fig:Lathentheat2scalar} are worth emphasising. First, in the strong transition regime, at small $x\lesssim0.05$, the strong transition expansion converges rapidly: at NLO the residual renormalisation scale dependence is significantly reduced in agreement with our analytical results of Eqs.~\eqref{eq:NLO_rg_cancellation_first} to \eqref{eq:NLO_rg_cancellation_last}. This is in contrast to e.g.\ the non-EFT-based computation of Ref.~\cite{Laine:2017hdk}, which numerically showed similar renormalisation scale uncertainty at one and two loops. The intermediate strength expansion, applied in the same region, shows a much larger shift between orders and a much wider band. At larger $x\gtrsim0.05$ the situation reverses, and it is the strong expansion that deteriorates or ceases to predict a first-order transition.
Second, there is a range $x\sim[0.03,0.07]$ where the two expansions agree within their uncertainty bands, illustrating the overlapping regimes of validity discussed in the previous sections.

The size of the first infrared correction is shown in Table~\ref{tab:two_scalar_latent_heat}. For example, at $x=0.002$ and $\bar\mu=\pi T$, the LO result $L/T_c^4= 0.250$ receives an NLO correction of $+0.010$ and an N$^2$LO correction of $+0.003$. The successive corrections are therefore small and decreasing at this strong transition point.
This convergence cannot be judged from the leading-order result alone; but is supported by the reduced scale dependence at NLO and by the smallness of higher-order corrections.

\begin{table}[t]
    \centering
    \begin{tabular}{c|ccc}
        $x$ & LO & NLO correction & N$^2$LO correction \\
        \hline
        $0.002$ & $0.250$ & $+0.010$ & $+0.003$ \\
        $0.020$ & $0.235$ & $+0.001$ & $+0.003$ \\
        $0.050$ & $0.154$ & $-0.025$ & $+0.003$
    \end{tabular}
    \caption{Order-by-order contributions to $L/T_c^4$ in the strong phase transition EFT at $\bar\mu=\pi T$.}
    \label{tab:two_scalar_latent_heat}
\end{table}

\begin{figure}[t]
    \centering
    \includegraphics[width=\linewidth]{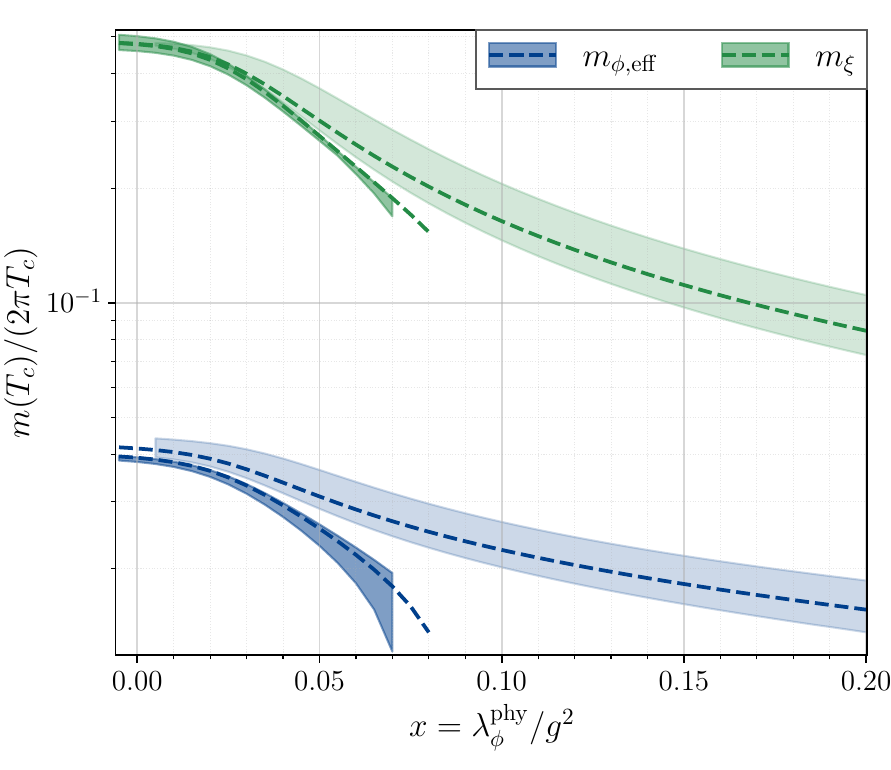}
    \caption{The masses $m_{\phi, \rm eff}$ (blue) and $m_\xi$ (green), evaluated at the broken minimum at $T_c$ and normalised to $2\pi T_c$, as a function of $x$. Here these quantities are computed in the strong transition EFT and the darker colour includes NLO terms. As the transition becomes stronger at small $x$, the $\xi$ field becomes heavier relative to the temperature, while the order parameter field remains much lighter than both $m_\xi$ and $2\pi T_c$.
    }
    \label{fig:two_scalar_masses}
\end{figure}

It is also important to check that the scale hierarchy assumed in constructing the EFT is actually present at the transition. Figure~\ref{fig:two_scalar_masses} shows $m_\xi(\phi_{\rm IR}^{\rm broken})/(2\pi T_c)$ and $m_{\rm eff}(\phi_{\rm IR}^{\rm broken})/(2\pi T_c)$, evaluated at the broken minimum at $T_c$, with $m_{\rm eff}^2=V_{\rm EFT}''$. As $x$ decreases and the transition becomes stronger, the heavy mass ratio grows to a few tenths, while the light mass ratio remains at a few times $10^{-2}$. Thus $m_\xi$ is of the order of the hard scale, whereas $m_{\rm eff}$ remains much smaller than both $m_\xi$ and $2\pi T_c$. Crucially, the order parameter remains light in the broken phase even for the strongest transitions shown, as argued in Sec.~\ref{sec:scale_hierarchies}. This is the hierarchy assumed in the construction: the field inducing the transition is hard, while the field retained in the EFT is soft.

Taken together, the reduced renormalisation scale dependence, the small higher-order corrections to the latent heat, and the mass hierarchy at the broken minimum
provide strong numerical evidence for the consistency of
the strong transition EFT in the small-$x$ regime. For this benchmark, the strongest transitions remain compatible with a light order parameter field and a systematically improvable perturbative expansion.

In the next section, we show that the same construction extends beyond the two-scalar example by applying the power counting to Yukawa, Abelian gauge, and non-Abelian gauge theories. For these models we derive the leading-order potentials and identify the complete sets of diagrams entering at NLO and N${}^2$LO, illustrating that the construction applies rather generically.

\section{Generalisation to other models}\label{sec:other_models}
Having shown explicitly in the two-scalar model how
the strong-transition power counting organises the EFT, we now showcase how the same procedure can be applied to other models. The purpose of this section is not to repeat the full higher-order calculation of Sec.~\ref{sec:strong_transitions} for every theory. Instead, we identify the hierarchy of scales, show how the diagrams for computing the EFT and effective potentials are selected by the power counting, and give the resulting leading-order potential. Higher orders can then be computed following exactly the same procedure as before. We first show a fermionic case, and then
two gauge cases, respectively Abelian and non-Abelian models. 

\subsection{Yukawa model} \label{sec:yukawa}
We first consider a real scalar field $\phi$ coupled to a Dirac fermion $\psi$, with Euclidean Lagrangian
\begin{equation}
\begin{aligned}
\mathcal{L}_\text{E}^{\rm Yukawa}
=&
\frac12(\partial_\mu\phi)^2
+ \bar\psi\bigl(\gamma_\mu\partial_\mu + \mu_\psi\bigr)\psi
\\
&+\sigma\phi
+ \frac12\mu_\phi^2\phi^2
+\frac{\kappa}{3!}\phi^3
+\frac{\lambda}{4!}\phi^4
+y\phi\bar\psi\psi .
\end{aligned}
\end{equation}
Here $\phi$ again plays the role of the order parameter, while the fermion
provides the fluctuations that generate the leading thermal contributions. The
scalar has no internal $\mathrm{Z}_2$ symmetry, so that the transition is not a
symmetry breaking one. By shifting the background field by a constant, we can fix $\sigma=0$. This model was also considered in
Sec.~\ref{sec:soft_supersoft_transitions}. 

The strong transition power counting is introduced similarly to the
two-scalar theory.
The key requirements are that: (i) thermal effects should modify the potential at
leading order, (ii) all the tree-level terms in the potential should be of the same order at the
transition, and (iii) the scalar should get a large enough background field to push the fermion
mass to the hard scale.
In terms of the formal power-counting parameter $\alpha$, this leads to
\begin{align} \label{eq:yukawa_power_counting}
\{\mu_\phi^2,\kappa,\lambda,\mu_\psi,y\}
\rightarrow
\{\alpha^2\mu_\phi^2,
\alpha^3 \kappa, \alpha^4\lambda,\mu_\psi,\alpha y\}.
\end{align}
In particular this contains the hierarchy $\lambda=O(y^4)$, comparable to
$\lambda_\phi=O(g^4)$ in the two-scalar model. As before, the small scalar self-coupling pushes the
broken minimum to large field values, so that
\begin{align}
    \phi_{\rm IR}=O(\alpha^{-1}T).
\end{align}
This is what makes the fermion behave as a hard field even though the Yukawa coupling is perturbatively small. The background-dependent masses take the form
\begin{equation} \label{eq:yukawa_masses}
\begin{aligned}
m_\phi^{2}(\phi_{\rm IR})
&=\mu_\phi^{2}+\kappa\phi_{\rm IR}
+\tfrac12\lambda\phi_{\rm IR}^{2},
\\[2mm]
m_\psi(\phi_{\rm IR})
&=\mu_\psi+y\phi_{\rm IR},
\end{aligned}
\end{equation}
and hence count as
\begin{align} \label{eq:yukawa_masses_couting}
    m_\phi^2=O(\alpha^2T^2),
    \qquad
    m_\psi=O(\alpha^0T).
\end{align}
The order-parameter field therefore remains soft, while the fermion has a hard
mass of order the temperature.
The scale hierarchy is therefore the same as in Table \ref{tab:hierarchy_tree_level} for intermediate-strength transitions in the Yukawa model, the only difference being that the field-dependence of the fermion mass becomes of leading order.
The full dependence on $m_\psi(\phi_{\rm IR})$ must therefore be retained in the propagator.
Just as for the two-scalar case, our power-counting relations ensure the additive
terms in Eq.~\eqref{eq:yukawa_masses} are of the same order of magnitude, but
this can be relaxed as long as Eq.~\eqref{eq:yukawa_masses_couting} holds.

Organising the diagrams by powers of $\alpha$, the leading-order potential takes the form
\begin{equation}
\begin{gathered}
\tikzset{every picture/.style={line width=0.75pt}} 
\begin{tikzpicture}[x=0.75pt,y=0.75pt,yscale=-1,xscale=1]
\draw  [color={rgb, 255:red, 245; green, 166; blue, 35 }  ,draw opacity=1 ] (140,30) .. controls (140,18.95) and (148.95,10) .. (160,10) .. controls (171.05,10) and (180,18.95) .. (180,30) .. controls (180,41.05) and (171.05,50) .. (160,50) .. controls (148.95,50) and (140,41.05) .. (140,30) -- cycle ;
\draw  [dash pattern={on 4.5pt off 4.5pt}] (200,30) .. controls (200,18.95) and (208.95,10) .. (220,10) .. controls (231.05,10) and (240,18.95) .. (240,30) .. controls (240,41.05) and (231.05,50) .. (220,50) .. controls (208.95,50) and (200,41.05) .. (200,30) -- cycle ;
\draw  [draw opacity=0] (219.49,10.01) .. controls (219.66,10) and (219.83,10) .. (220,10) .. controls (222.12,10) and (224.16,10.33) .. (226.08,10.94) -- (220,30) -- cycle ; \draw [color={rgb, 255:red, 0; green, 0; blue, 0 }  ,draw opacity=1 ]   (219.49,10.01) .. controls (219.66,10) and (219.83,10) .. (220,10) .. controls (221.08,10) and (222.14,10.09) .. (223.18,10.25) ; \draw [shift={(226.08,10.94)}, rotate = 180.45] [fill={rgb, 255:red, 0; green, 0; blue, 0 }  ,fill opacity=1 ][line width=0.08]  [draw opacity=0] (8.93,-4.29) -- (0,0) -- (8.93,4.29) -- cycle    ; 

\draw (65,29) node    {$V_{\text{LO}} =V_{\text{tree}}+V_{\rm LO}^{\rm ct}$};
\draw (189.68,29.2) node    {$+$};
\draw (126,28.5) node    {$+$};
\end{tikzpicture}
\end{gathered}
\end{equation}
where the scalar loop denotes the massless hard fluctuations of $\phi$, while
the second diagram denotes the fermionic determinant. 
Both one-loop diagrams are of $O(\alpha^0 T^4)$,
and must therefore be included at leading order. The former is a pure
constant, while the latter is background-field dependent, since
$m_\psi(\phi_{\rm IR})=O(\alpha^0 T)$.

The diagrams that contribute at the next order are 
\begin{equation}
\begin{gathered}
\tikzset{every picture/.style={line width=0.75pt}} 
\begin{tikzpicture}[x=0.75pt,y=0.75pt,yscale=-1,xscale=1]
\draw  [color={rgb, 255:red, 245; green, 166; blue, 35 }  ,draw opacity=1 ] (100,50) .. controls (100,38.95) and (108.95,30) .. (120,30) .. controls (131.05,30) and (140,38.95) .. (140,50) .. controls (140,61.05) and (131.05,70) .. (120,70) .. controls (108.95,70) and (100,61.05) .. (100,50) -- cycle ;
\draw    (115,25) -- (125,35) ;
\draw    (115,35) -- (125,25) ;
\draw  [color={rgb, 255:red, 245; green, 166; blue, 35 }  ,draw opacity=1 ] (130,105) .. controls (130,93.95) and (138.95,85) .. (150,85) .. controls (161.05,85) and (170,93.95) .. (170,105) .. controls (170,116.05) and (161.05,125) .. (150,125) .. controls (138.95,125) and (130,116.05) .. (130,105) -- cycle ;
\draw  [fill={rgb, 255:red, 255; green, 255; blue, 255 }  ,fill opacity=1 ] (146.72,81.23) .. controls (148.8,79.41) and (151.96,79.64) .. (153.77,81.72) .. controls (155.59,83.8) and (155.36,86.96) .. (153.28,88.77) .. controls (151.2,90.59) and (148.04,90.36) .. (146.23,88.28) .. controls (144.41,86.2) and (144.64,83.04) .. (146.72,81.23) -- cycle ;
\draw    (146.46,81.46) -- (153.54,88.54) ;
\draw    (153.54,81.46) -- (146.46,88.54) ;

\draw  [dash pattern={on 4.5pt off 4.5pt}] (190,105) .. controls (190,93.95) and (198.95,85) .. (210,85) .. controls (221.05,85) and (230,93.95) .. (230,105) .. controls (230,116.05) and (221.05,125) .. (210,125) .. controls (198.95,125) and (190,116.05) .. (190,105) -- cycle ;
\draw  [fill={rgb, 255:red, 255; green, 255; blue, 255 }  ,fill opacity=1 ] (206.72,81.23) .. controls (208.8,79.42) and (211.96,79.64) .. (213.77,81.72) .. controls (215.59,83.8) and (215.36,86.96) .. (213.28,88.77) .. controls (211.2,90.59) and (208.04,90.36) .. (206.23,88.28) .. controls (204.41,86.2) and (204.64,83.04) .. (206.72,81.23) -- cycle ;
\draw    (206.46,81.46) -- (213.54,88.54) ;
\draw    (213.54,81.46) -- (206.46,88.54) ;

\draw  [dash pattern={on 4.5pt off 4.5pt}] (160,50) .. controls (160,38.95) and (168.95,30) .. (180,30) .. controls (191.05,30) and (200,38.95) .. (200,50) .. controls (200,61.05) and (191.05,70) .. (180,70) .. controls (168.95,70) and (160,61.05) .. (160,50) -- cycle ;
\draw [color={rgb, 255:red, 245; green, 166; blue, 35 }  ,draw opacity=1 ]   (160,50) -- (200,50) ;
\draw  [draw opacity=0] (179.49,30.01) .. controls (179.66,30) and (179.83,30) .. (180,30) .. controls (182.12,30) and (184.16,30.33) .. (186.08,30.94) -- (180,50) -- cycle ; \draw [color={rgb, 255:red, 0; green, 0; blue, 0 }  ,draw opacity=1 ]   (179.49,30.01) .. controls (179.66,30) and (179.83,30) .. (180,30) .. controls (181.08,30) and (182.14,30.09) .. (183.18,30.25) ; \draw [shift={(186.08,30.94)}, rotate = 180.45] [fill={rgb, 255:red, 0; green, 0; blue, 0 }  ,fill opacity=1 ][line width=0.08]  [draw opacity=0] (8.93,-4.29) -- (0,0) -- (8.93,4.29) -- cycle    ; 
\draw  [draw opacity=0] (182.64,69.83) .. controls (181.77,69.94) and (180.89,70) .. (180,70) .. controls (178.06,70) and (176.19,69.72) .. (174.42,69.21) -- (180,50) -- cycle ; \draw [color={rgb, 255:red, 0; green, 0; blue, 0 }  ,draw opacity=1 ]   (182.64,69.83) .. controls (181.77,69.94) and (180.89,70) .. (180,70) .. controls (179.1,70) and (178.21,69.94) .. (177.34,69.83) ; \draw [shift={(174.42,69.21)}, rotate = 358.94] [fill={rgb, 255:red, 0; green, 0; blue, 0 }  ,fill opacity=1 ][line width=0.08]  [draw opacity=0] (8.93,-4.29) -- (0,0) -- (8.93,4.29) -- cycle    ; 
\draw  [draw opacity=0] (212.64,124.83) .. controls (211.77,124.94) and (210.89,125) .. (210,125) .. controls (208.06,125) and (206.19,124.72) .. (204.42,124.21) -- (210,105) -- cycle ; \draw [color={rgb, 255:red, 0; green, 0; blue, 0 }  ,draw opacity=1 ]   (212.64,124.83) .. controls (211.77,124.94) and (210.89,125) .. (210,125) .. controls (209.1,125) and (208.21,124.94) .. (207.34,124.83) ; \draw [shift={(204.42,124.21)}, rotate = 358.94] [fill={rgb, 255:red, 0; green, 0; blue, 0 }  ,fill opacity=1 ][line width=0.08]  [draw opacity=0] (8.93,-4.29) -- (0,0) -- (8.93,4.29) -- cycle    ; 

\draw (149.68,49.2) node    {$+$};
\draw (179.68,104.2) node    {$+$};
\draw (119.68,104.2) node    {$+$};
\draw (51,49) node    {$V_{\text{NLO}} =V_{\rm NLO}^{\rm ct}+$};

\end{tikzpicture}
\end{gathered}
\end{equation}
The same power counting continues to organise the calculation beyond leading
order. Their evaluation follows
the methodology of the two-scalar example.

Collecting the leading terms, the EFT potential is
\begin{align}
V_{\rm LO}
=
V_{\rm tree}
+\frac12 J^0(0)
-\frac{n_\text{f}}{2}J_\text{F}^0(m_\psi),
\end{align}
where $J_\text{F}^0$ is the fermionic sum-integral (see Eq.~\ref{eq:app_JF0}, and Eq.~\ref{eq:yukawaLOapp}), with fermionic Matsubara frequencies $P_0=(2n+1)\pi T$, and $n_\text{f}$ counts the fermionic degrees of freedom. For a single Dirac fermion $n_\text{f}=4$.

The full mass-dependent fermionic thermal function appears
at leading order,
giving
an $O(\alpha^0 T^4)$
background-dependent contribution to the potential, in direct analogy with the
hard $\xi$ contribution of the two-scalar model.

Similarly to the two-scalar model, one-loop fluctuations within the EFT
contribute to the effective potential at N${}^{2}$LO. The result takes the same
form as Eq.~\eqref{eq:VN2LO_two_scalar}, after shifting away the temperature-dependent tadpole.

\begin{figure}[t]
    \centering
    \includegraphics[width=\linewidth]{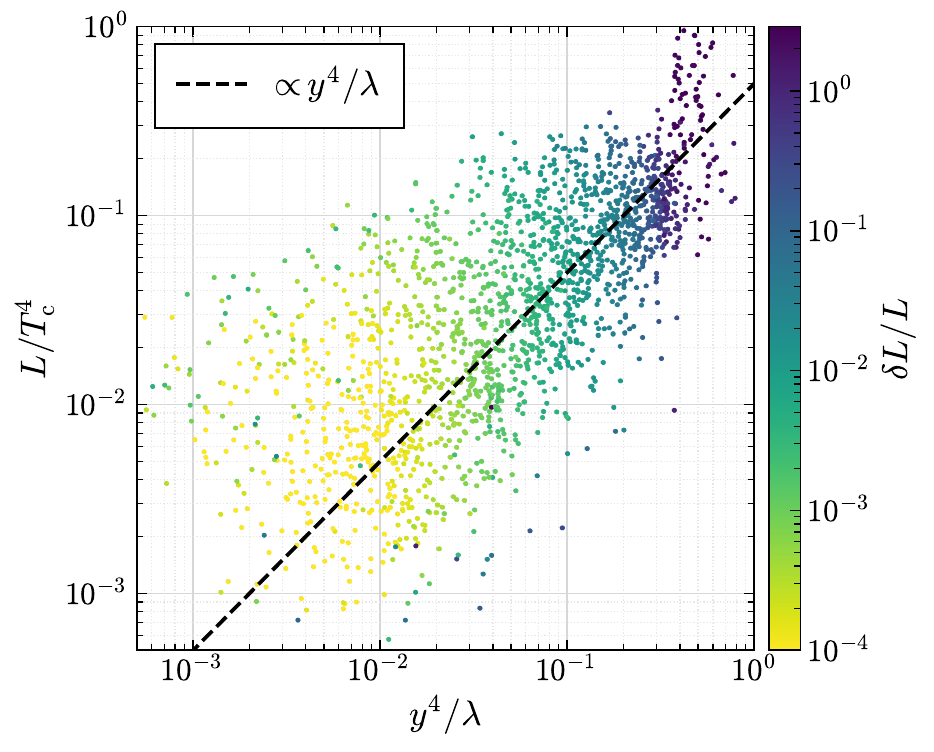}
    \caption{Scatter plot of the latent heat versus $y^4/\lambda$ in the Yukawa
    model. The data is from Refs.~\cite{Gould:2023jbz, gould_2023_10276044}
    using the EFT for intermediate strength transitions. The colour of
    each point gives the relative theoretical uncertainty $\delta L/L$.
    The data shows that the strongest transitions are restricted to $y^4/\lambda\gtrsim 0.1$, the region of parameter space identified by our analysis, and that these points suffer from large uncertainties using the EFT for intermediate strength transitions.
    }
    \label{fig:xie_yukawa}
\end{figure}

The region of parameter space with strong phase transitions identified in
Eq.~\eqref{eq:yukawa_power_counting} is a small region of the multidimensional
parameter space of this model. One might imagine that it is possible to find
other regions of parameter space with strong phase transitions. To investigate
this, in Fig.~\ref{fig:xie_yukawa} we have plotted the results of a 20,000 point
random parameter scan of this model carried out in Ref.~\cite{Gould:2023jbz}.
We have included all parameter points for which results are available
at three different renormalisation scales, so that the variation over these scales
can be used to estimate the theoretical uncertainty.
The points with the largest latent heat all have
$y^4/\lambda \gtrsim 0.1$ numerically, suggesting that
Eq.~\eqref{eq:yukawa_power_counting} correctly identifies the region of
parameter space with the strongest phase transitions.

The analysis of Ref.~\cite{Gould:2023jbz} was carried out within the EFT for intermediate strength transitions. It was found that the strongest transitions with $y^4/\lambda=O(1)$ suffered from the largest theoretical uncertainties, identified through uncancelled renormalisation scale dependence. The colour of each point on Fig.~\ref{fig:xie_yukawa} gives its relative theoretical uncertainty, $\delta L/L$. Our EFT for strong transitions shows an order-by-order cancellation of renormalisation scale dependence in this parameter space region, suggesting a reanalysis using it could resolve the issue. We leave this for future work.

\subsection{Abelian Higgs model}

Having considered the fermionic case, we now turn to gauge fields. We take a $\mathrm{U}(1)$ gauge field coupled to a complex scalar, with Euclidean Lagrangian
\begin{equation}
\begin{aligned}
\mathcal{L}^{U(1)}_E
=&\frac14 F_{\mu\nu}F_{\mu\nu}
 +(D_\mu\phi)^*(D_\mu\phi)
 -\mu_\phi^2\phi^*\phi
 +\lambda(\phi^*\phi)^2\\
&+\frac{1}{2\xi}(\partial_\mu A_\mu)^2
 +\partial_\mu \bar c\,\partial_\mu c ,
\end{aligned}
\end{equation}
and work in Landau gauge $\xi\to 0_+$, for which the multiplicative determinant anomaly is absent~\cite{Balui:2025yvd}.
The covariant derivative is $D_\mu\phi = \partial_\mu \phi + i g A_\mu \phi$, and the field strength tensor is $F_{\mu\nu} = \partial_\mu A_\nu - \partial_\nu A_\mu$ as usual.

The strong-transition power counting in this model is
\begin{align}
\{\mu_\phi^2,\lambda,g^2\}
\rightarrow
\{\alpha^2\mu_\phi^2,\alpha^4\lambda,
  \alpha^2g^2\}
.
\end{align}
This leads to $\phi_{\rm IR}=O(\alpha^{-1}T)$ as required for a strong
transition, as well as to the condition $\lambda=O(g^4)$.
The background dependent masses are
\begin{equation}
\begin{aligned}
m_H^2(\phi_{\rm IR})
 &=-\mu_\phi^2+3\lambda\phi_{\rm IR}^2,
\\[1mm]
m_G^2(\phi_{\rm IR})
 &=-\mu_\phi^2+\lambda\phi_{\rm IR}^2,
\\[1mm]
m_A^2(\phi_{\rm IR})
 &=g^2\phi_{\rm IR}^2 ,
\end{aligned}
\end{equation}
and therefore, barring further cancellations,
\begin{align}
m_H^2=O(\alpha^2T^2),\quad
m_G^2=O(\alpha^2T^2),\quad
m_A^2=O(\alpha^0T^2).
\end{align}
The Higgs and Goldstone masses therefore enter as perturbative insertions, while
the full background dependent gauge mass is retained in the propagator. The mass of the transverse gauge field shifts parametrically between phases, but its dominant effects arise at larger field values where it is heavier~\cite{Gould:2021ccf}.
For general gauge fixing parameter $\xi$, the Goldstone and the longitudinal component of the gauge field both have nonzero $\xi$-dependent masses, which in Landau gauge become $m_G^2$ and $0$ respectively~\cite{Ekstedt:2020abj}. The ghost fields decouple.
The hierarchies of scale are summarised in Table~\ref{tab:hierarchy_strong_u1}.

\begin{table}[t]
    \centering
    \begin{tabular}{c|c}
        $m_{\rm eff}$ & Fields \\
        \hline
        $\pi T$         & $A_\mu$, $\phi^\text{UV}$ \\
        $g T$           & $\phi^\text{IR}$ 
    \end{tabular}
    \caption{Scale hierarchies for strong transitions in the Abelian Higgs model, with $\phi_\text{IR}=O(T/g)$.}
    \label{tab:hierarchy_strong_u1}
\end{table}

Here for $m_A\gg gT$ it is important to note that there is no need to resum the
thermal corrections (Debye screening) to the gauge field propagator, as these effects are
subleading and so can be treated perturbatively, arising order-by-order in the
$\alpha$ expansion. As a consequence, the whole
$A_\mu$ field can be treated together, without the need to separate out $A_0$.

The leading order diagrams for this model are
\begin{equation}
\label{eq:VLO_Abelian_diagrams}
\begin{gathered}
\tikzset{every picture/.style={line width=0.75pt}} 
\begin{tikzpicture}[x=0.75pt,y=0.75pt,yscale=-1,xscale=1]
\draw  [color={rgb, 255:red, 245; green, 166; blue, 35 }  ,draw opacity=1 ] (200,30) .. controls (200,18.95) and (208.95,10) .. (220,10) .. controls (231.05,10) and (240,18.95) .. (240,30) .. controls (240,41.05) and (231.05,50) .. (220,50) .. controls (208.95,50) and (200,41.05) .. (200,30) -- cycle ;
\draw  [color={rgb, 255:red, 245; green, 166; blue, 35 }  ,draw opacity=1 ][dash pattern={on 4.5pt off 4.5pt}] (260,30) .. controls (260,18.95) and (268.95,10) .. (280,10) .. controls (291.05,10) and (300,18.95) .. (300,30) .. controls (300,41.05) and (291.05,50) .. (280,50) .. controls (268.95,50) and (260,41.05) .. (260,30) -- cycle ;
\draw  [dash pattern={on 0.84pt off 2.51pt}][line width=0.75]  (230,69.99) .. controls (230,58.94) and (238.95,49.99) .. (250,49.99) .. controls (261.05,49.99) and (270,58.94) .. (270,69.99) .. controls (270,81.04) and (261.05,89.99) .. (250,89.99) .. controls (238.95,89.99) and (230,81.04) .. (230,69.99) -- cycle ;
\draw    (169.94,70) .. controls (168.5,67.97) and (168.8,66.31) .. (170.84,65.02) .. controls (172.95,64.17) and (173.62,62.68) .. (172.87,60.53) .. controls (172.54,58.06) and (173.57,56.71) .. (175.97,56.47) .. controls (178.25,56.62) and (179.52,55.56) .. (179.77,53.29) .. controls (180.45,50.98) and (181.92,50.24) .. (184.18,51.08) .. controls (186.11,52.26) and (187.77,51.91) .. (189.16,50.03) .. controls (190.96,48.38) and (192.62,48.5) .. (194.13,50.41) .. controls (195.08,52.42) and (196.6,52.97) .. (198.71,52.06) .. controls (201.09,51.53) and (202.5,52.46) .. (202.93,54.87) .. controls (202.94,57.17) and (204.06,58.36) .. (206.29,58.44) .. controls (208.66,59.03) and (209.48,60.48) .. (208.75,62.81) .. controls (207.65,64.79) and (208.05,66.4) .. (209.95,67.63) -- (209.99,70.15) ;
\draw    (169.94,70) .. controls (171.71,71.47) and (171.93,73.15) .. (170.62,75.02) .. controls (169.64,77.25) and (170.29,78.8) .. (172.57,79.65) .. controls (174.82,80.06) and (175.82,81.4) .. (175.56,83.67) .. controls (175.71,86.09) and (176.98,87.16) .. (179.39,86.87) .. controls (181.56,86.19) and (183.05,86.91) .. (183.84,89.04) .. controls (185.11,91.1) and (186.73,91.41) .. (188.69,89.98) .. controls (190.2,88.28) and (191.86,88.12) .. (193.67,89.51) .. controls (195.86,90.59) and (197.44,90.01) .. (198.41,87.77) .. controls (198.88,85.58) and (200.26,84.68) .. (202.54,85.07) .. controls (204.98,85.09) and (206.15,83.88) .. (206.06,81.45) .. controls (205.56,79.22) and (206.4,77.81) .. (208.58,77.21) .. controls (210.73,76.12) and (211.16,74.52) .. (209.87,72.41) -- (209.94,70) ;
\draw  [draw opacity=0] (249.49,50) .. controls (249.66,49.99) and (249.83,49.99) .. (250,49.99) .. controls (252.12,49.99) and (254.16,50.32) .. (256.08,50.93) -- (250,69.99) -- cycle ; \draw [color={rgb, 255:red, 0; green, 0; blue, 0 }  ,draw opacity=1 ]   (249.49,50) .. controls (249.66,49.99) and (249.83,49.99) .. (250,49.99) .. controls (251.08,49.99) and (252.14,50.08) .. (253.18,50.24) ; \draw [shift={(256.08,50.93)}, rotate = 180.45] [fill={rgb, 255:red, 0; green, 0; blue, 0 }  ,fill opacity=1 ][line width=0.08]  [draw opacity=0] (8.93,-4.29) -- (0,0) -- (8.93,4.29) -- cycle    ; 

\draw (137,29.5) node    {$V_{\text{LO}} =V_{\text{tree}} +V_{\text{LO}}^{\text{ct}} +$};
\draw (249.68,29.2) node    {$+$};
\draw (160,68.49) node    {$+$};
\draw (220,68.49) node    {$+$};

\end{tikzpicture}
\end{gathered}
\end{equation}
Here the solid orange line denotes the Higgs, the dashed orange line the Goldstone, the wavy line denotes the gauge field, and the dotted line the ghost. Note that the longitudinal gauge field contributes to the potential at one-loop, but not at higher loops in Landau gauge as its propagator is proportional to $\xi$~\cite{Laine:1994zq}. After the cancellation of $1/\epsilon$ poles, Eq.~\eqref{eq:VLO_Abelian_diagrams} may be written explicitly as
\begin{align}
V_{\rm LO}
=
V_{\rm tree}
+\frac12 J^0(0)
+\frac32 J^0(m_A)
-\frac22 J^{-1}(m_A).
\end{align}
The dependence on $m_A(\phi_{\rm IR})$ and temperature in the gauge-field sum-integral drives the scalar through the transition.
The explicit expression is given in Eq.~\ref{eq:abelianLOapp}.

In the vicinity of the transition, the strong-transition power counting,
\begin{align}
m_{H,G}^2=O(\alpha^2T^2),\quad
g^2=O(\alpha^2),\quad
\bigl(g^2\phi_{\rm IR}\bigr)^2=O(\alpha^2T^2),
\end{align}
implies that diagrams adding a single
scalar mass insertion,
a single
gauge--scalar quartic vertex, or
two background-induced cubic
vertices all contribute at the same order. Contrary to a loop expansion, but in agreement with the other models: one-loop mass insertions and
genuine two-loop diagrams are included on the same footing.

The complete set of contributions at NLO is 
\begin{widetext}
\vspace{-0.35\baselineskip}
    \begin{equation}
    \begin{gathered}
\tikzset{every picture/.style={line width=0.75pt}} 
\begin{tikzpicture}[x=0.75pt,y=0.75pt,yscale=-1,xscale=1]
\draw  [color={rgb, 255:red, 245; green, 166; blue, 35 }  ,draw opacity=1 ] (240,50) .. controls (251.05,50) and (260,58.95) .. (260,70) .. controls (260,81.05) and (251.05,90) .. (240,90) .. controls (228.95,90) and (220,81.05) .. (220,70) .. controls (220,58.95) and (228.95,50) .. (240,50) -- cycle ;
\draw  [color={rgb, 255:red, 245; green, 166; blue, 35 }  ,draw opacity=1 ][dash pattern={on 4.5pt off 4.5pt}] (160,70) .. controls (160,58.95) and (168.95,50) .. (180,50) .. controls (191.05,50) and (200,58.95) .. (200,70) .. controls (200,81.05) and (191.05,90) .. (180,90) .. controls (168.95,90) and (160,81.05) .. (160,70) -- cycle ;
\draw  [color={rgb, 255:red, 245; green, 166; blue, 35 }  ,draw opacity=1 ] (100,50) .. controls (100,38.95) and (108.95,30) .. (120,30) .. controls (131.05,30) and (140,38.95) .. (140,50) .. controls (140,61.05) and (131.05,70) .. (120,70) .. controls (108.95,70) and (100,61.05) .. (100,50) -- cycle ;
\draw    (115,25) -- (125,35) ;
\draw    (115,35) -- (125,25) ;
\draw  [color={rgb, 255:red, 245; green, 166; blue, 35 }  ,draw opacity=1 ] (129.92,120) .. controls (129.92,108.95) and (138.87,100) .. (149.92,100) .. controls (160.96,100) and (169.92,108.95) .. (169.92,120) .. controls (169.92,131.05) and (160.96,140) .. (149.92,140) .. controls (138.87,140) and (129.92,131.05) .. (129.92,120) -- cycle ;
\draw  [fill={rgb, 255:red, 255; green, 255; blue, 255 }  ,fill opacity=1 ] (146.64,96.23) .. controls (148.72,94.41) and (151.88,94.64) .. (153.69,96.72) .. controls (155.5,98.8) and (155.28,101.96) .. (153.2,103.77) .. controls (151.11,105.59) and (147.95,105.36) .. (146.14,103.28) .. controls (144.33,101.2) and (144.55,98.04) .. (146.64,96.23) -- cycle ;
\draw [fill={rgb, 255:red, 255; green, 255; blue, 255 }  ,fill opacity=1 ]   (146.38,96.46) -- (153.45,103.54) ;
\draw [fill={rgb, 255:red, 255; green, 255; blue, 255 }  ,fill opacity=1 ]   (153.45,96.46) -- (146.38,103.54) ;

\draw  [color={rgb, 255:red, 245; green, 166; blue, 35 }  ,draw opacity=1 ][dash pattern={on 4.5pt off 4.5pt}] (189.92,120) .. controls (189.92,108.95) and (198.87,100) .. (209.92,100) .. controls (220.96,100) and (229.92,108.95) .. (229.92,120) .. controls (229.92,131.05) and (220.96,140) .. (209.92,140) .. controls (198.87,140) and (189.92,131.05) .. (189.92,120) -- cycle ;
\draw  [fill={rgb, 255:red, 255; green, 255; blue, 255 }  ,fill opacity=1 ] (206.64,96.23) .. controls (208.72,94.41) and (211.88,94.64) .. (213.69,96.72) .. controls (215.5,98.8) and (215.28,101.96) .. (213.2,103.77) .. controls (211.11,105.59) and (207.95,105.36) .. (206.14,103.28) .. controls (204.33,101.2) and (204.55,98.04) .. (206.64,96.23) -- cycle ;
\draw [fill={rgb, 255:red, 255; green, 255; blue, 255 }  ,fill opacity=1 ]   (206.38,96.46) -- (213.45,103.54) ;
\draw [fill={rgb, 255:red, 255; green, 255; blue, 255 }  ,fill opacity=1 ]   (213.45,96.46) -- (206.38,103.54) ;

\draw    (340,50) .. controls (338.56,47.97) and (338.86,46.31) .. (340.9,45.03) .. controls (343.01,44.18) and (343.68,42.69) .. (342.93,40.54) .. controls (342.6,38.07) and (343.63,36.72) .. (346.03,36.48) .. controls (348.31,36.63) and (349.58,35.57) .. (349.84,33.3) .. controls (350.51,30.99) and (351.98,30.25) .. (354.24,31.09) .. controls (356.17,32.27) and (357.83,31.92) .. (359.22,30.04) .. controls (361.02,28.39) and (362.68,28.51) .. (364.19,30.42) .. controls (365.14,32.43) and (366.66,32.98) .. (368.77,32.07) .. controls (371.15,31.54) and (372.56,32.47) .. (372.99,34.88) .. controls (373,37.18) and (374.12,38.37) .. (376.35,38.45) .. controls (378.72,39.04) and (379.54,40.49) .. (378.82,42.82) .. controls (377.71,44.8) and (378.11,46.41) .. (380.01,47.64) -- (380.06,50) ;
\draw    (340,50) .. controls (341.77,51.48) and (341.99,53.16) .. (340.68,55.03) .. controls (339.7,57.26) and (340.35,58.81) .. (342.63,59.66) .. controls (344.88,60.07) and (345.88,61.41) .. (345.63,63.68) .. controls (345.77,66.1) and (347.04,67.17) .. (349.45,66.88) .. controls (351.62,66.2) and (353.11,66.92) .. (353.9,69.05) .. controls (355.17,71.11) and (356.79,71.42) .. (358.75,69.99) .. controls (360.26,68.29) and (361.92,68.13) .. (363.73,69.52) .. controls (365.92,70.6) and (367.5,70.02) .. (368.47,67.78) .. controls (368.94,65.59) and (370.32,64.69) .. (372.6,65.08) .. controls (375.04,65.1) and (376.21,63.89) .. (376.12,61.46) .. controls (375.62,59.23) and (376.46,57.82) .. (378.64,57.22) .. controls (380.79,56.13) and (381.22,54.53) .. (379.94,52.42) -- (380,50) ;
\draw [color={rgb, 255:red, 245; green, 166; blue, 35 }  ,draw opacity=1 ] [dash pattern={on 4.5pt off 4.5pt}]  (340,50) -- (380,50) ;
\draw    (280,50) .. controls (278.56,47.97) and (278.86,46.31) .. (280.9,45.03) .. controls (283.01,44.18) and (283.68,42.69) .. (282.93,40.54) .. controls (282.6,38.07) and (283.63,36.72) .. (286.03,36.48) .. controls (288.31,36.63) and (289.58,35.57) .. (289.84,33.3) .. controls (290.51,30.99) and (291.98,30.25) .. (294.24,31.09) .. controls (296.17,32.27) and (297.83,31.92) .. (299.22,30.04) .. controls (301.01,28.39) and (302.66,28.51) .. (304.17,30.42) .. controls (305.18,32.45) and (306.74,33.02) .. (308.83,32.12) .. controls (311.2,31.6) and (312.57,32.54) .. (312.96,34.94) .. controls (312.92,37.24) and (314.05,38.48) .. (316.34,38.66) .. controls (318.63,39.18) and (319.42,40.6) .. (318.69,42.92) .. controls (317.6,44.93) and (318.02,46.55) .. (319.94,47.78) -- (320.06,50) ;
\draw    (280,50) .. controls (281.77,51.48) and (281.99,53.16) .. (280.68,55.03) .. controls (279.7,57.26) and (280.35,58.81) .. (282.63,59.66) .. controls (284.88,60.07) and (285.88,61.41) .. (285.63,63.68) .. controls (285.77,66.1) and (287.04,67.17) .. (289.45,66.88) .. controls (291.62,66.2) and (293.11,66.92) .. (293.9,69.05) .. controls (295.17,71.11) and (296.79,71.42) .. (298.75,69.99) .. controls (300.26,68.29) and (301.92,68.13) .. (303.73,69.52) .. controls (305.92,70.6) and (307.5,70.02) .. (308.47,67.78) .. controls (308.94,65.59) and (310.32,64.69) .. (312.6,65.08) .. controls (315.04,65.1) and (316.21,63.89) .. (316.12,61.46) .. controls (315.62,59.23) and (316.46,57.82) .. (318.64,57.22) .. controls (320.79,56.13) and (321.22,54.53) .. (319.94,52.42) -- (320,50) ;
\draw [color={rgb, 255:red, 245; green, 166; blue, 35 }  ,draw opacity=1 ]   (280.08,50) -- (320.08,50) ;
\draw  [color={rgb, 255:red, 245; green, 166; blue, 35 }  ,draw opacity=1 ] (460,50) .. controls (460,38.95) and (468.95,30) .. (480,30) .. controls (491.05,30) and (500,38.95) .. (500,50) .. controls (500,61.05) and (491.05,70) .. (480,70) .. controls (468.95,70) and (460,61.05) .. (460,50) -- cycle ;
\draw [color={rgb, 255:red, 0; green, 0; blue, 0 }  ,draw opacity=1 ]   (460,50) .. controls (461.67,48.33) and (463.33,48.33) .. (465,50) .. controls (466.67,51.67) and (468.33,51.67) .. (470,50) .. controls (471.67,48.33) and (473.33,48.33) .. (475,50) .. controls (476.67,51.67) and (478.33,51.67) .. (480,50) .. controls (481.67,48.33) and (483.33,48.33) .. (485,50) .. controls (486.67,51.67) and (488.33,51.67) .. (490,50) .. controls (491.67,48.33) and (493.33,48.33) .. (495,50) .. controls (496.67,51.67) and (498.33,51.67) .. (500,50) -- (500,50) ;
\draw  [color={rgb, 255:red, 245; green, 166; blue, 35 }  ,draw opacity=1 ][dash pattern={on 4.5pt off 4.5pt}] (520,50) .. controls (520,38.95) and (528.95,30) .. (540,30) .. controls (551.05,30) and (560,38.95) .. (560,50) .. controls (560,61.05) and (551.05,70) .. (540,70) .. controls (528.95,70) and (520,61.05) .. (520,50) -- cycle ;
\draw [color={rgb, 255:red, 0; green, 0; blue, 0 }  ,draw opacity=1 ]   (520,50) .. controls (521.67,48.33) and (523.33,48.33) .. (525,50) .. controls (526.67,51.67) and (528.33,51.67) .. (530,50) .. controls (531.67,48.33) and (533.33,48.33) .. (535,50) .. controls (536.67,51.67) and (538.33,51.67) .. (540,50) .. controls (541.67,48.33) and (543.33,48.33) .. (545,50) .. controls (546.67,51.67) and (548.33,51.67) .. (550,50) .. controls (551.67,48.33) and (553.33,48.33) .. (555,50) .. controls (556.67,51.67) and (558.33,51.67) .. (560,50) -- (560,50) ;
\draw    (400,50) .. controls (401.77,51.47) and (401.99,53.15) .. (400.68,55.03) .. controls (399.7,57.26) and (400.35,58.81) .. (402.63,59.66) .. controls (404.88,60.06) and (405.88,61.4) .. (405.63,63.68) .. controls (405.77,66.1) and (407.04,67.17) .. (409.45,66.88) .. controls (411.62,66.2) and (413.11,66.92) .. (413.9,69.05) .. controls (415.17,71.11) and (416.79,71.42) .. (418.75,69.98) .. controls (420.26,68.29) and (421.92,68.13) .. (423.73,69.51) .. controls (425.92,70.59) and (427.5,70.01) .. (428.46,67.76) .. controls (428.93,65.56) and (430.3,64.65) .. (432.58,65.04) .. controls (435.02,65.05) and (436.19,63.84) .. (436.09,61.41) .. controls (435.6,59.18) and (436.44,57.77) .. (438.62,57.17) .. controls (440.77,56.1) and (441.2,54.5) .. (439.92,52.39) -- (440,50) ;
\draw [color={rgb, 255:red, 245; green, 166; blue, 35 }  ,draw opacity=1 ]   (400,50) -- (440,50) ;
\draw  [draw opacity=0][dash pattern={on 4.5pt off 4.5pt}] (400.01,49.41) .. controls (400.32,38.64) and (409.15,30) .. (420,30) .. controls (431.05,30) and (440,38.95) .. (440,50) -- (420,50) -- cycle ; \draw  [color={rgb, 255:red, 245; green, 166; blue, 35 }  ,draw opacity=1 ][dash pattern={on 4.5pt off 4.5pt}] (400.01,49.41) .. controls (400.32,38.64) and (409.15,30) .. (420,30) .. controls (431.05,30) and (440,38.95) .. (440,50) ;  
\draw    (180.04,9.96) .. controls (182.07,8.52) and (183.73,8.82) .. (185.02,10.86) .. controls (185.86,12.97) and (187.36,13.64) .. (189.51,12.89) .. controls (191.98,12.56) and (193.33,13.6) .. (193.57,16) .. controls (193.42,18.27) and (194.48,19.54) .. (196.74,19.8) .. controls (199.06,20.47) and (199.8,21.94) .. (198.95,24.2) .. controls (197.78,26.13) and (198.13,27.79) .. (200.01,29.18) .. controls (201.66,30.98) and (201.53,32.64) .. (199.62,34.16) .. controls (197.62,35.11) and (197.07,36.63) .. (197.98,38.73) .. controls (198.51,41.12) and (197.58,42.52) .. (195.17,42.95) .. controls (192.86,42.96) and (191.67,44.08) .. (191.59,46.31) .. controls (191.01,48.68) and (189.56,49.5) .. (187.23,48.78) .. controls (185.25,47.68) and (183.64,48.08) .. (182.4,49.97) -- (179.89,50.02) ;
\draw    (180.04,9.96) .. controls (178.57,11.73) and (176.89,11.96) .. (175.01,10.65) .. controls (172.78,9.66) and (171.23,10.31) .. (170.38,12.59) .. controls (169.98,14.85) and (168.64,15.85) .. (166.37,15.59) .. controls (163.94,15.74) and (162.88,17.01) .. (163.17,19.42) .. controls (163.84,21.59) and (163.12,23.07) .. (160.99,23.86) .. controls (158.93,25.13) and (158.62,26.75) .. (160.06,28.71) .. controls (161.76,30.18) and (161.96,31.81) .. (160.65,33.61) .. controls (159.82,35.97) and (160.57,37.49) .. (162.9,38.17) .. controls (165.13,38.34) and (166.21,39.59) .. (166.12,41.92) .. controls (166.36,44.35) and (167.64,45.4) .. (169.97,45.07) .. controls (172.24,44.56) and (173.68,45.43) .. (174.31,47.69) .. controls (175.1,49.9) and (176.61,50.54) .. (178.86,49.61) -- (180.14,49.99) ;
\draw    (240,10) .. controls (242.03,8.56) and (243.69,8.86) .. (244.97,10.9) .. controls (245.82,13.01) and (247.31,13.68) .. (249.46,12.93) .. controls (251.93,12.6) and (253.28,13.63) .. (253.52,16.03) .. controls (253.37,18.31) and (254.43,19.58) .. (256.7,19.84) .. controls (259.01,20.51) and (259.75,21.98) .. (258.91,24.24) .. controls (257.73,26.17) and (258.08,27.83) .. (259.96,29.22) .. controls (261.61,31.03) and (261.46,32.68) .. (259.53,34.16) .. controls (257.47,35.12) and (256.85,36.67) .. (257.66,38.8) .. controls (258.16,41.11) and (257.2,42.43) .. (254.77,42.78) .. controls (252.45,42.73) and (251.22,43.83) .. (251.09,46.08) .. controls (250.5,48.44) and (249.06,49.27) .. (246.75,48.56) .. controls (244.68,47.53) and (243.05,47.99) .. (241.86,49.93) -- (240,50.06) ;
\draw    (240,10) .. controls (238.52,11.77) and (236.84,11.99) .. (234.97,10.68) .. controls (232.74,9.7) and (231.19,10.35) .. (230.34,12.63) .. controls (229.93,14.88) and (228.59,15.88) .. (226.32,15.63) .. controls (223.9,15.77) and (222.83,17.04) .. (223.12,19.45) .. controls (223.8,21.62) and (223.08,23.11) .. (220.95,23.9) .. controls (218.89,25.17) and (218.58,26.79) .. (220.01,28.75) .. controls (221.71,30.25) and (221.87,31.9) .. (220.5,33.71) .. controls (219.45,35.91) and (220.06,37.47) .. (222.32,38.38) .. controls (224.57,38.87) and (225.51,40.21) .. (225.12,42.42) .. controls (225.15,44.85) and (226.37,46.01) .. (228.8,45.88) .. controls (231.03,45.38) and (232.45,46.22) .. (233.04,48.41) .. controls (234.15,50.6) and (235.76,51.08) .. (237.85,49.84) -- (240.14,50) ;
\draw    (270.02,140) .. controls (268.07,141.52) and (266.39,141.27) .. (264.97,139.26) .. controls (264.13,137.19) and (262.62,136.55) .. (260.45,137.32) .. controls (258.04,137.72) and (256.68,136.71) .. (256.37,134.29) .. controls (256.54,132.06) and (255.48,130.82) .. (253.21,130.58) .. controls (250.88,129.92) and (250.12,128.41) .. (250.92,126.06) .. controls (252.14,124.19) and (251.84,122.57) .. (250.01,121.2) .. controls (248.44,119.28) and (248.65,117.64) .. (250.63,116.27) .. controls (252.71,115.35) and (253.34,113.82) .. (252.53,111.68) .. controls (252.06,109.29) and (253.04,107.91) .. (255.45,107.53) .. controls (257.7,107.59) and (258.89,106.47) .. (259.02,104.16) .. controls (259.52,101.79) and (260.97,100.92) .. (263.37,101.53) .. controls (265.34,102.56) and (266.89,102.1) .. (268.02,100.14) -- (270.16,100) ;
\draw    (270.01,139.94) .. controls (271.52,138.21) and (273.19,138) .. (275,139.31) .. controls (277.23,140.25) and (278.73,139.57) .. (279.5,137.26) .. controls (279.83,134.99) and (281.15,133.94) .. (283.46,134.11) .. controls (285.9,133.89) and (286.94,132.61) .. (286.58,130.28) .. controls (285.95,128.07) and (286.71,126.58) .. (288.86,125.83) .. controls (290.98,124.55) and (291.35,122.9) .. (289.98,120.87) .. controls (288.3,119.39) and (288.13,117.74) .. (289.48,115.93) .. controls (290.48,113.74) and (289.84,112.24) .. (287.57,111.41) .. controls (285.3,110.98) and (284.32,109.64) .. (284.65,107.41) .. controls (284.55,104.95) and (283.32,103.83) .. (280.95,104.04) .. controls (278.69,104.53) and (277.24,103.68) .. (276.61,101.48) .. controls (275.56,99.32) and (273.99,98.87) .. (271.88,100.12) -- (270.02,100) ;
\draw  [fill={rgb, 255:red, 255; green, 255; blue, 255 }  ,fill opacity=1 ] (266.74,96.23) .. controls (268.82,94.41) and (271.98,94.64) .. (273.79,96.72) .. controls (275.6,98.8) and (275.38,101.96) .. (273.3,103.77) .. controls (271.21,105.59) and (268.06,105.36) .. (266.24,103.28) .. controls (264.43,101.2) and (264.65,98.04) .. (266.74,96.23) -- cycle ;
\draw [fill={rgb, 255:red, 255; green, 255; blue, 255 }  ,fill opacity=1 ]   (266.48,96.46) -- (273.55,103.54) ;
\draw [fill={rgb, 255:red, 255; green, 255; blue, 255 }  ,fill opacity=1 ]   (273.55,96.46) -- (266.48,103.54) ;

\draw (150.68,49.2) node    {$+$};
\draw (210.68,49.2) node    {$+$};
\draw (180.6,118.2) node    {$+$};
\draw (113.92,110.4) node [anchor=north west][inner sep=0.75pt]    {$+$};
\draw (50,49) node    {$V_{\text{NLO}} =V_{\text{NLO}}^{\rm ct}+$};
\draw (267.68,49.2) node    {$+$};
\draw (330.68,49.2) node    {$+$};
\draw (390.68,49.2) node    {$+$};
\draw (450.68,49.2) node    {$+$};
\draw (510.68,49.2) node    {$+$};
\draw (240.6,119.2) node    {$+$};

\end{tikzpicture}
    \end{gathered}
    \end{equation}
\vspace{-0.55\baselineskip}
\end{widetext}
Every diagram above is suppressed by exactly $\alpha^2$ relative to $V_{\rm
LO}$. Diagrams containing additional mass insertions, additional
gauge vertices, or the scalar self-interaction are parametrically higher order.

The N${}^{2}$LO potential arises from the one-loop vacuum diagrams within
the three-dimensional EFT. The EFT describes the light complex scalar field $\phi_\text{IR}$, and has a global
$\mathrm{U}(1)$ symmetry.
So, there are two one-loop vacuum diagrams, one each from
the Higgs and Goldstone
bosons, with masses $V_{\rm EFT}''$ and $V_{\rm EFT}'/v$ respectively.

We have not directly investigated gauge dependence here. However, as the method produces an asymptotic expansion in powers of $\alpha$ for observables, the uniqueness of the coefficients of such an asymptotic expansion should result in order-by-order gauge invariance~\cite{Fukuda:1975di, Bender:1999box, Ekstedt:2020abj, Hirvonen:2021zej}.

\subsection{Non-Abelian Higgs model}
Finally, we consider the non-Abelian Georgi--Glashow model, an
$\mathrm{SU}(2)\simeq \mathrm{SO}(3)$
gauge theory coupled to a real scalar in the adjoint representation,
\begin{equation}
\begin{aligned}
\mathcal{L}_\text{E}^{\rm SU(2)}
=&\frac14 F_{\mu\nu}^aF_{\mu\nu}^a
 +\frac12(D_\mu\phi)^a(D_\mu\phi)^a
 +V(\phi)\notag\\
&+\frac{1}{2\xi}(\partial_\mu A_\mu^a)^2
 +\partial_\mu\bar c^a(D_\mu c)^a,
\end{aligned}
\end{equation}
with
\begin{equation}
V(\phi)=-\frac12\mu_\phi^2\,\phi^a\phi^a
+\frac14\lambda(\phi^a\phi^a)^2.
\end{equation}
Here $a=1,2,3$ labels the adjoint representation, and
\begin{align}
F_{\mu\nu}^a
&=\partial_\mu A_\nu^a-\partial_\nu A_\mu^a
 +g\epsilon^{abc}A_\mu^bA_\nu^c,\\
(D_\mu\phi)^a
&=\partial_\mu\phi^a+g\epsilon^{abc}A_\mu^b\phi^c.
\end{align}
Again, we work in Landau gauge $\xi\to 0_+$.
The power counting is the same as the Abelian case,
\begin{align}
\{\mu_\phi^2,\lambda,g^2\}
\rightarrow
\{\alpha^2\mu_\phi^2,\alpha^4\lambda,\alpha^2g^2\},
\qquad \alpha\rightarrow0^+,
\end{align}
and implies that $\phi_{\rm IR}=O(\alpha^{-1}T)$. The new ingredient is the colour structure of the gauge sector.
We choose the background field along the first adjoint direction,
\begin{align}
\phi_{\rm IR}^a
=\phi_{\rm IR}\,\delta^{a1}
=(\phi_{\rm IR},0,0),
\end{align}
breaking $\mathrm{SU}(2)\to \mathrm{U}(1)\simeq \mathrm{SO}(2)$. The gauge-field mass matrix generated by this background is
\begin{align}
(M_A^2)^{ab}
=g^2\left(\phi_{\rm IR}^2\delta^{ab}
-\phi_{\rm IR}^a\phi_{\rm IR}^b\right),
\end{align}
with eigenvalues
\begin{align}
m_{A^2}^2=m_{A^3}^2
=g^2\phi_{\rm IR}^2\equiv m_A^2,
\qquad m_{A^1}^2=0.
\end{align}
Thus the two gauge fields associated with the broken generators are hard for
$v=O(T/\alpha)$, while the gauge field of the unbroken $\mathrm{U}(1)$ remains
massless. As in the Abelian model, the masses of the heavy gauge fields
shift parametrically between phases, with their dominant effects arising when they
are heaviest~\cite{Gould:2021ccf}.

The scalar fluctuations separate into one Higgs and two Goldstone modes, with masses
\begin{equation}
\begin{aligned}
m_H^2(\phi_{\rm IR})&=-\mu_\phi^2+3\lambda\phi_{\rm IR}^2,\\
m_G^2(\phi_{\rm IR})&=-\mu_\phi^2+\lambda\phi_{\rm IR}^2,\\
\end{aligned}
\end{equation}
so that
\begin{align}
m_H^2,m_G^2=O(\alpha^2T^2),
\qquad m_A^2=O(\alpha^0T^2).
\end{align}
As in the Abelian theory, the scalar masses are treated as perturbative insertions, whereas the full mass of the heavy gauge fields is kept in the propagator.
As in the Abelian model, $m_A$ is much larger than the Debye mass for $A^2$ and $A^3$, so that thermal corrections to the gauge field propagator are subdominant for these components,
and do not need resumming.
However, as $A^1$ does not get a (hard) background-dependent mass, it does require thermal resummation.
The hierarchies of scale in this model are summarised in Table~\ref{tab:hierarchy_strong_su2}.

\begin{table}[t]
    \centering
    \begin{tabular}{c|c}
        $m_{\rm eff}$ & Fields \\
        \hline
        $\pi T$         & $A^{1,\text{UV}}_\mu$, $A^2_\mu$, $A^3_\mu$, $\phi^\text{UV}$ \\
        $g T$           & $A^{1,\text{IR}}_0$, $\phi^\text{IR}$ \\
        $m_\gamma$           & $A^{1,\text{IR}}_i$
    \end{tabular}
    \caption{Scale hierarchies for strong transitions in the  Georgi--Glashow model, with $\phi^1_\text{IR}=O(T/g)$. The mass $m_\gamma$ is proportional to the magnetic monopole density~\cite{Niemi:2022bjg}, which lies at the ultrasoft scale or below.}
    \label{tab:hierarchy_strong_su2}
\end{table}

In Landau gauge, collecting the scalar, gauge and ghost contributions gives the leading potential,
\begin{align}
V_{\rm LO}
=V_{\rm tree}+\frac{3}{2}J^0(0)+\frac{6}{2}J^0(m_A)-\frac{4}{2}J^{-1}(m_A).
\end{align}
Here the $J^0(0)$ term collects the massless contributions, and
the two broken gauge directions give
$J^0(m_A)$ and $J^{-1}(m_A)$. For the explicit expression please see Eq.~\ref{eq:SU(2)LOapp}.

The NLO correction contains the Abelian diagrams of the previous subsection, now
with the appropriate $SU(2)$ multiplicities. 
In addition, there are the following additional diagrams involving vector self-interactions and
vector--ghost interactions
\begin{equation}
\begin{gathered}
\tikzset{every picture/.style={line width=0.75pt}} 
\begin{tikzpicture}[x=0.75pt,y=0.75pt,yscale=-1,xscale=1]
\draw    (120,50) .. controls (121.62,48.29) and (123.32,48.15) .. (125.1,49.57) .. controls (127.17,50.74) and (128.72,50.19) .. (129.77,47.94) .. controls (130.24,45.71) and (131.58,44.76) .. (133.81,45.09) .. controls (136.28,44.86) and (137.32,43.57) .. (136.94,41.21) .. controls (136.21,39.01) and (136.9,37.49) .. (138.99,36.65) .. controls (141.03,35.38) and (141.35,33.73) .. (139.94,31.69) .. controls (138.3,30.13) and (138.21,28.49) .. (139.68,26.78) .. controls (140.83,24.65) and (140.29,23.08) .. (138.08,22.07) .. controls (135.85,21.5) and (134.94,20.08) .. (135.35,17.81) .. controls (135.46,15.45) and (134.3,14.3) .. (131.86,14.35) .. controls (129.61,14.78) and (128.19,13.9) .. (127.62,11.69) .. controls (126.61,9.5) and (125.01,9) .. (122.84,10.19) .. controls (121.16,11.72) and (119.51,11.72) .. (117.9,10.19) .. controls (115.75,8.97) and (114.15,9.44) .. (113.08,11.59) .. controls (112.43,13.78) and (111,14.61) .. (108.77,14.08) .. controls (106.35,13.91) and (105.11,15.04) .. (105.06,17.47) .. controls (105.39,19.74) and (104.44,21.1) .. (102.21,21.56) .. controls (99.96,22.47) and (99.38,24) .. (100.45,26.17) .. controls (101.85,27.94) and (101.71,29.58) .. (100.02,31.09) .. controls (98.53,33) and (98.78,34.65) .. (100.76,36.05) .. controls (102.82,37.06) and (103.41,38.61) .. (102.52,40.71) .. controls (102.05,43.16) and (103.02,44.55) .. (105.41,44.87) .. controls (107.65,44.69) and (108.93,45.72) .. (109.25,47.95) .. controls (110.2,50.24) and (111.76,50.83) .. (113.94,49.73) .. controls (115.69,48.3) and (117.38,48.41) .. (119.01,50.08) -- cycle ;
\draw    (159.92,50) .. controls (158.48,47.97) and (158.78,46.31) .. (160.82,45.03) .. controls (162.93,44.18) and (163.6,42.69) .. (162.85,40.54) .. controls (162.52,38.07) and (163.55,36.72) .. (165.95,36.48) .. controls (168.23,36.63) and (169.5,35.57) .. (169.75,33.3) .. controls (170.43,30.99) and (171.9,30.25) .. (174.15,31.09) .. controls (176.09,32.27) and (177.75,31.92) .. (179.14,30.04) .. controls (180.93,28.39) and (182.59,28.51) .. (184.11,30.42) .. controls (185.06,32.43) and (186.58,32.98) .. (188.69,32.07) .. controls (191.07,31.54) and (192.48,32.47) .. (192.91,34.88) .. controls (192.92,37.18) and (194.04,38.37) .. (196.27,38.45) .. controls (198.64,39.04) and (199.46,40.49) .. (198.73,42.82) .. controls (197.63,44.8) and (198.03,46.41) .. (199.93,47.64) -- (199.97,50.16) ;
\draw    (159.92,50) .. controls (161.68,51.48) and (161.91,53.16) .. (160.6,55.03) .. controls (159.62,57.26) and (160.27,58.81) .. (162.55,59.66) .. controls (164.8,60.07) and (165.8,61.41) .. (165.54,63.68) .. controls (165.69,66.1) and (166.96,67.17) .. (169.37,66.88) .. controls (171.54,66.2) and (173.03,66.92) .. (173.82,69.05) .. controls (175.09,71.11) and (176.71,71.42) .. (178.67,69.99) .. controls (180.18,68.29) and (181.84,68.13) .. (183.65,69.52) .. controls (185.84,70.6) and (187.42,70.02) .. (188.39,67.78) .. controls (188.86,65.59) and (190.24,64.69) .. (192.51,65.08) .. controls (194.95,65.1) and (196.13,63.89) .. (196.04,61.46) .. controls (195.54,59.23) and (196.38,57.82) .. (198.56,57.22) .. controls (200.71,56.13) and (201.14,54.53) .. (199.85,52.42) -- (199.92,50) ;
\draw [color={rgb, 255:red, 0; green, 0; blue, 0 }  ,draw opacity=1 ]   (160,50) .. controls (161.67,48.33) and (163.33,48.33) .. (165,50) .. controls (166.67,51.67) and (168.33,51.67) .. (170,50) .. controls (171.67,48.33) and (173.33,48.33) .. (175,50) .. controls (176.67,51.67) and (178.33,51.67) .. (180,50) .. controls (181.67,48.33) and (183.33,48.33) .. (185,50) .. controls (186.67,51.67) and (188.33,51.67) .. (190,50) .. controls (191.67,48.33) and (193.33,48.33) .. (195,50) .. controls (196.67,51.67) and (198.33,51.67) .. (200,50) -- (200,50) ;
\draw  [color={rgb, 255:red, 0; green, 0; blue, 0 }  ,draw opacity=1 ][dash pattern={on 0.84pt off 2.51pt}] (220,50) .. controls (220,38.95) and (228.95,30) .. (240,30) .. controls (251.05,30) and (260,38.95) .. (260,50) .. controls (260,61.05) and (251.05,70) .. (240,70) .. controls (228.95,70) and (220,61.05) .. (220,50) -- cycle ;
\draw [color={rgb, 255:red, 0; green, 0; blue, 0 }  ,draw opacity=1 ]   (220,50) .. controls (221.67,48.33) and (223.33,48.33) .. (225,50) .. controls (226.67,51.67) and (228.33,51.67) .. (230,50) .. controls (231.67,48.33) and (233.33,48.33) .. (235,50) .. controls (236.67,51.67) and (238.33,51.67) .. (240,50) .. controls (241.67,48.33) and (243.33,48.33) .. (245,50) .. controls (246.67,51.67) and (248.33,51.67) .. (250,50) .. controls (251.67,48.33) and (253.33,48.33) .. (255,50) .. controls (256.67,51.67) and (258.33,51.67) .. (260,50) -- (260,50) ;
\draw  [draw opacity=0] (239.49,30.01) .. controls (239.66,30) and (239.83,30) .. (240,30) .. controls (242.12,30) and (244.16,30.33) .. (246.08,30.94) -- (240,50) -- cycle ; \draw [color={rgb, 255:red, 0; green, 0; blue, 0 }  ,draw opacity=1 ]   (239.49,30.01) .. controls (239.66,30) and (239.83,30) .. (240,30) .. controls (241.08,30) and (242.14,30.09) .. (243.18,30.25) ; \draw [shift={(246.08,30.94)}, rotate = 180.45] [fill={rgb, 255:red, 0; green, 0; blue, 0 }  ,fill opacity=1 ][line width=0.08]  [draw opacity=0] (8.93,-4.29) -- (0,0) -- (8.93,4.29) -- cycle    ; 
\draw  [draw opacity=0] (242.63,69.83) .. controls (241.77,69.94) and (240.89,70) .. (240,70) .. controls (238.06,70) and (236.19,69.72) .. (234.42,69.21) -- (240,50) -- cycle ; \draw [color={rgb, 255:red, 0; green, 0; blue, 0 }  ,draw opacity=1 ]   (242.63,69.83) .. controls (241.77,69.94) and (240.89,70) .. (240,70) .. controls (239.1,70) and (238.21,69.94) .. (237.34,69.83) ; \draw [shift={(234.42,69.21)}, rotate = 358.94] [fill={rgb, 255:red, 0; green, 0; blue, 0 }  ,fill opacity=1 ][line width=0.08]  [draw opacity=0] (8.93,-4.29) -- (0,0) -- (8.93,4.29) -- cycle    ; 
\draw    (120,50) .. controls (121.98,48.49) and (123.68,48.71) .. (125.1,50.68) .. controls (126.05,52.71) and (127.6,53.29) .. (129.77,52.44) .. controls (132.07,51.89) and (133.42,52.85) .. (133.81,55.3) .. controls (133.63,57.53) and (134.67,58.8) .. (136.94,59.1) .. controls (139.25,59.89) and (139.95,61.41) .. (139.04,63.67) .. controls (137.81,65.61) and (138.11,67.22) .. (139.95,68.5) .. controls (141.62,70.31) and (141.5,72.01) .. (139.59,73.59) .. controls (137.56,74.55) and (136.98,76.07) .. (137.87,78.16) .. controls (138.38,80.52) and (137.44,81.89) .. (135.04,82.28) .. controls (132.73,82.28) and (131.51,83.43) .. (131.36,85.73) .. controls (130.83,88.1) and (129.39,88.96) .. (127.06,88.3) .. controls (125.03,87.3) and (123.47,87.79) .. (122.39,89.76) .. controls (120.72,91.59) and (119.03,91.61) .. (117.31,89.83) .. controls (116.22,87.87) and (114.67,87.4) .. (112.66,88.41) .. controls (110.32,89.04) and (108.9,88.17) .. (108.39,85.78) .. controls (108.28,83.47) and (107.07,82.29) .. (104.75,82.22) .. controls (102.42,81.85) and (101.52,80.48) .. (102.06,78.11) .. controls (102.97,76.02) and (102.41,74.47) .. (100.38,73.46) .. controls (98.48,71.85) and (98.36,70.2) .. (100.01,68.51) .. controls (101.76,66.92) and (101.9,65.23) .. (100.45,63.45) .. controls (99.26,61.31) and (99.77,59.74) .. (101.99,58.75) .. controls (104.2,58.44) and (105.25,57.17) .. (105.15,54.94) .. controls (105.77,52.52) and (107.21,51.69) .. (109.47,52.44) .. controls (111.53,53.47) and (113.14,52.96) .. (114.31,50.93) .. controls (115.74,48.96) and (117.37,48.68) .. (119.19,50.1) -- cycle ;

\draw (70,49) node    {$V_{\text{NLO}} \supset $};
\draw (147.68,49.2) node    {$+$};
\draw (210.68,49.2) node    {$+$};

\end{tikzpicture}
\end{gathered}
\end{equation}
These and the scalar insertion diagrams and gauge--scalar diagrams inherited
from the Abelian theory are all suppressed by $\alpha^2$ relative to $V_{\rm
LO}$. Again, the power counting fixes the calculation unambiguously.

The N${}^2$LO corrections arise from one-loop vacuum diagrams within the EFT.
The three-dimensional EFT is particularly interesting in this case. It
contains two neutral real
scalars $\phi_\text{IR}^1$ and $A_{0,\text{IR}}^1$, a charged complex scalar $(\phi_\text{IR}^2+i\phi_\text{IR}^3)/\sqrt{2}$ coupled to the $\mathrm{U}(1)$ vector field $A^1_{i,\text{IR}}$, plus
corresponding ghosts. Note that $A_{0,\text{IR}}^1$ also gains
self-interactions
through its couplings to the heavy gauge fields, but this is not
needed to compute the potential for $\phi_\text{IR}^1$ at N${}^2$LO.
The contribution of $A^1_{i,\text{IR}}$ is zero at one-loop, as it remains massless at this order. It receives a small non-perturbative mass proportional to the magnetic monopole density~\cite{Niemi:2022bjg}.

\subsection{Extensions of the Standard Model}

In realistic particle physics models, which contain and extend the Standard
Model, there are inevitably a large number of masses and couplings. Determining
appropriate power-counting relations to construct the relevant strong phase
transition EFT may then seem like a hopelessly difficult task, even if the model
is weakly coupled at zero temperature. However, our
analysis suggests that the appropriate power-counting is captured by a simple recipe, as follows.

We denote by $\phi$ the order parameter field, or collection of fields. We assume that $\phi$ has some self-couplings as well as some cross-couplings to other fields. Then the conditions on the couplings determining when the strong transition EFT describes a putative phase transition are as follows:
\begin{enumerate}
    \item $\phi$ self-couplings are anomalously small; \label{enum:intermediate_and_strong_a}
    \item or, a $\phi$ cross-coupling is anomalously large. \label{enum:intermediate_and_strong_b}
\end{enumerate}
For both conditions, the relevant ratio is the quartic $\phi$ self-coupling divided by the largest $\phi$ cross-coupling. In other words, the coupling $\lambda_\phi$ in our two-scalar example maps to the quartic $\phi$ self-coupling, and the coupling $g^2$ maps to the
quartic
$\phi$ cross-coupling. The strong phase transition scenario requires this ratio to be sufficiently small.

We have mostly used the phrasing of condition~\ref{enum:intermediate_and_strong_a} above, but condition~\ref{enum:intermediate_and_strong_b} is symmetric, as only the ratio of couplings matters. Condition~\ref{enum:intermediate_and_strong_b} arises naturally for scalar extensions of the Standard Model, as the Higgs self-coupling is essentially fixed~\cite{Ramsey-Musolf:2019lsf, Niemi:2024vzw}. For instance, with a new scalar $S$, coupling to the Standard Model through
\begin{align}
    \mathcal{L} \supset \frac{1}{2}a_2 H^\dagger H S^2 ,
\end{align}
a strong phase transition would occur for $\lambda \sim a_2^{2}/\pi^2$. The $O(1)$ numerical coefficient in this relation is model dependent.

All this assumes the order parameter does undergo some transition, e.g.\ that the vacuum mass squared parameter is negative. Further, the field masses must not be so large as to give only exponentially suppressed contributions $\propto e^{-m/T}$ at the transition. For the order parameter field, this holds almost automatically, following the argument of Sec.~\ref{sec:scale_hierarchies}. However, for the fields coupled to the order parameter, those fields with masses $m\gg T$ can be completely dropped when considering the phase transition, as their thermal contributions are exponentially small.

\section{Discussion}
\label{sec:discussion}

Finding the appropriate expansion is not merely a calculational step; it is part of identifying the physics. This is perhaps the central lesson of the present work. Rather than beginning by asking how many loops to calculate, we begin by determining how couplings, masses and background fields must scale in the regime of interest. Expressing this power counting through an auxiliary parameter $\alpha$ defines a perturbative expansion adapted to that regime: it fixes which contributions must enter at leading order, which quantities must be retained unexpanded, and which corrections are suppressed. The hierarchy therefore determines the expansion, fixing the necessary resummations and truncations of the loop expansion~\cite{Lofgren:2023sep}.

We have used this approach to classify thermal EFTs by the strength of the phase transitions that they can faithfully capture. The well known thermal EFTs in the literature describe either weak or intermediate strength transitions, and we identify a new class of EFTs describing strong phase transitions. The latent heat in these classes of EFTs satisfy
\begin{align*}
    0 \leq\  &L_\text{weak}\ \ll O(g^2T^4) ,\\
    O(g^4T^4) \ll\ &L_\text{intermediate}\ \ll O(T^4) ,\\
    O(g^2T^4) \ll\ &L_\text{strong}\ \leq O(T^4) .
\end{align*}
Importantly, the ranges of validity overlap, however as one approaches the edge of validity of one class of EFTs, higher dimensional operators become increasingly important, until eventually a new class of EFT is required.
Within a strong transition EFT, calculations of phase transition parameters remain under perturbative control, as long as the model is weakly coupled at zero temperature.
Ref.~\cite{Camargo-Molina:2024sde} carried out a similar power-counting analysis of phase transition parameters for the Standard Model EFT.

Going further, we have explicitly constructed
EFTs for strong first-order phase transitions. In this case, the fields inducing the transition become hard while the order parameter field remains soft. The tree level potential and the leading hard-scale loop contributions
belong to the same perturbative order. The physics responsible for the transition is therefore present already at leading order, while subsequent corrections are selected by their powers of $\alpha$, rather than loop number. The result is a systematically improvable perturbative description constructed for the strong-transition regime, and not an extrapolation of an expansion appropriate to weaker transitions.

In the two-scalar model, we have carried out the hard-scale matching through NLO and included the first infrared correction at N$^2$LO. The cancellation of renormalisation scale dependence
and the decreasing successive corrections to the latent heat provide analytical and numerical tests of the validity of the expansion. In the benchmark studied, the strong transition EFT performs better where its hierarchy is realised, while the intermediate strength EFT becomes more appropriate as this hierarchy weakens. Their overlapping regimes of validity show that these are complementary descriptions, rather than interchangeable approximations over the entire phase diagram.

The same construction applies to much more general classes of theories, including those with scalars, fermions and gauge fields. We illustrated this by constructing the leading order potentials and NLO diagram sets for a Yukawa model, the Abelian Higgs model, and the Georgi-Glashow model. These include both symmetry-breaking and non-symmetry breaking transitions.
For symmetry-breaking transitions, such as in the two-scalar and gauge-Higgs models, the different classes of EFTs occur for the following relative coupling strengths
\begin{align*}
    \text{Weak:}& &\lambda &\sim g^2, \\
    \text{Intermediate:}& &\lambda &\sim g^3/\pi, \\
    \text{Strong:}& &\lambda &\sim g^4/\pi^2,
\end{align*}
with $\lambda$ representing the
quartic
self-coupling of the order parameter field, and $g^2$ representing the largest cross-coupling.
For non-symmetry-breaking transitions with a tree-level barrier, such as in the Yukawa model or a 2-step transition in the two-scalar model, we have instead
\begin{align*}
    \text{Intermediate:}& &\lambda &\sim g^2, \\
    \text{Semi-strong:}& &\lambda &\sim g^3/\pi, \\
    \text{Strong:}& &\lambda &\sim g^4/\pi^2.
\end{align*}
We did not study EFTs for semi-strong transitions in this paper, for details see Refs~\cite{Gould:2023ovu, Camargo-Molina:2024sde}, but in brief the order parameter field has a soft-scale mass and its background is of order $v\sim \sqrt{\pi}T/\sqrt{g}$, pushing the mass of a field coupled to it to the \textit{semi-soft scale} $\sim \sqrt{g\pi}T$, and yielding a typical latent heat of $L_\text{semi-strong}=O(g T^4)$.

For model building, the useful criterion for discovering strong phase transitions is the hierarchy $\lambda \sim g^4/\pi^2$, which applies irrespective of symmetry breaking.
This can be realised either with a small value of $\lambda$ or a large value of $g^2$. Whether a given parameter point realises a strong phase transition then depends on its mass spectrum and phase structure.

Classically conformal models are an apparently separate class of models that have been identified in the literature as leading to strongly first-order phase transitions~\cite{Espinosa:2008kw, Jinno:2016knw, Prokopec:2018tnq}. However, these models also require the power counting relation $\lambda\sim g^4/\pi^2$, suggesting a connection with the strong transition EFTs introduced here, which we leave for future work to develop.

Our class of strong transition EFTs are particularly relevant for gravitational wave experiments, for which the strongest transitions yield the largest amplitude signals with the best prospects of observation. Further, previous studies utilising thermal EFTs (of the intermediate strength class) have found the largest theoretical uncertainties for those parameter points with the strongest signals~\cite{Gould:2023jbz, Gould:2024jjt}; see also Fig.~\ref{fig:xie_yukawa}. This is expected from our analysis, as the high-temperature expansion underlying the intermediate strength EFTs should eventually break down as the transition strength increases.
Our construction makes the phase transition parameters systematically improvable within the strong transition hierarchy. This addresses an essential part of the prediction, rather than assuming that a large signal inferred from a leading order potential is already a reliable one.

The bubble nucleation rate is computable within a strong transition EFT using the EFT approach to thermal bubble nucleation~\cite{Gould:2021ccf}. As only modes with energies parametrically larger than the order parameter mass are integrated out, the derivative expansion is under perturbative control. For the two-scalar model, we have computed the leading derivative corrections to the EFT, which appear at NLO when evaluated on the critical bubble. An important caveat is that as the system supercools, the shape of the effective potential changes and a different set of power-counting relations may hold at the nucleation or percolation temperature, as compared to the critical temperature~\cite{Kierkla:2023von}.

Two recent works~\cite{Navarrete:2025yxy, Klose:2026pux} presented alternative approaches to the study of strong phase transitions. Ref.~\cite{Navarrete:2025yxy} constructed an effective potential which is not based on the construction of an EFT, but which agrees order by order with our strong-transition effective potential in the regime of strong phase transitions. Their potential has the advantage that it applies to a wider range of temperatures, but the disadvantage that it can only be applied to study the free energy in homogeneous phases. Applying it to inhomogeneous phenomena such as bubble nucleation would lead to an uncontrolled derivative expansion, because all modes are integrated out in the construction of Ref.~\cite{Navarrete:2025yxy}. Ref.~\cite{Klose:2026pux} came to a number of identical conclusions to us regarding the appropriate power counting relations for strong phase transitions, but we seem to differ in our final effective potentials.

Finally, for community-wide phenomenological applications, automation is an important milestone. The software package DRalgo~\cite{Ekstedt:2022bff} is able to construct thermal EFTs for generic UV models in the weak and intermediate strength classes. For constructing such EFTs, the high-temperature approximation applies all the necessary sum-integrals, in which case the two-loop sum integrals can be reduced to products of one-loop sum integrals by integration by parts~\cite{Nishimura:2012ee, Gil:2026cqz}. Generalising these automations to include strong transition EFTs, and hence sum integrals outside the high-temperature approximation, is an important next step.

\begin{acknowledgements}
The authors would like to thank
L.~Gil,
J.~Hirvonen,
J.~Löfgren,
T.V.I.~Tenkanen,
and C.~Xie
for useful discussions.
In particular, O.G.\ would like to thank C.~Xie for pointing out the trends in the data underlying Fig.~\ref{fig:xie_yukawa}.
O.G.\ and F-S.R.\ were supported by a Royal Society Dorothy Hodgkin Fellowship, grant number DHF\textbackslash{}R1\textbackslash{}221001.
P.M.S.\ acknowledges partial support from the STFC Consolidated Grant ST/X000672/1.
\end{acknowledgements}

\section*{Data availability}
The data underlying Figs.~\ref{fig:Lathentheat2scalar} and \ref{fig:two_scalar_masses} are available from Ref.~\cite{rassouli_2026_23172314}, and that for Fig.~\ref{fig:xie_yukawa} is based on existing public data from Ref.~\cite{gould_2023_10276044}.

\bibliography{refs}

\begin{thebibliography}{83}%
\makeatletter
\providecommand \@ifxundefined [1]{%
 \@ifx{#1\undefined}
}%
\providecommand \@ifnum [1]{%
 \ifnum #1\expandafter \@firstoftwo
 \else \expandafter \@secondoftwo
 \fi
}%
\providecommand \@ifx [1]{%
 \ifx #1\expandafter \@firstoftwo
 \else \expandafter \@secondoftwo
 \fi
}%
\providecommand \natexlab [1]{#1}%
\providecommand \enquote  [1]{``#1''}%
\providecommand \bibnamefont  [1]{#1}%
\providecommand \bibfnamefont [1]{#1}%
\providecommand \citenamefont [1]{#1}%
\providecommand \href@noop [0]{\@secondoftwo}%
\providecommand \href [0]{\begingroup \@sanitize@url \@href}%
\providecommand \@href[1]{\@@startlink{#1}\@@href}%
\providecommand \@@href[1]{\endgroup#1\@@endlink}%
\providecommand \@sanitize@url [0]{\catcode `\\12\catcode `\$12\catcode `\&12\catcode `\#12\catcode `\^12\catcode `\_12\catcode `\%12\relax}%
\providecommand \@@startlink[1]{}%
\providecommand \@@endlink[0]{}%
\providecommand \url  [0]{\begingroup\@sanitize@url \@url }%
\providecommand \@url [1]{\endgroup\@href {#1}{\urlprefix }}%
\providecommand \urlprefix  [0]{URL }%
\providecommand \Eprint [0]{\href }%
\providecommand \doibase [0]{https://doi.org/}%
\providecommand \selectlanguage [0]{\@gobble}%
\providecommand \bibinfo  [0]{\@secondoftwo}%
\providecommand \bibfield  [0]{\@secondoftwo}%
\providecommand \translation [1]{[#1]}%
\providecommand \BibitemOpen [0]{}%
\providecommand \bibitemStop [0]{}%
\providecommand \bibitemNoStop [0]{.\EOS\space}%
\providecommand \EOS [0]{\spacefactor3000\relax}%
\providecommand \BibitemShut  [1]{\csname bibitem#1\endcsname}%
\let\auto@bib@innerbib\@empty
\bibitem [{\citenamefont {Kapusta}(1979)}]{Kapusta:1979fh}%
  \BibitemOpen
  \bibfield  {author} {\bibinfo {author} {\bibfnamefont {J.~I.}\ \bibnamefont {Kapusta}},\ }\bibfield  {title} {\bibinfo {title} {{Quantum Chromodynamics at High Temperature}},\ }\href {https://doi.org/10.1016/0550-3213(79)90146-9} {\bibfield  {journal} {\bibinfo  {journal} {Nucl. Phys. B}\ }\textbf {\bibinfo {volume} {148}},\ \bibinfo {pages} {461} (\bibinfo {year} {1979})}\BibitemShut {NoStop}%
\bibitem [{\citenamefont {Linde}(1980)}]{Linde:1980ts}%
  \BibitemOpen
  \bibfield  {author} {\bibinfo {author} {\bibfnamefont {A.~D.}\ \bibnamefont {Linde}},\ }\bibfield  {title} {\bibinfo {title} {{Infrared Problem in Thermodynamics of the Yang-Mills Gas}},\ }\href {https://doi.org/10.1016/0370-2693(80)90769-8} {\bibfield  {journal} {\bibinfo  {journal} {Phys. Lett. B}\ }\textbf {\bibinfo {volume} {96}},\ \bibinfo {pages} {289} (\bibinfo {year} {1980})}\BibitemShut {NoStop}%
\bibitem [{\citenamefont {Carena}\ \emph {et~al.}(2020)\citenamefont {Carena}, \citenamefont {Liu},\ and\ \citenamefont {Wang}}]{Carena:2019une}%
  \BibitemOpen
  \bibfield  {author} {\bibinfo {author} {\bibfnamefont {M.}~\bibnamefont {Carena}}, \bibinfo {author} {\bibfnamefont {Z.}~\bibnamefont {Liu}},\ and\ \bibinfo {author} {\bibfnamefont {Y.}~\bibnamefont {Wang}},\ }\bibfield  {title} {\bibinfo {title} {{Electroweak phase transition with spontaneous Z$_{2}$-breaking}},\ }\href {https://doi.org/10.1007/JHEP08(2020)107} {\bibfield  {journal} {\bibinfo  {journal} {JHEP}\ }\textbf {\bibinfo {volume} {08}},\ \bibinfo {pages} {107}},\ \Eprint {https://arxiv.org/abs/1911.10206} {arXiv:1911.10206 [hep-ph]} \BibitemShut {NoStop}%
\bibitem [{\citenamefont {Croon}\ \emph {et~al.}(2021)\citenamefont {Croon}, \citenamefont {Gould}, \citenamefont {Schicho}, \citenamefont {Tenkanen},\ and\ \citenamefont {White}}]{Croon:2020cgk}%
  \BibitemOpen
  \bibfield  {author} {\bibinfo {author} {\bibfnamefont {D.}~\bibnamefont {Croon}}, \bibinfo {author} {\bibfnamefont {O.}~\bibnamefont {Gould}}, \bibinfo {author} {\bibfnamefont {P.}~\bibnamefont {Schicho}}, \bibinfo {author} {\bibfnamefont {T.~V.~I.}\ \bibnamefont {Tenkanen}},\ and\ \bibinfo {author} {\bibfnamefont {G.}~\bibnamefont {White}},\ }\bibfield  {title} {\bibinfo {title} {{Theoretical uncertainties for cosmological first-order phase transitions}},\ }\href {https://doi.org/10.1007/JHEP04(2021)055} {\bibfield  {journal} {\bibinfo  {journal} {JHEP}\ }\textbf {\bibinfo {volume} {04}},\ \bibinfo {pages} {055}},\ \Eprint {https://arxiv.org/abs/2009.10080} {arXiv:2009.10080 [hep-ph]} \BibitemShut {NoStop}%
\bibitem [{\citenamefont {Gould}\ and\ \citenamefont {Tenkanen}(2021)}]{Gould:2021oba}%
  \BibitemOpen
  \bibfield  {author} {\bibinfo {author} {\bibfnamefont {O.}~\bibnamefont {Gould}}\ and\ \bibinfo {author} {\bibfnamefont {T.~V.~I.}\ \bibnamefont {Tenkanen}},\ }\bibfield  {title} {\bibinfo {title} {{On the perturbative expansion at high temperature and implications for cosmological phase transitions}},\ }\href {https://doi.org/10.1007/JHEP06(2021)069} {\bibfield  {journal} {\bibinfo  {journal} {JHEP}\ }\textbf {\bibinfo {volume} {06}},\ \bibinfo {pages} {069}},\ \Eprint {https://arxiv.org/abs/2104.04399} {arXiv:2104.04399 [hep-ph]} \BibitemShut {NoStop}%
\bibitem [{\citenamefont {Lewicki}\ \emph {et~al.}(2024)\citenamefont {Lewicki}, \citenamefont {Merchand}, \citenamefont {Sagunski}, \citenamefont {Schicho},\ and\ \citenamefont {Schmitt}}]{Lewicki:2024xan}%
  \BibitemOpen
  \bibfield  {author} {\bibinfo {author} {\bibfnamefont {M.}~\bibnamefont {Lewicki}}, \bibinfo {author} {\bibfnamefont {M.}~\bibnamefont {Merchand}}, \bibinfo {author} {\bibfnamefont {L.}~\bibnamefont {Sagunski}}, \bibinfo {author} {\bibfnamefont {P.}~\bibnamefont {Schicho}},\ and\ \bibinfo {author} {\bibfnamefont {D.}~\bibnamefont {Schmitt}},\ }\bibfield  {title} {\bibinfo {title} {{Impact of theoretical uncertainties on model parameter reconstruction from GW signals sourced by cosmological phase transitions}},\ }\href {https://doi.org/10.1103/PhysRevD.110.023538} {\bibfield  {journal} {\bibinfo  {journal} {Phys. Rev. D}\ }\textbf {\bibinfo {volume} {110}},\ \bibinfo {pages} {023538} (\bibinfo {year} {2024})},\ \Eprint {https://arxiv.org/abs/2403.03769} {arXiv:2403.03769 [hep-ph]} \BibitemShut {NoStop}%
\bibitem [{\citenamefont {Appelquist}\ and\ \citenamefont {Pisarski}(1981)}]{Appelquist:1981vg}%
  \BibitemOpen
  \bibfield  {author} {\bibinfo {author} {\bibfnamefont {T.}~\bibnamefont {Appelquist}}\ and\ \bibinfo {author} {\bibfnamefont {R.~D.}\ \bibnamefont {Pisarski}},\ }\bibfield  {title} {\bibinfo {title} {{High-Temperature Yang-Mills Theories and Three-Dimensional Quantum Chromodynamics}},\ }\href {https://doi.org/10.1103/PhysRevD.23.2305} {\bibfield  {journal} {\bibinfo  {journal} {Phys. Rev. D}\ }\textbf {\bibinfo {volume} {23}},\ \bibinfo {pages} {2305} (\bibinfo {year} {1981})}\BibitemShut {NoStop}%
\bibitem [{\citenamefont {Nadkarni}(1983)}]{Nadkarni:1982kb}%
  \BibitemOpen
  \bibfield  {author} {\bibinfo {author} {\bibfnamefont {S.}~\bibnamefont {Nadkarni}},\ }\bibfield  {title} {\bibinfo {title} {{Dimensional Reduction in Hot QCD}},\ }\href {https://doi.org/10.1103/PhysRevD.27.917} {\bibfield  {journal} {\bibinfo  {journal} {Phys. Rev. D}\ }\textbf {\bibinfo {volume} {27}},\ \bibinfo {pages} {917} (\bibinfo {year} {1983})}\BibitemShut {NoStop}%
\bibitem [{\citenamefont {Farakos}\ \emph {et~al.}(1994)\citenamefont {Farakos}, \citenamefont {Kajantie}, \citenamefont {Rummukainen},\ and\ \citenamefont {Shaposhnikov}}]{Farakos:1994kx}%
  \BibitemOpen
  \bibfield  {author} {\bibinfo {author} {\bibfnamefont {K.}~\bibnamefont {Farakos}}, \bibinfo {author} {\bibfnamefont {K.}~\bibnamefont {Kajantie}}, \bibinfo {author} {\bibfnamefont {K.}~\bibnamefont {Rummukainen}},\ and\ \bibinfo {author} {\bibfnamefont {M.~E.}\ \bibnamefont {Shaposhnikov}},\ }\bibfield  {title} {\bibinfo {title} {{3-D physics and the electroweak phase transition: Perturbation theory}},\ }\href {https://doi.org/10.1016/0550-3213(94)90173-2} {\bibfield  {journal} {\bibinfo  {journal} {Nucl. Phys. B}\ }\textbf {\bibinfo {volume} {425}},\ \bibinfo {pages} {67} (\bibinfo {year} {1994})},\ \Eprint {https://arxiv.org/abs/hep-ph/9404201} {arXiv:hep-ph/9404201} \BibitemShut {NoStop}%
\bibitem [{\citenamefont {Kajantie}\ \emph {et~al.}(1996{\natexlab{a}})\citenamefont {Kajantie}, \citenamefont {Laine}, \citenamefont {Rummukainen},\ and\ \citenamefont {Shaposhnikov}}]{Kajantie:1995dw}%
  \BibitemOpen
  \bibfield  {author} {\bibinfo {author} {\bibfnamefont {K.}~\bibnamefont {Kajantie}}, \bibinfo {author} {\bibfnamefont {M.}~\bibnamefont {Laine}}, \bibinfo {author} {\bibfnamefont {K.}~\bibnamefont {Rummukainen}},\ and\ \bibinfo {author} {\bibfnamefont {M.~E.}\ \bibnamefont {Shaposhnikov}},\ }\bibfield  {title} {\bibinfo {title} {{Generic rules for high temperature dimensional reduction and their application to the standard model}},\ }\href {https://doi.org/10.1016/0550-3213(95)00549-8} {\bibfield  {journal} {\bibinfo  {journal} {Nucl. Phys. B}\ }\textbf {\bibinfo {volume} {458}},\ \bibinfo {pages} {90} (\bibinfo {year} {1996}{\natexlab{a}})},\ \Eprint {https://arxiv.org/abs/hep-ph/9508379} {arXiv:hep-ph/9508379} \BibitemShut {NoStop}%
\bibitem [{\citenamefont {Braaten}\ and\ \citenamefont {Nieto}(1995)}]{Braaten:1995cm}%
  \BibitemOpen
  \bibfield  {author} {\bibinfo {author} {\bibfnamefont {E.}~\bibnamefont {Braaten}}\ and\ \bibinfo {author} {\bibfnamefont {A.}~\bibnamefont {Nieto}},\ }\bibfield  {title} {\bibinfo {title} {{Effective field theory approach to high temperature thermodynamics}},\ }\href {https://doi.org/10.1103/PhysRevD.51.6990} {\bibfield  {journal} {\bibinfo  {journal} {Phys. Rev. D}\ }\textbf {\bibinfo {volume} {51}},\ \bibinfo {pages} {6990} (\bibinfo {year} {1995})},\ \Eprint {https://arxiv.org/abs/hep-ph/9501375} {arXiv:hep-ph/9501375} \BibitemShut {NoStop}%
\bibitem [{\citenamefont {Braaten}\ and\ \citenamefont {Nieto}(1996)}]{Braaten:1995jr}%
  \BibitemOpen
  \bibfield  {author} {\bibinfo {author} {\bibfnamefont {E.}~\bibnamefont {Braaten}}\ and\ \bibinfo {author} {\bibfnamefont {A.}~\bibnamefont {Nieto}},\ }\bibfield  {title} {\bibinfo {title} {{Free energy of QCD at high temperature}},\ }\href {https://doi.org/10.1103/PhysRevD.53.3421} {\bibfield  {journal} {\bibinfo  {journal} {Phys. Rev. D}\ }\textbf {\bibinfo {volume} {53}},\ \bibinfo {pages} {3421} (\bibinfo {year} {1996})},\ \Eprint {https://arxiv.org/abs/hep-ph/9510408} {arXiv:hep-ph/9510408} \BibitemShut {NoStop}%
\bibitem [{\citenamefont {Kajantie}\ \emph {et~al.}(1996{\natexlab{b}})\citenamefont {Kajantie}, \citenamefont {Laine}, \citenamefont {Rummukainen},\ and\ \citenamefont {Shaposhnikov}}]{Kajantie:1996mn}%
  \BibitemOpen
  \bibfield  {author} {\bibinfo {author} {\bibfnamefont {K.}~\bibnamefont {Kajantie}}, \bibinfo {author} {\bibfnamefont {M.}~\bibnamefont {Laine}}, \bibinfo {author} {\bibfnamefont {K.}~\bibnamefont {Rummukainen}},\ and\ \bibinfo {author} {\bibfnamefont {M.~E.}\ \bibnamefont {Shaposhnikov}},\ }\bibfield  {title} {\bibinfo {title} {{Is there a~ hot electroweak phase transition at $m_H \gtrsim m_W$?}},\ }\href {https://doi.org/10.1103/PhysRevLett.77.2887} {\bibfield  {journal} {\bibinfo  {journal} {Phys. Rev. Lett.}\ }\textbf {\bibinfo {volume} {77}},\ \bibinfo {pages} {2887} (\bibinfo {year} {1996}{\natexlab{b}})},\ \Eprint {https://arxiv.org/abs/hep-ph/9605288} {arXiv:hep-ph/9605288} \BibitemShut {NoStop}%
\bibitem [{\citenamefont {D'Onofrio}\ and\ \citenamefont {Rummukainen}(2016)}]{DOnofrio:2015gop}%
  \BibitemOpen
  \bibfield  {author} {\bibinfo {author} {\bibfnamefont {M.}~\bibnamefont {D'Onofrio}}\ and\ \bibinfo {author} {\bibfnamefont {K.}~\bibnamefont {Rummukainen}},\ }\bibfield  {title} {\bibinfo {title} {{Standard model cross-over on the lattice}},\ }\href {https://doi.org/10.1103/PhysRevD.93.025003} {\bibfield  {journal} {\bibinfo  {journal} {Phys. Rev. D}\ }\textbf {\bibinfo {volume} {93}},\ \bibinfo {pages} {025003} (\bibinfo {year} {2016})},\ \Eprint {https://arxiv.org/abs/1508.07161} {arXiv:1508.07161 [hep-ph]} \BibitemShut {NoStop}%
\bibitem [{\citenamefont {Ramsey-Musolf}\ \emph {et~al.}(2024)\citenamefont {Ramsey-Musolf}, \citenamefont {Tenkanen},\ and\ \citenamefont {Tran}}]{Ramsey-Musolf:2024ykk}%
  \BibitemOpen
  \bibfield  {author} {\bibinfo {author} {\bibfnamefont {M.~J.}\ \bibnamefont {Ramsey-Musolf}}, \bibinfo {author} {\bibfnamefont {T.~V.~I.}\ \bibnamefont {Tenkanen}},\ and\ \bibinfo {author} {\bibfnamefont {V.~Q.}\ \bibnamefont {Tran}},\ }\bibfield  {title} {\bibinfo {title} {{Refining Gravitational Wave and Collider Physics Dialogue via Singlet Scalar Extension}},\ }\href@noop {} {\bibfield  {journal} {\bibinfo  {journal} {arXiv}\ } (\bibinfo {year} {2024})},\ \Eprint {https://arxiv.org/abs/2409.17554} {arXiv:2409.17554 [hep-ph]} \BibitemShut {NoStop}%
\bibitem [{\citenamefont {Gould}\ and\ \citenamefont {Saffin}(2025)}]{Gould:2024jjt}%
  \BibitemOpen
  \bibfield  {author} {\bibinfo {author} {\bibfnamefont {O.}~\bibnamefont {Gould}}\ and\ \bibinfo {author} {\bibfnamefont {P.~M.}\ \bibnamefont {Saffin}},\ }\bibfield  {title} {\bibinfo {title} {{Perturbative gravitational wave predictions for the real-scalar extended Standard Model}},\ }\href {https://doi.org/10.1007/JHEP03(2025)105} {\bibfield  {journal} {\bibinfo  {journal} {JHEP}\ }\textbf {\bibinfo {volume} {03}},\ \bibinfo {pages} {105}},\ \Eprint {https://arxiv.org/abs/2411.08951} {arXiv:2411.08951 [hep-ph]} \BibitemShut {NoStop}%
\bibitem [{\citenamefont {Chala}\ \emph {et~al.}(2024)\citenamefont {Chala}, \citenamefont {Criado}, \citenamefont {Gil},\ and\ \citenamefont {Miras}}]{Chala:2024xll}%
  \BibitemOpen
  \bibfield  {author} {\bibinfo {author} {\bibfnamefont {M.}~\bibnamefont {Chala}}, \bibinfo {author} {\bibfnamefont {J.~C.}\ \bibnamefont {Criado}}, \bibinfo {author} {\bibfnamefont {L.}~\bibnamefont {Gil}},\ and\ \bibinfo {author} {\bibfnamefont {J.~L.}\ \bibnamefont {Miras}},\ }\bibfield  {title} {\bibinfo {title} {{Higher-order-operator corrections to phase-transition parameters in dimensional reduction}},\ }\href {https://doi.org/10.1007/JHEP10(2024)025} {\bibfield  {journal} {\bibinfo  {journal} {JHEP}\ }\textbf {\bibinfo {volume} {10}},\ \bibinfo {pages} {025}},\ \Eprint {https://arxiv.org/abs/2406.02667} {arXiv:2406.02667 [hep-ph]} \BibitemShut {NoStop}%
\bibitem [{\citenamefont {Bernardo}\ \emph {et~al.}(2025)\citenamefont {Bernardo}, \citenamefont {Klose}, \citenamefont {Schicho},\ and\ \citenamefont {Tenkanen}}]{Bernardo:2025vkz}%
  \BibitemOpen
  \bibfield  {author} {\bibinfo {author} {\bibfnamefont {F.}~\bibnamefont {Bernardo}}, \bibinfo {author} {\bibfnamefont {P.}~\bibnamefont {Klose}}, \bibinfo {author} {\bibfnamefont {P.}~\bibnamefont {Schicho}},\ and\ \bibinfo {author} {\bibfnamefont {T.~V.~I.}\ \bibnamefont {Tenkanen}},\ }\bibfield  {title} {\bibinfo {title} {{Higher-dimensional operators at finite-temperature affect gravitational-wave predictions}},\ }\href@noop {} {\bibfield  {journal} {\bibinfo  {journal} {arXiv}\ } (\bibinfo {year} {2025})},\ \Eprint {https://arxiv.org/abs/2503.18904} {arXiv:2503.18904 [hep-ph]} \BibitemShut {NoStop}%
\bibitem [{\citenamefont {Chala}\ \emph {et~al.}(2025)\citenamefont {Chala}, \citenamefont {Gil},\ and\ \citenamefont {Ren}}]{Chala:2025oul}%
  \BibitemOpen
  \bibfield  {author} {\bibinfo {author} {\bibfnamefont {M.}~\bibnamefont {Chala}}, \bibinfo {author} {\bibfnamefont {L.}~\bibnamefont {Gil}},\ and\ \bibinfo {author} {\bibfnamefont {Z.}~\bibnamefont {Ren}},\ }\bibfield  {title} {\bibinfo {title} {{Phase Transitions in Dimensional Reduction up to Three Loops}},\ }\href@noop {} {\bibfield  {journal} {\bibinfo  {journal} {arXiv}\ } (\bibinfo {year} {2025})},\ \Eprint {https://arxiv.org/abs/2505.14335} {arXiv:2505.14335 [hep-ph]} \BibitemShut {NoStop}%
\bibitem [{\citenamefont {Kajantie}\ \emph {et~al.}(1996{\natexlab{c}})\citenamefont {Kajantie}, \citenamefont {Laine}, \citenamefont {Rummukainen},\ and\ \citenamefont {Shaposhnikov}}]{Kajantie:1995kf}%
  \BibitemOpen
  \bibfield  {author} {\bibinfo {author} {\bibfnamefont {K.}~\bibnamefont {Kajantie}}, \bibinfo {author} {\bibfnamefont {M.}~\bibnamefont {Laine}}, \bibinfo {author} {\bibfnamefont {K.}~\bibnamefont {Rummukainen}},\ and\ \bibinfo {author} {\bibfnamefont {M.~E.}\ \bibnamefont {Shaposhnikov}},\ }\bibfield  {title} {\bibinfo {title} {{The Electroweak phase transition: A Nonperturbative analysis}},\ }\href {https://doi.org/10.1016/0550-3213(96)00052-1} {\bibfield  {journal} {\bibinfo  {journal} {Nucl. Phys. B}\ }\textbf {\bibinfo {volume} {466}},\ \bibinfo {pages} {189} (\bibinfo {year} {1996}{\natexlab{c}})},\ \Eprint {https://arxiv.org/abs/hep-lat/9510020} {arXiv:hep-lat/9510020} \BibitemShut {NoStop}%
\bibitem [{\citenamefont {Arnold}\ and\ \citenamefont {Espinosa}(1993)}]{Arnold:1992rz}%
  \BibitemOpen
  \bibfield  {author} {\bibinfo {author} {\bibfnamefont {P.~B.}\ \bibnamefont {Arnold}}\ and\ \bibinfo {author} {\bibfnamefont {O.}~\bibnamefont {Espinosa}},\ }\bibfield  {title} {\bibinfo {title} {{The Effective potential and first order phase transitions: Beyond leading-order}},\ }\href {https://doi.org/10.1103/PhysRevD.47.3546} {\bibfield  {journal} {\bibinfo  {journal} {Phys. Rev. D}\ }\textbf {\bibinfo {volume} {47}},\ \bibinfo {pages} {3546} (\bibinfo {year} {1993})},\ \bibinfo {note} {[Erratum: Phys.Rev.D 50, 6662 (1994)]},\ \Eprint {https://arxiv.org/abs/hep-ph/9212235} {arXiv:hep-ph/9212235} \BibitemShut {NoStop}%
\bibitem [{\citenamefont {Ekstedt}\ \emph {et~al.}(2022)\citenamefont {Ekstedt}, \citenamefont {Gould},\ and\ \citenamefont {L\"ofgren}}]{Ekstedt:2022zro}%
  \BibitemOpen
  \bibfield  {author} {\bibinfo {author} {\bibfnamefont {A.}~\bibnamefont {Ekstedt}}, \bibinfo {author} {\bibfnamefont {O.}~\bibnamefont {Gould}},\ and\ \bibinfo {author} {\bibfnamefont {J.}~\bibnamefont {L\"ofgren}},\ }\bibfield  {title} {\bibinfo {title} {{Radiative first-order phase transitions to next-to-next-to-leading order}},\ }\href {https://doi.org/10.1103/PhysRevD.106.036012} {\bibfield  {journal} {\bibinfo  {journal} {Phys. Rev. D}\ }\textbf {\bibinfo {volume} {106}},\ \bibinfo {pages} {036012} (\bibinfo {year} {2022})},\ \bibinfo {note} {[Erratum: Phys.Rev.D 110, 019901 (2024)]},\ \Eprint {https://arxiv.org/abs/2205.07241} {arXiv:2205.07241 [hep-ph]} \BibitemShut {NoStop}%
\bibitem [{\citenamefont {Gould}\ and\ \citenamefont {Tenkanen}(2024)}]{Gould:2023ovu}%
  \BibitemOpen
  \bibfield  {author} {\bibinfo {author} {\bibfnamefont {O.}~\bibnamefont {Gould}}\ and\ \bibinfo {author} {\bibfnamefont {T.~V.~I.}\ \bibnamefont {Tenkanen}},\ }\bibfield  {title} {\bibinfo {title} {{Perturbative effective field theory expansions for cosmological phase transitions}},\ }\href {https://doi.org/10.1007/JHEP01(2024)048} {\bibfield  {journal} {\bibinfo  {journal} {JHEP}\ }\textbf {\bibinfo {volume} {01}},\ \bibinfo {pages} {048}},\ \Eprint {https://arxiv.org/abs/2309.01672} {arXiv:2309.01672 [hep-ph]} \BibitemShut {NoStop}%
\bibitem [{\citenamefont {Niemi}\ \emph {et~al.}(2021{\natexlab{a}})\citenamefont {Niemi}, \citenamefont {Ramsey-Musolf}, \citenamefont {Tenkanen},\ and\ \citenamefont {Weir}}]{Niemi:2020hto}%
  \BibitemOpen
  \bibfield  {author} {\bibinfo {author} {\bibfnamefont {L.}~\bibnamefont {Niemi}}, \bibinfo {author} {\bibfnamefont {M.~J.}\ \bibnamefont {Ramsey-Musolf}}, \bibinfo {author} {\bibfnamefont {T.~V.~I.}\ \bibnamefont {Tenkanen}},\ and\ \bibinfo {author} {\bibfnamefont {D.~J.}\ \bibnamefont {Weir}},\ }\bibfield  {title} {\bibinfo {title} {{Thermodynamics of a Two-Step Electroweak Phase Transition}},\ }\href {https://doi.org/10.1103/PhysRevLett.126.171802} {\bibfield  {journal} {\bibinfo  {journal} {Phys. Rev. Lett.}\ }\textbf {\bibinfo {volume} {126}},\ \bibinfo {pages} {171802} (\bibinfo {year} {2021}{\natexlab{a}})},\ \Eprint {https://arxiv.org/abs/2005.11332} {arXiv:2005.11332 [hep-ph]} \BibitemShut {NoStop}%
\bibitem [{\citenamefont {Niemi}\ \emph {et~al.}(2021{\natexlab{b}})\citenamefont {Niemi}, \citenamefont {Schicho},\ and\ \citenamefont {Tenkanen}}]{Niemi:2021qvp}%
  \BibitemOpen
  \bibfield  {author} {\bibinfo {author} {\bibfnamefont {L.}~\bibnamefont {Niemi}}, \bibinfo {author} {\bibfnamefont {P.}~\bibnamefont {Schicho}},\ and\ \bibinfo {author} {\bibfnamefont {T.~V.~I.}\ \bibnamefont {Tenkanen}},\ }\bibfield  {title} {\bibinfo {title} {{Singlet-assisted electroweak phase transition at two loops}},\ }\href {https://doi.org/10.1103/PhysRevD.103.115035} {\bibfield  {journal} {\bibinfo  {journal} {Phys. Rev. D}\ }\textbf {\bibinfo {volume} {103}},\ \bibinfo {pages} {115035} (\bibinfo {year} {2021}{\natexlab{b}})},\ \bibinfo {note} {[Erratum: Phys.Rev.D 109, 039902 (2024)]},\ \Eprint {https://arxiv.org/abs/2103.07467} {arXiv:2103.07467 [hep-ph]} \BibitemShut {NoStop}%
\bibitem [{\citenamefont {Gould}\ and\ \citenamefont {Xie}(2023{\natexlab{a}})}]{Gould:2023jbz}%
  \BibitemOpen
  \bibfield  {author} {\bibinfo {author} {\bibfnamefont {O.}~\bibnamefont {Gould}}\ and\ \bibinfo {author} {\bibfnamefont {C.}~\bibnamefont {Xie}},\ }\bibfield  {title} {\bibinfo {title} {{Higher orders for cosmological phase transitions: a global study in a Yukawa model}},\ }\href {https://doi.org/10.1007/JHEP12(2023)049} {\bibfield  {journal} {\bibinfo  {journal} {JHEP}\ }\textbf {\bibinfo {volume} {12}},\ \bibinfo {pages} {049}},\ \Eprint {https://arxiv.org/abs/2310.02308} {arXiv:2310.02308 [hep-ph]} \BibitemShut {NoStop}%
\bibitem [{\citenamefont {L\"ofgren}(2023)}]{Lofgren:2023sep}%
  \BibitemOpen
  \bibfield  {author} {\bibinfo {author} {\bibfnamefont {J.}~\bibnamefont {L\"ofgren}},\ }\bibfield  {title} {\bibinfo {title} {{Stop comparing resummation methods}},\ }\href {https://doi.org/10.1088/1361-6471/ad074b} {\bibfield  {journal} {\bibinfo  {journal} {J. Phys. G}\ }\textbf {\bibinfo {volume} {50}},\ \bibinfo {pages} {125008} (\bibinfo {year} {2023})},\ \Eprint {https://arxiv.org/abs/2301.05197} {arXiv:2301.05197 [hep-ph]} \BibitemShut {NoStop}%
\bibitem [{\citenamefont {Camargo-Molina}\ \emph {et~al.}(2024)\citenamefont {Camargo-Molina}, \citenamefont {Enberg},\ and\ \citenamefont {L\"ofgren}}]{Camargo-Molina:2024sde}%
  \BibitemOpen
  \bibfield  {author} {\bibinfo {author} {\bibfnamefont {E.}~\bibnamefont {Camargo-Molina}}, \bibinfo {author} {\bibfnamefont {R.}~\bibnamefont {Enberg}},\ and\ \bibinfo {author} {\bibfnamefont {J.}~\bibnamefont {L\"ofgren}},\ }\bibfield  {title} {\bibinfo {title} {{A Catalog of First-Order Electroweak Phase Transitions in the Standard Model Effective Field Theory}},\ }\href@noop {} {\bibfield  {journal} {\bibinfo  {journal} {arXiv}\ } (\bibinfo {year} {2024})},\ \Eprint {https://arxiv.org/abs/2410.23210} {arXiv:2410.23210 [hep-ph]} \BibitemShut {NoStop}%
\bibitem [{\citenamefont {Espinosa}\ \emph {et~al.}(2008)\citenamefont {Espinosa}, \citenamefont {Konstandin}, \citenamefont {No},\ and\ \citenamefont {Quiros}}]{Espinosa:2008kw}%
  \BibitemOpen
  \bibfield  {author} {\bibinfo {author} {\bibfnamefont {J.~R.}\ \bibnamefont {Espinosa}}, \bibinfo {author} {\bibfnamefont {T.}~\bibnamefont {Konstandin}}, \bibinfo {author} {\bibfnamefont {J.~M.}\ \bibnamefont {No}},\ and\ \bibinfo {author} {\bibfnamefont {M.}~\bibnamefont {Quiros}},\ }\bibfield  {title} {\bibinfo {title} {{Some Cosmological Implications of Hidden Sectors}},\ }\href {https://doi.org/10.1103/PhysRevD.78.123528} {\bibfield  {journal} {\bibinfo  {journal} {Phys. Rev. D}\ }\textbf {\bibinfo {volume} {78}},\ \bibinfo {pages} {123528} (\bibinfo {year} {2008})},\ \Eprint {https://arxiv.org/abs/0809.3215} {arXiv:0809.3215 [hep-ph]} \BibitemShut {NoStop}%
\bibitem [{\citenamefont {Jinno}\ and\ \citenamefont {Takimoto}(2017)}]{Jinno:2016knw}%
  \BibitemOpen
  \bibfield  {author} {\bibinfo {author} {\bibfnamefont {R.}~\bibnamefont {Jinno}}\ and\ \bibinfo {author} {\bibfnamefont {M.}~\bibnamefont {Takimoto}},\ }\bibfield  {title} {\bibinfo {title} {{Probing a classically conformal B-L model with gravitational waves}},\ }\href {https://doi.org/10.1103/PhysRevD.95.015020} {\bibfield  {journal} {\bibinfo  {journal} {Phys. Rev. D}\ }\textbf {\bibinfo {volume} {95}},\ \bibinfo {pages} {015020} (\bibinfo {year} {2017})},\ \Eprint {https://arxiv.org/abs/1604.05035} {arXiv:1604.05035 [hep-ph]} \BibitemShut {NoStop}%
\bibitem [{\citenamefont {Prokopec}\ \emph {et~al.}(2019)\citenamefont {Prokopec}, \citenamefont {Rezacek},\ and\ \citenamefont {{\'S}wie{\.z}ewska}}]{Prokopec:2018tnq}%
  \BibitemOpen
  \bibfield  {author} {\bibinfo {author} {\bibfnamefont {T.}~\bibnamefont {Prokopec}}, \bibinfo {author} {\bibfnamefont {J.}~\bibnamefont {Rezacek}},\ and\ \bibinfo {author} {\bibfnamefont {B.}~\bibnamefont {{\'S}wie{\.z}ewska}},\ }\bibfield  {title} {\bibinfo {title} {{Gravitational waves from conformal symmetry breaking}},\ }\href {https://doi.org/10.1088/1475-7516/2019/02/009} {\bibfield  {journal} {\bibinfo  {journal} {JCAP}\ }\textbf {\bibinfo {volume} {02}},\ \bibinfo {pages} {009}},\ \Eprint {https://arxiv.org/abs/1809.11129} {arXiv:1809.11129 [hep-ph]} \BibitemShut {NoStop}%
\bibitem [{\citenamefont {Gould}\ and\ \citenamefont {Hirvonen}(2021)}]{Gould:2021ccf}%
  \BibitemOpen
  \bibfield  {author} {\bibinfo {author} {\bibfnamefont {O.}~\bibnamefont {Gould}}\ and\ \bibinfo {author} {\bibfnamefont {J.}~\bibnamefont {Hirvonen}},\ }\bibfield  {title} {\bibinfo {title} {{Effective field theory approach to thermal bubble nucleation}},\ }\href {https://doi.org/10.1103/PhysRevD.104.096015} {\bibfield  {journal} {\bibinfo  {journal} {Phys. Rev. D}\ }\textbf {\bibinfo {volume} {104}},\ \bibinfo {pages} {096015} (\bibinfo {year} {2021})},\ \Eprint {https://arxiv.org/abs/2108.04377} {arXiv:2108.04377 [hep-ph]} \BibitemShut {NoStop}%
\bibitem [{\citenamefont {Quiros}(1999)}]{Quiros:1999jp}%
  \BibitemOpen
  \bibfield  {author} {\bibinfo {author} {\bibfnamefont {M.}~\bibnamefont {Quiros}},\ }\bibfield  {title} {\bibinfo {title} {{Finite temperature field theory and phase transitions}},\ }in\ \href@noop {} {\emph {\bibinfo {booktitle} {{ICTP Summer School in High-Energy Physics and Cosmology}}}}\ (\bibinfo {year} {1999})\ pp.\ \bibinfo {pages} {187--259},\ \Eprint {https://arxiv.org/abs/hep-ph/9901312} {arXiv:hep-ph/9901312} \BibitemShut {NoStop}%
\bibitem [{\citenamefont {Wainwright}(2012)}]{Wainwright:2011kj}%
  \BibitemOpen
  \bibfield  {author} {\bibinfo {author} {\bibfnamefont {C.~L.}\ \bibnamefont {Wainwright}},\ }\bibfield  {title} {\bibinfo {title} {{CosmoTransitions: Computing Cosmological Phase Transition Temperatures and Bubble Profiles with Multiple Fields}},\ }\href {https://doi.org/10.1016/j.cpc.2012.04.004} {\bibfield  {journal} {\bibinfo  {journal} {Comput. Phys. Commun.}\ }\textbf {\bibinfo {volume} {183}},\ \bibinfo {pages} {2006} (\bibinfo {year} {2012})},\ \Eprint {https://arxiv.org/abs/1109.4189} {arXiv:1109.4189 [hep-ph]} \BibitemShut {NoStop}%
\bibitem [{\citenamefont {Caprini}\ \emph {et~al.}(2020)\citenamefont {Caprini} \emph {et~al.}}]{Caprini:2019egz}%
  \BibitemOpen
  \bibfield  {author} {\bibinfo {author} {\bibfnamefont {C.}~\bibnamefont {Caprini}} \emph {et~al.},\ }\bibfield  {title} {\bibinfo {title} {{Detecting gravitational waves from cosmological phase transitions with LISA: an update}},\ }\href {https://doi.org/10.1088/1475-7516/2020/03/024} {\bibfield  {journal} {\bibinfo  {journal} {JCAP}\ }\textbf {\bibinfo {volume} {03}},\ \bibinfo {pages} {024}},\ \Eprint {https://arxiv.org/abs/1910.13125} {arXiv:1910.13125 [astro-ph.CO]} \BibitemShut {NoStop}%
\bibitem [{\citenamefont {Athron}\ \emph {et~al.}(2024{\natexlab{a}})\citenamefont {Athron}, \citenamefont {Fowlie}, \citenamefont {Lu}, \citenamefont {Morris}, \citenamefont {Wu}, \citenamefont {Wu},\ and\ \citenamefont {Xu}}]{Athron:2023mer}%
  \BibitemOpen
  \bibfield  {author} {\bibinfo {author} {\bibfnamefont {P.}~\bibnamefont {Athron}}, \bibinfo {author} {\bibfnamefont {A.}~\bibnamefont {Fowlie}}, \bibinfo {author} {\bibfnamefont {C.-T.}\ \bibnamefont {Lu}}, \bibinfo {author} {\bibfnamefont {L.}~\bibnamefont {Morris}}, \bibinfo {author} {\bibfnamefont {L.}~\bibnamefont {Wu}}, \bibinfo {author} {\bibfnamefont {Y.}~\bibnamefont {Wu}},\ and\ \bibinfo {author} {\bibfnamefont {Z.}~\bibnamefont {Xu}},\ }\bibfield  {title} {\bibinfo {title} {{Can Supercooled Phase Transitions Explain the Gravitational Wave Background Observed by Pulsar Timing Arrays?}},\ }\href {https://doi.org/10.1103/PhysRevLett.132.221001} {\bibfield  {journal} {\bibinfo  {journal} {Phys. Rev. Lett.}\ }\textbf {\bibinfo {volume} {132}},\ \bibinfo {pages} {221001} (\bibinfo {year} {2024}{\natexlab{a}})},\ \Eprint {https://arxiv.org/abs/2306.17239} {arXiv:2306.17239 [hep-ph]} \BibitemShut {NoStop}%
\bibitem [{\citenamefont {Coleman}\ and\ \citenamefont {Weinberg}(1973)}]{Coleman:1973jx}%
  \BibitemOpen
  \bibfield  {author} {\bibinfo {author} {\bibfnamefont {S.~R.}\ \bibnamefont {Coleman}}\ and\ \bibinfo {author} {\bibfnamefont {E.~J.}\ \bibnamefont {Weinberg}},\ }\bibfield  {title} {\bibinfo {title} {{Radiative Corrections as the Origin of Spontaneous Symmetry Breaking}},\ }\href {https://doi.org/10.1103/PhysRevD.7.1888} {\bibfield  {journal} {\bibinfo  {journal} {Phys. Rev. D}\ }\textbf {\bibinfo {volume} {7}},\ \bibinfo {pages} {1888} (\bibinfo {year} {1973})}\BibitemShut {NoStop}%
\bibitem [{\citenamefont {Dolan}\ and\ \citenamefont {Jackiw}(1974)}]{Dolan:1973qd}%
  \BibitemOpen
  \bibfield  {author} {\bibinfo {author} {\bibfnamefont {L.}~\bibnamefont {Dolan}}\ and\ \bibinfo {author} {\bibfnamefont {R.}~\bibnamefont {Jackiw}},\ }\bibfield  {title} {\bibinfo {title} {{Symmetry Behavior at Finite Temperature}},\ }\href {https://doi.org/10.1103/PhysRevD.9.3320} {\bibfield  {journal} {\bibinfo  {journal} {Phys. Rev. D}\ }\textbf {\bibinfo {volume} {9}},\ \bibinfo {pages} {3320} (\bibinfo {year} {1974})}\BibitemShut {NoStop}%
\bibitem [{\citenamefont {Parwani}(1992)}]{Parwani:1991gq}%
  \BibitemOpen
  \bibfield  {author} {\bibinfo {author} {\bibfnamefont {R.~R.}\ \bibnamefont {Parwani}},\ }\bibfield  {title} {\bibinfo {title} {{Resummation in a hot scalar field theory}},\ }\href {https://doi.org/10.1103/PhysRevD.45.4695} {\bibfield  {journal} {\bibinfo  {journal} {Phys. Rev. D}\ }\textbf {\bibinfo {volume} {45}},\ \bibinfo {pages} {4695} (\bibinfo {year} {1992})},\ \bibinfo {note} {[Erratum: Phys.Rev.D 48, 5965 (1993)]},\ \Eprint {https://arxiv.org/abs/hep-ph/9204216} {arXiv:hep-ph/9204216} \BibitemShut {NoStop}%
\bibitem [{\citenamefont {Gould}\ \emph {et~al.}(ming)\citenamefont {Gould}, \citenamefont {Hirvonen},\ and\ \citenamefont {Tenkanen}}]{Gould:2026forthcoming}%
  \BibitemOpen
  \bibfield  {author} {\bibinfo {author} {\bibfnamefont {O.}~\bibnamefont {Gould}}, \bibinfo {author} {\bibfnamefont {J.}~\bibnamefont {Hirvonen}},\ and\ \bibinfo {author} {\bibfnamefont {T.}~\bibnamefont {Tenkanen}},\ }\bibfield  {title} {\bibinfo {title} {{Effective field theories are almost inevitable for cosmological phase transitions}},\ }\href@noop {} {\  (\bibinfo {year} {forthcoming})}\BibitemShut {NoStop}%
\bibitem [{\citenamefont {Laine}\ and\ \citenamefont {Vuorinen}(2016)}]{Laine:2016hma}%
  \BibitemOpen
  \bibfield  {author} {\bibinfo {author} {\bibfnamefont {M.}~\bibnamefont {Laine}}\ and\ \bibinfo {author} {\bibfnamefont {A.}~\bibnamefont {Vuorinen}},\ }\href {https://doi.org/10.1007/978-3-319-31933-9} {\emph {\bibinfo {title} {{Basics of Thermal Field Theory}}}},\ Vol.\ \bibinfo {volume} {925}\ (\bibinfo  {publisher} {Springer},\ \bibinfo {year} {2016})\ \Eprint {https://arxiv.org/abs/1701.01554} {arXiv:1701.01554 [hep-ph]} \BibitemShut {NoStop}%
\bibitem [{\citenamefont {Greiner}\ and\ \citenamefont {Muller}(1997)}]{Greiner:1996dx}%
  \BibitemOpen
  \bibfield  {author} {\bibinfo {author} {\bibfnamefont {C.}~\bibnamefont {Greiner}}\ and\ \bibinfo {author} {\bibfnamefont {B.}~\bibnamefont {Muller}},\ }\bibfield  {title} {\bibinfo {title} {{Classical fields near thermal equilibrium}},\ }\href {https://doi.org/10.1103/PhysRevD.55.1026} {\bibfield  {journal} {\bibinfo  {journal} {Phys. Rev. D}\ }\textbf {\bibinfo {volume} {55}},\ \bibinfo {pages} {1026} (\bibinfo {year} {1997})},\ \Eprint {https://arxiv.org/abs/hep-th/9605048} {arXiv:hep-th/9605048} \BibitemShut {NoStop}%
\bibitem [{\citenamefont {Jantzen}(2011)}]{Jantzen:2011nz}%
  \BibitemOpen
  \bibfield  {author} {\bibinfo {author} {\bibfnamefont {B.}~\bibnamefont {Jantzen}},\ }\bibfield  {title} {\bibinfo {title} {{Foundation and generalization of the expansion by regions}},\ }\href {https://doi.org/10.1007/JHEP12(2011)076} {\bibfield  {journal} {\bibinfo  {journal} {JHEP}\ }\textbf {\bibinfo {volume} {12}},\ \bibinfo {pages} {076}},\ \Eprint {https://arxiv.org/abs/1111.2589} {arXiv:1111.2589 [hep-ph]} \BibitemShut {NoStop}%
\bibitem [{\citenamefont {Hirvonen}(2022)}]{Hirvonen:2022jba}%
  \BibitemOpen
  \bibfield  {author} {\bibinfo {author} {\bibfnamefont {J.}~\bibnamefont {Hirvonen}},\ }\bibfield  {title} {\bibinfo {title} {{Intuitive method for constructing effective field theories}},\ }\href@noop {} {\bibfield  {journal} {\bibinfo  {journal} {arXiv}\ } (\bibinfo {year} {2022})},\ \Eprint {https://arxiv.org/abs/2205.02687} {arXiv:2205.02687 [hep-ph]} \BibitemShut {NoStop}%
\bibitem [{\citenamefont {Chala}(2026)}]{Chala:2026wjp}%
  \BibitemOpen
  \bibfield  {author} {\bibinfo {author} {\bibfnamefont {M.}~\bibnamefont {Chala}},\ }\bibfield  {title} {\bibinfo {title} {{Thermal masses in the Standard Model at three loops}},\ }\href@noop {} {\  (\bibinfo {year} {2026})},\ \Eprint {https://arxiv.org/abs/2609.33448} {arXiv:2609.33448 [hep-ph]} \BibitemShut {NoStop}%
\bibitem [{\citenamefont {Burgess}(2020)}]{Burgess:2020tbq}%
  \BibitemOpen
  \bibfield  {author} {\bibinfo {author} {\bibfnamefont {C.~P.}\ \bibnamefont {Burgess}},\ }\href {https://doi.org/10.1017/9781139048040} {\emph {\bibinfo {title} {{Introduction to Effective Field Theory}}}}\ (\bibinfo  {publisher} {Cambridge University Press},\ \bibinfo {year} {2020})\BibitemShut {NoStop}%
\bibitem [{\citenamefont {Manohar}\ and\ \citenamefont {Georgi}(1984)}]{Manohar:1983md}%
  \BibitemOpen
  \bibfield  {author} {\bibinfo {author} {\bibfnamefont {A.}~\bibnamefont {Manohar}}\ and\ \bibinfo {author} {\bibfnamefont {H.}~\bibnamefont {Georgi}},\ }\bibfield  {title} {\bibinfo {title} {{Chiral Quarks and the Nonrelativistic Quark Model}},\ }\href {https://doi.org/10.1016/0550-3213(84)90231-1} {\bibfield  {journal} {\bibinfo  {journal} {Nucl. Phys. B}\ }\textbf {\bibinfo {volume} {234}},\ \bibinfo {pages} {189} (\bibinfo {year} {1984})}\BibitemShut {NoStop}%
\bibitem [{\citenamefont {Burgess}(2004)}]{Burgess:2003jk}%
  \BibitemOpen
  \bibfield  {author} {\bibinfo {author} {\bibfnamefont {C.~P.}\ \bibnamefont {Burgess}},\ }\bibfield  {title} {\bibinfo {title} {{Quantum gravity in everyday life: General relativity as an effective field theory}},\ }\href {https://doi.org/10.12942/lrr-2004-5} {\bibfield  {journal} {\bibinfo  {journal} {Living Rev. Rel.}\ }\textbf {\bibinfo {volume} {7}},\ \bibinfo {pages} {5} (\bibinfo {year} {2004})},\ \Eprint {https://arxiv.org/abs/gr-qc/0311082} {arXiv:gr-qc/0311082} \BibitemShut {NoStop}%
\bibitem [{\citenamefont {Gavela}\ \emph {et~al.}(2016)\citenamefont {Gavela}, \citenamefont {Jenkins}, \citenamefont {Manohar},\ and\ \citenamefont {Merlo}}]{Gavela:2016bzc}%
  \BibitemOpen
  \bibfield  {author} {\bibinfo {author} {\bibfnamefont {B.~M.}\ \bibnamefont {Gavela}}, \bibinfo {author} {\bibfnamefont {E.~E.}\ \bibnamefont {Jenkins}}, \bibinfo {author} {\bibfnamefont {A.~V.}\ \bibnamefont {Manohar}},\ and\ \bibinfo {author} {\bibfnamefont {L.}~\bibnamefont {Merlo}},\ }\bibfield  {title} {\bibinfo {title} {{Analysis of General Power Counting Rules in Effective Field Theory}},\ }\href {https://doi.org/10.1140/epjc/s10052-016-4332-1} {\bibfield  {journal} {\bibinfo  {journal} {Eur. Phys. J. C}\ }\textbf {\bibinfo {volume} {76}},\ \bibinfo {pages} {485} (\bibinfo {year} {2016})},\ \Eprint {https://arxiv.org/abs/1601.07551} {arXiv:1601.07551 [hep-ph]} \BibitemShut {NoStop}%
\bibitem [{\citenamefont {Weinberg}(1968)}]{Weinberg:1968de}%
  \BibitemOpen
  \bibfield  {author} {\bibinfo {author} {\bibfnamefont {S.}~\bibnamefont {Weinberg}},\ }\bibfield  {title} {\bibinfo {title} {{Nonlinear realizations of chiral symmetry}},\ }\href {https://doi.org/10.1103/PhysRev.166.1568} {\bibfield  {journal} {\bibinfo  {journal} {Phys. Rev.}\ }\textbf {\bibinfo {volume} {166}},\ \bibinfo {pages} {1568} (\bibinfo {year} {1968})}\BibitemShut {NoStop}%
\bibitem [{\citenamefont {Weinberg}(1979)}]{Weinberg:1978kz}%
  \BibitemOpen
  \bibfield  {author} {\bibinfo {author} {\bibfnamefont {S.}~\bibnamefont {Weinberg}},\ }\bibfield  {title} {\bibinfo {title} {{Phenomenological Lagrangians}},\ }\href {https://doi.org/10.1016/0378-4371(79)90223-1} {\bibfield  {journal} {\bibinfo  {journal} {Physica A}\ }\textbf {\bibinfo {volume} {96}},\ \bibinfo {pages} {327} (\bibinfo {year} {1979})}\BibitemShut {NoStop}%
\bibitem [{\citenamefont {Brivio}\ \emph {et~al.}(2026)\citenamefont {Brivio}, \citenamefont {Gr{\"o}ber},\ and\ \citenamefont {Schmid}}]{Brivio:2025yrr}%
  \BibitemOpen
  \bibfield  {author} {\bibinfo {author} {\bibfnamefont {I.}~\bibnamefont {Brivio}}, \bibinfo {author} {\bibfnamefont {R.}~\bibnamefont {Gr{\"o}ber}},\ and\ \bibinfo {author} {\bibfnamefont {K.}~\bibnamefont {Schmid}},\ }\bibfield  {title} {\bibinfo {title} {{The Art of Counting: a reappraisal of the HEFT expansion}},\ }\href {https://doi.org/10.1007/JHEP04(2026)202} {\bibfield  {journal} {\bibinfo  {journal} {JHEP}\ }\textbf {\bibinfo {volume} {04}},\ \bibinfo {pages} {202}},\ \Eprint {https://arxiv.org/abs/2511.23410} {arXiv:2511.23410 [hep-ph]} \BibitemShut {NoStop}%
\bibitem [{\citenamefont {Weinberg}(2008)}]{Weinberg:2008hq}%
  \BibitemOpen
  \bibfield  {author} {\bibinfo {author} {\bibfnamefont {S.}~\bibnamefont {Weinberg}},\ }\bibfield  {title} {\bibinfo {title} {{Effective Field Theory for Inflation}},\ }\href {https://doi.org/10.1103/PhysRevD.77.123541} {\bibfield  {journal} {\bibinfo  {journal} {Phys. Rev. D}\ }\textbf {\bibinfo {volume} {77}},\ \bibinfo {pages} {123541} (\bibinfo {year} {2008})},\ \Eprint {https://arxiv.org/abs/0804.4291} {arXiv:0804.4291 [hep-th]} \BibitemShut {NoStop}%
\bibitem [{\citenamefont {Burgess}(2017)}]{Burgess:2017ytm}%
  \BibitemOpen
  \bibfield  {author} {\bibinfo {author} {\bibfnamefont {C.~P.}\ \bibnamefont {Burgess}},\ }\bibfield  {title} {\bibinfo {title} {{Intro to Effective Field Theories and Inflation}},\ }\href@noop {} {\  (\bibinfo {year} {2017})},\ \Eprint {https://arxiv.org/abs/1711.10592} {arXiv:1711.10592 [hep-th]} \BibitemShut {NoStop}%
\bibitem [{\citenamefont {Gould}(2021)}]{Gould:2021dzl}%
  \BibitemOpen
  \bibfield  {author} {\bibinfo {author} {\bibfnamefont {O.}~\bibnamefont {Gould}},\ }\bibfield  {title} {\bibinfo {title} {{Real scalar phase transitions: a nonperturbative analysis}},\ }\href {https://doi.org/10.1007/JHEP04(2021)057} {\bibfield  {journal} {\bibinfo  {journal} {JHEP}\ }\textbf {\bibinfo {volume} {04}},\ \bibinfo {pages} {057}},\ \Eprint {https://arxiv.org/abs/2101.05528} {arXiv:2101.05528 [hep-ph]} \BibitemShut {NoStop}%
\bibitem [{\citenamefont {Ekstedt}\ and\ \citenamefont {L{\"o}fgren}(2020)}]{Ekstedt:2020abj}%
  \BibitemOpen
  \bibfield  {author} {\bibinfo {author} {\bibfnamefont {A.}~\bibnamefont {Ekstedt}}\ and\ \bibinfo {author} {\bibfnamefont {J.}~\bibnamefont {L{\"o}fgren}},\ }\bibfield  {title} {\bibinfo {title} {{A Critical Look at the Electroweak Phase Transition}},\ }\href {https://doi.org/10.1007/JHEP12(2020)136} {\bibfield  {journal} {\bibinfo  {journal} {JHEP}\ }\textbf {\bibinfo {volume} {12}},\ \bibinfo {pages} {136}},\ \Eprint {https://arxiv.org/abs/2006.12614} {arXiv:2006.12614 [hep-ph]} \BibitemShut {NoStop}%
\bibitem [{\citenamefont {Laine}\ and\ \citenamefont {Rummukainen}(1999)}]{Laine:1998jb}%
  \BibitemOpen
  \bibfield  {author} {\bibinfo {author} {\bibfnamefont {M.}~\bibnamefont {Laine}}\ and\ \bibinfo {author} {\bibfnamefont {K.}~\bibnamefont {Rummukainen}},\ }\bibfield  {title} {\bibinfo {title} {{What's new with the electroweak phase transition?}},\ }\href {https://doi.org/10.1016/S0920-5632(99)85017-8} {\bibfield  {journal} {\bibinfo  {journal} {Nucl. Phys. B Proc. Suppl.}\ }\textbf {\bibinfo {volume} {73}},\ \bibinfo {pages} {180} (\bibinfo {year} {1999})},\ \Eprint {https://arxiv.org/abs/hep-lat/9809045} {arXiv:hep-lat/9809045} \BibitemShut {NoStop}%
\bibitem [{\citenamefont {Gould}\ \emph {et~al.}(2022)\citenamefont {Gould}, \citenamefont {G{\"u}yer},\ and\ \citenamefont {Rummukainen}}]{Gould:2022ran}%
  \BibitemOpen
  \bibfield  {author} {\bibinfo {author} {\bibfnamefont {O.}~\bibnamefont {Gould}}, \bibinfo {author} {\bibfnamefont {S.}~\bibnamefont {G{\"u}yer}},\ and\ \bibinfo {author} {\bibfnamefont {K.}~\bibnamefont {Rummukainen}},\ }\bibfield  {title} {\bibinfo {title} {{First-order electroweak phase transitions: A nonperturbative update}},\ }\href {https://doi.org/10.1103/PhysRevD.106.114507} {\bibfield  {journal} {\bibinfo  {journal} {Phys. Rev. D}\ }\textbf {\bibinfo {volume} {106}},\ \bibinfo {pages} {114507} (\bibinfo {year} {2022})},\ \bibinfo {note} {[Erratum: Phys.Rev.D 110, 119903 (2024)]},\ \Eprint {https://arxiv.org/abs/2205.07238} {arXiv:2205.07238 [hep-lat]} \BibitemShut {NoStop}%
\bibitem [{\citenamefont {Rummukainen}\ \emph {et~al.}(1998)\citenamefont {Rummukainen}, \citenamefont {Tsypin}, \citenamefont {Kajantie}, \citenamefont {Laine},\ and\ \citenamefont {Shaposhnikov}}]{Rummukainen:1998as}%
  \BibitemOpen
  \bibfield  {author} {\bibinfo {author} {\bibfnamefont {K.}~\bibnamefont {Rummukainen}}, \bibinfo {author} {\bibfnamefont {M.}~\bibnamefont {Tsypin}}, \bibinfo {author} {\bibfnamefont {K.}~\bibnamefont {Kajantie}}, \bibinfo {author} {\bibfnamefont {M.}~\bibnamefont {Laine}},\ and\ \bibinfo {author} {\bibfnamefont {M.~E.}\ \bibnamefont {Shaposhnikov}},\ }\bibfield  {title} {\bibinfo {title} {{The Universality class of the electroweak theory}},\ }\href {https://doi.org/10.1016/S0550-3213(98)00494-5} {\bibfield  {journal} {\bibinfo  {journal} {Nucl. Phys. B}\ }\textbf {\bibinfo {volume} {532}},\ \bibinfo {pages} {283} (\bibinfo {year} {1998})},\ \Eprint {https://arxiv.org/abs/hep-lat/9805013} {arXiv:hep-lat/9805013} \BibitemShut {NoStop}%
\bibitem [{\citenamefont {Andersen}\ \emph {et~al.}(2018)\citenamefont {Andersen}, \citenamefont {Gorda}, \citenamefont {Helset}, \citenamefont {Niemi}, \citenamefont {Tenkanen}, \citenamefont {Tranberg}, \citenamefont {Vuorinen},\ and\ \citenamefont {Weir}}]{Andersen:2017ika}%
  \BibitemOpen
  \bibfield  {author} {\bibinfo {author} {\bibfnamefont {J.~O.}\ \bibnamefont {Andersen}}, \bibinfo {author} {\bibfnamefont {T.}~\bibnamefont {Gorda}}, \bibinfo {author} {\bibfnamefont {A.}~\bibnamefont {Helset}}, \bibinfo {author} {\bibfnamefont {L.}~\bibnamefont {Niemi}}, \bibinfo {author} {\bibfnamefont {T.~V.~I.}\ \bibnamefont {Tenkanen}}, \bibinfo {author} {\bibfnamefont {A.}~\bibnamefont {Tranberg}}, \bibinfo {author} {\bibfnamefont {A.}~\bibnamefont {Vuorinen}},\ and\ \bibinfo {author} {\bibfnamefont {D.~J.}\ \bibnamefont {Weir}},\ }\bibfield  {title} {\bibinfo {title} {{Nonperturbative Analysis of the Electroweak Phase Transition in the Two Higgs Doublet Model}},\ }\href {https://doi.org/10.1103/PhysRevLett.121.191802} {\bibfield  {journal} {\bibinfo  {journal} {Phys. Rev. Lett.}\ }\textbf {\bibinfo {volume} {121}},\ \bibinfo {pages} {191802} (\bibinfo {year} {2018})},\ \Eprint {https://arxiv.org/abs/1711.09849} {arXiv:1711.09849 [hep-ph]} \BibitemShut {NoStop}%
\bibitem [{\citenamefont {Gorda}\ \emph {et~al.}(2019)\citenamefont {Gorda}, \citenamefont {Helset}, \citenamefont {Niemi}, \citenamefont {Tenkanen},\ and\ \citenamefont {Weir}}]{Gorda:2018hvi}%
  \BibitemOpen
  \bibfield  {author} {\bibinfo {author} {\bibfnamefont {T.}~\bibnamefont {Gorda}}, \bibinfo {author} {\bibfnamefont {A.}~\bibnamefont {Helset}}, \bibinfo {author} {\bibfnamefont {L.}~\bibnamefont {Niemi}}, \bibinfo {author} {\bibfnamefont {T.~V.~I.}\ \bibnamefont {Tenkanen}},\ and\ \bibinfo {author} {\bibfnamefont {D.~J.}\ \bibnamefont {Weir}},\ }\bibfield  {title} {\bibinfo {title} {{Three-dimensional effective theories for the two Higgs doublet model at high temperature}},\ }\href {https://doi.org/10.1007/JHEP02(2019)081} {\bibfield  {journal} {\bibinfo  {journal} {JHEP}\ }\textbf {\bibinfo {volume} {02}},\ \bibinfo {pages} {081}},\ \Eprint {https://arxiv.org/abs/1802.05056} {arXiv:1802.05056 [hep-ph]} \BibitemShut {NoStop}%
\bibitem [{\citenamefont {Bender}\ and\ \citenamefont {Orszag}(1999)}]{Bender:1999box}%
  \BibitemOpen
  \bibfield  {author} {\bibinfo {author} {\bibfnamefont {C.~M.}\ \bibnamefont {Bender}}\ and\ \bibinfo {author} {\bibfnamefont {S.~A.}\ \bibnamefont {Orszag}},\ }\href {https://doi.org/10.1007/978-1-4757-3069-2} {\emph {\bibinfo {title} {{Advanced Mathematical Methods for Scientists and Engineers I}}}}\ (\bibinfo  {publisher} {Springer},\ \bibinfo {year} {1999})\BibitemShut {NoStop}%
\bibitem [{\citenamefont {Laine}\ \emph {et~al.}(2017)\citenamefont {Laine}, \citenamefont {Meyer},\ and\ \citenamefont {Nardini}}]{Laine:2017hdk}%
  \BibitemOpen
  \bibfield  {author} {\bibinfo {author} {\bibfnamefont {M.}~\bibnamefont {Laine}}, \bibinfo {author} {\bibfnamefont {M.}~\bibnamefont {Meyer}},\ and\ \bibinfo {author} {\bibfnamefont {G.}~\bibnamefont {Nardini}},\ }\bibfield  {title} {\bibinfo {title} {{Thermal phase transition with full 2-loop effective potential}},\ }\href {https://doi.org/10.1016/j.nuclphysb.2017.04.023} {\bibfield  {journal} {\bibinfo  {journal} {Nucl. Phys. B}\ }\textbf {\bibinfo {volume} {920}},\ \bibinfo {pages} {565} (\bibinfo {year} {2017})},\ \Eprint {https://arxiv.org/abs/1702.07479} {arXiv:1702.07479 [hep-ph]} \BibitemShut {NoStop}%
\bibitem [{\citenamefont {Moss}\ \emph {et~al.}(1992)\citenamefont {Moss}, \citenamefont {Toms},\ and\ \citenamefont {Wright}}]{Moss:1985ve}%
  \BibitemOpen
  \bibfield  {author} {\bibinfo {author} {\bibfnamefont {I.~G.}\ \bibnamefont {Moss}}, \bibinfo {author} {\bibfnamefont {D.~J.}\ \bibnamefont {Toms}},\ and\ \bibinfo {author} {\bibfnamefont {W.~A.}\ \bibnamefont {Wright}},\ }\bibfield  {title} {\bibinfo {title} {{The Effective Action at Finite Temperature}},\ }\href {https://doi.org/10.1103/PhysRevD.46.1671} {\bibfield  {journal} {\bibinfo  {journal} {Phys. Rev. D}\ }\textbf {\bibinfo {volume} {46}},\ \bibinfo {pages} {1671} (\bibinfo {year} {1992})}\BibitemShut {NoStop}%
\bibitem [{\citenamefont {Ekstedt}(2022)}]{Ekstedt:2021kyx}%
  \BibitemOpen
  \bibfield  {author} {\bibinfo {author} {\bibfnamefont {A.}~\bibnamefont {Ekstedt}},\ }\bibfield  {title} {\bibinfo {title} {{Higher-order corrections to the bubble-nucleation rate at finite temperature}},\ }\href {https://doi.org/10.1140/epjc/s10052-022-10130-5} {\bibfield  {journal} {\bibinfo  {journal} {Eur. Phys. J. C}\ }\textbf {\bibinfo {volume} {82}},\ \bibinfo {pages} {173} (\bibinfo {year} {2022})},\ \Eprint {https://arxiv.org/abs/2104.11804} {arXiv:2104.11804 [hep-ph]} \BibitemShut {NoStop}%
\bibitem [{\citenamefont {Hirvonen}\ \emph {et~al.}(2022)\citenamefont {Hirvonen}, \citenamefont {L{\"o}fgren}, \citenamefont {Ramsey-Musolf}, \citenamefont {Schicho},\ and\ \citenamefont {Tenkanen}}]{Hirvonen:2021zej}%
  \BibitemOpen
  \bibfield  {author} {\bibinfo {author} {\bibfnamefont {J.}~\bibnamefont {Hirvonen}}, \bibinfo {author} {\bibfnamefont {J.}~\bibnamefont {L{\"o}fgren}}, \bibinfo {author} {\bibfnamefont {M.~J.}\ \bibnamefont {Ramsey-Musolf}}, \bibinfo {author} {\bibfnamefont {P.}~\bibnamefont {Schicho}},\ and\ \bibinfo {author} {\bibfnamefont {T.~V.~I.}\ \bibnamefont {Tenkanen}},\ }\bibfield  {title} {\bibinfo {title} {{Computing the gauge-invariant bubble nucleation rate in finite temperature effective field theory}},\ }\href {https://doi.org/10.1007/JHEP07(2022)135} {\bibfield  {journal} {\bibinfo  {journal} {JHEP}\ }\textbf {\bibinfo {volume} {07}},\ \bibinfo {pages} {135}},\ \Eprint {https://arxiv.org/abs/2112.08912} {arXiv:2112.08912 [hep-ph]} \BibitemShut {NoStop}%
\bibitem [{\citenamefont {Kierkla}\ \emph {et~al.}(2025)\citenamefont {Kierkla}, \citenamefont {Schicho}, \citenamefont {Swiezewska}, \citenamefont {Tenkanen},\ and\ \citenamefont {van~de Vis}}]{Kierkla:2025qyz}%
  \BibitemOpen
  \bibfield  {author} {\bibinfo {author} {\bibfnamefont {M.}~\bibnamefont {Kierkla}}, \bibinfo {author} {\bibfnamefont {P.}~\bibnamefont {Schicho}}, \bibinfo {author} {\bibfnamefont {B.}~\bibnamefont {Swiezewska}}, \bibinfo {author} {\bibfnamefont {T.~V.~I.}\ \bibnamefont {Tenkanen}},\ and\ \bibinfo {author} {\bibfnamefont {J.}~\bibnamefont {van~de Vis}},\ }\bibfield  {title} {\bibinfo {title} {{Finite-temperature bubble nucleation with shifting scale hierarchies}},\ }\href@noop {} {\bibfield  {journal} {\bibinfo  {journal} {arXiv}\ } (\bibinfo {year} {2025})},\ \Eprint {https://arxiv.org/abs/2503.13597} {arXiv:2503.13597 [hep-ph]} \BibitemShut {NoStop}%
\bibitem [{\citenamefont {Gould}\ and\ \citenamefont {Xie}(2023{\natexlab{b}})}]{gould_2023_10276044}%
  \BibitemOpen
  \bibfield  {author} {\bibinfo {author} {\bibfnamefont {O.}~\bibnamefont {Gould}}\ and\ \bibinfo {author} {\bibfnamefont {C.}~\bibnamefont {Xie}},\ }\bibfield  {title} {\bibinfo {title} {Higher orders for cosmological phase transitions: a global study in a yukawa model},\ }\href {https://doi.org/10.5281/zenodo.10276044} {10.5281/zenodo.10276044} (\bibinfo {year} {2023}{\natexlab{b}})\BibitemShut {NoStop}%
\bibitem [{\citenamefont {Balui}\ \emph {et~al.}(2025)\citenamefont {Balui}, \citenamefont {Biswas}, \citenamefont {Chakrabortty}, \citenamefont {Dey}, \citenamefont {Englert},\ and\ \citenamefont {Mohanty}}]{Balui:2025yvd}%
  \BibitemOpen
  \bibfield  {author} {\bibinfo {author} {\bibfnamefont {D.}~\bibnamefont {Balui}}, \bibinfo {author} {\bibfnamefont {T.}~\bibnamefont {Biswas}}, \bibinfo {author} {\bibfnamefont {J.}~\bibnamefont {Chakrabortty}}, \bibinfo {author} {\bibfnamefont {D.}~\bibnamefont {Dey}}, \bibinfo {author} {\bibfnamefont {C.}~\bibnamefont {Englert}},\ and\ \bibinfo {author} {\bibfnamefont {S.}~\bibnamefont {Mohanty}},\ }\bibfield  {title} {\bibinfo {title} {{Gauge choices, infrared pitfalls, and thermal effects in effective potentials}},\ }\href {https://doi.org/10.1103/drsd-wfns} {\bibfield  {journal} {\bibinfo  {journal} {Phys. Rev. D}\ }\textbf {\bibinfo {volume} {112}},\ \bibinfo {pages} {056022} (\bibinfo {year} {2025})},\ \Eprint {https://arxiv.org/abs/2507.22706} {arXiv:2507.22706 [hep-th]} \BibitemShut {NoStop}%
\bibitem [{\citenamefont {Laine}(1995)}]{Laine:1994zq}%
  \BibitemOpen
  \bibfield  {author} {\bibinfo {author} {\bibfnamefont {M.}~\bibnamefont {Laine}},\ }\bibfield  {title} {\bibinfo {title} {{Gauge dependence of the high temperature two loop effective potential for the Higgs field}},\ }\href {https://doi.org/10.1103/PhysRevD.51.4525} {\bibfield  {journal} {\bibinfo  {journal} {Phys. Rev. D}\ }\textbf {\bibinfo {volume} {51}},\ \bibinfo {pages} {4525} (\bibinfo {year} {1995})},\ \Eprint {https://arxiv.org/abs/hep-ph/9411252} {arXiv:hep-ph/9411252} \BibitemShut {NoStop}%
\bibitem [{\citenamefont {Fukuda}\ and\ \citenamefont {Kugo}(1976)}]{Fukuda:1975di}%
  \BibitemOpen
  \bibfield  {author} {\bibinfo {author} {\bibfnamefont {R.}~\bibnamefont {Fukuda}}\ and\ \bibinfo {author} {\bibfnamefont {T.}~\bibnamefont {Kugo}},\ }\bibfield  {title} {\bibinfo {title} {{Gauge Invariance in the Effective Action and Potential}},\ }\href {https://doi.org/10.1103/PhysRevD.13.3469} {\bibfield  {journal} {\bibinfo  {journal} {Phys. Rev. D}\ }\textbf {\bibinfo {volume} {13}},\ \bibinfo {pages} {3469} (\bibinfo {year} {1976})}\BibitemShut {NoStop}%
\bibitem [{\citenamefont {Niemi}\ \emph {et~al.}(2023)\citenamefont {Niemi}, \citenamefont {Rummukainen}, \citenamefont {Sepp{\"a}},\ and\ \citenamefont {Weir}}]{Niemi:2022bjg}%
  \BibitemOpen
  \bibfield  {author} {\bibinfo {author} {\bibfnamefont {L.}~\bibnamefont {Niemi}}, \bibinfo {author} {\bibfnamefont {K.}~\bibnamefont {Rummukainen}}, \bibinfo {author} {\bibfnamefont {R.}~\bibnamefont {Sepp{\"a}}},\ and\ \bibinfo {author} {\bibfnamefont {D.~J.}\ \bibnamefont {Weir}},\ }\bibfield  {title} {\bibinfo {title} {{Infrared physics of the 3D SU(2) adjoint Higgs model at the crossover transition}},\ }\href {https://doi.org/10.1007/JHEP02(2023)212} {\bibfield  {journal} {\bibinfo  {journal} {JHEP}\ }\textbf {\bibinfo {volume} {02}},\ \bibinfo {pages} {212}},\ \Eprint {https://arxiv.org/abs/2206.14487} {arXiv:2206.14487 [hep-lat]} \BibitemShut {NoStop}%
\bibitem [{\citenamefont {Ramsey-Musolf}(2020)}]{Ramsey-Musolf:2019lsf}%
  \BibitemOpen
  \bibfield  {author} {\bibinfo {author} {\bibfnamefont {M.~J.}\ \bibnamefont {Ramsey-Musolf}},\ }\bibfield  {title} {\bibinfo {title} {{The electroweak phase transition: a collider target}},\ }\href {https://doi.org/10.1007/JHEP09(2020)179} {\bibfield  {journal} {\bibinfo  {journal} {JHEP}\ }\textbf {\bibinfo {volume} {09}},\ \bibinfo {pages} {179}},\ \Eprint {https://arxiv.org/abs/1912.07189} {arXiv:1912.07189 [hep-ph]} \BibitemShut {NoStop}%
\bibitem [{\citenamefont {Niemi}\ and\ \citenamefont {Tenkanen}(2025)}]{Niemi:2024vzw}%
  \BibitemOpen
  \bibfield  {author} {\bibinfo {author} {\bibfnamefont {L.}~\bibnamefont {Niemi}}\ and\ \bibinfo {author} {\bibfnamefont {T.~V.~I.}\ \bibnamefont {Tenkanen}},\ }\bibfield  {title} {\bibinfo {title} {{Investigating two-loop effects for first-order electroweak phase transitions}},\ }\href {https://doi.org/10.1103/PhysRevD.111.075034} {\bibfield  {journal} {\bibinfo  {journal} {Phys. Rev. D}\ }\textbf {\bibinfo {volume} {111}},\ \bibinfo {pages} {075034} (\bibinfo {year} {2025})},\ \Eprint {https://arxiv.org/abs/2408.15912} {arXiv:2408.15912 [hep-ph]} \BibitemShut {NoStop}%
\bibitem [{\citenamefont {Kierkla}\ \emph {et~al.}(2024)\citenamefont {Kierkla}, \citenamefont {Swiezewska}, \citenamefont {Tenkanen},\ and\ \citenamefont {van~de Vis}}]{Kierkla:2023von}%
  \BibitemOpen
  \bibfield  {author} {\bibinfo {author} {\bibfnamefont {M.}~\bibnamefont {Kierkla}}, \bibinfo {author} {\bibfnamefont {B.}~\bibnamefont {Swiezewska}}, \bibinfo {author} {\bibfnamefont {T.~V.~I.}\ \bibnamefont {Tenkanen}},\ and\ \bibinfo {author} {\bibfnamefont {J.}~\bibnamefont {van~de Vis}},\ }\bibfield  {title} {\bibinfo {title} {{Gravitational waves from supercooled phase transitions: dimensional transmutation meets dimensional reduction}},\ }\href {https://doi.org/10.1007/JHEP02(2024)234} {\bibfield  {journal} {\bibinfo  {journal} {JHEP}\ }\textbf {\bibinfo {volume} {02}},\ \bibinfo {pages} {234}},\ \Eprint {https://arxiv.org/abs/2312.12413} {arXiv:2312.12413 [hep-ph]} \BibitemShut {NoStop}%
\bibitem [{\citenamefont {Navarrete}\ \emph {et~al.}(2026)\citenamefont {Navarrete}, \citenamefont {Paatelainen}, \citenamefont {Sepp{\"a}nen},\ and\ \citenamefont {Tenkanen}}]{Navarrete:2025yxy}%
  \BibitemOpen
  \bibfield  {author} {\bibinfo {author} {\bibfnamefont {P.}~\bibnamefont {Navarrete}}, \bibinfo {author} {\bibfnamefont {R.}~\bibnamefont {Paatelainen}}, \bibinfo {author} {\bibfnamefont {K.}~\bibnamefont {Sepp{\"a}nen}},\ and\ \bibinfo {author} {\bibfnamefont {T.~V.~I.}\ \bibnamefont {Tenkanen}},\ }\bibfield  {title} {\bibinfo {title} {{Cosmological phase transitions without high-temperature expansions}},\ }\href {https://doi.org/10.1007/JHEP01(2026)113} {\bibfield  {journal} {\bibinfo  {journal} {JHEP}\ }\textbf {\bibinfo {volume} {01}},\ \bibinfo {pages} {113}},\ \Eprint {https://arxiv.org/abs/2507.07014} {arXiv:2507.07014 [hep-ph]} \BibitemShut {NoStop}%
\bibitem [{\citenamefont {Klose}(2026)}]{Klose:2026pux}%
  \BibitemOpen
  \bibfield  {author} {\bibinfo {author} {\bibfnamefont {P.}~\bibnamefont {Klose}},\ }\bibfield  {title} {\bibinfo {title} {{Thermal Resummation for Very Strong First-order Phase Transitions}},\ }\href@noop {} {\  (\bibinfo {year} {2026})},\ \Eprint {https://arxiv.org/abs/2607.21252} {arXiv:2607.21252 [hep-ph]} \BibitemShut {NoStop}%
\bibitem [{\citenamefont {Ekstedt}\ \emph {et~al.}(2023)\citenamefont {Ekstedt}, \citenamefont {Schicho},\ and\ \citenamefont {Tenkanen}}]{Ekstedt:2022bff}%
  \BibitemOpen
  \bibfield  {author} {\bibinfo {author} {\bibfnamefont {A.}~\bibnamefont {Ekstedt}}, \bibinfo {author} {\bibfnamefont {P.}~\bibnamefont {Schicho}},\ and\ \bibinfo {author} {\bibfnamefont {T.~V.~I.}\ \bibnamefont {Tenkanen}},\ }\bibfield  {title} {\bibinfo {title} {{DRalgo: A package for effective field theory approach for thermal phase transitions}},\ }\href {https://doi.org/10.1016/j.cpc.2023.108725} {\bibfield  {journal} {\bibinfo  {journal} {Comput. Phys. Commun.}\ }\textbf {\bibinfo {volume} {288}},\ \bibinfo {pages} {108725} (\bibinfo {year} {2023})},\ \Eprint {https://arxiv.org/abs/2205.08815} {arXiv:2205.08815 [hep-ph]} \BibitemShut {NoStop}%
\bibitem [{\citenamefont {Nishimura}\ and\ \citenamefont {Schroder}(2012)}]{Nishimura:2012ee}%
  \BibitemOpen
  \bibfield  {author} {\bibinfo {author} {\bibfnamefont {M.}~\bibnamefont {Nishimura}}\ and\ \bibinfo {author} {\bibfnamefont {Y.}~\bibnamefont {Schroder}},\ }\bibfield  {title} {\bibinfo {title} {{IBP methods at finite temperature}},\ }\href {https://doi.org/10.1007/JHEP09(2012)051} {\bibfield  {journal} {\bibinfo  {journal} {JHEP}\ }\textbf {\bibinfo {volume} {09}},\ \bibinfo {pages} {051}},\ \Eprint {https://arxiv.org/abs/1207.4042} {arXiv:1207.4042 [hep-ph]} \BibitemShut {NoStop}%
\bibitem [{\citenamefont {Gil}\ \emph {et~al.}(2026)\citenamefont {Gil}, \citenamefont {L{\'o}pez~Miras},\ and\ \citenamefont {Moreno-S{\'a}nchez}}]{Gil:2026cqz}%
  \BibitemOpen
  \bibfield  {author} {\bibinfo {author} {\bibfnamefont {L.}~\bibnamefont {Gil}}, \bibinfo {author} {\bibfnamefont {J.}~\bibnamefont {L{\'o}pez~Miras}},\ and\ \bibinfo {author} {\bibfnamefont {A.}~\bibnamefont {Moreno-S{\'a}nchez}},\ }\bibfield  {title} {\bibinfo {title} {{SIRENA -- Sum-Integral REductioN Algorithm}},\ }\href@noop {} {\  (\bibinfo {year} {2026})},\ \Eprint {https://arxiv.org/abs/2605.06775} {arXiv:2605.06775 [hep-ph]} \BibitemShut {NoStop}%
\bibitem [{\citenamefont {Rassouli}(2026)}]{rassouli_2026_23172314}%
  \BibitemOpen
  \bibfield  {author} {\bibinfo {author} {\bibfnamefont {F.-S.}\ \bibnamefont {Rassouli}},\ }\bibfield  {title} {\bibinfo {title} {Two-scalar lo/nlo latent-heat scan},\ }\href {https://doi.org/10.5281/zenodo.23172314} {10.5281/zenodo.23172314} (\bibinfo {year} {2026})\BibitemShut {NoStop}%
\bibitem [{\citenamefont {Laine}\ and\ \citenamefont {Losada}(2000)}]{Laine:2000kv}%
  \BibitemOpen
  \bibfield  {author} {\bibinfo {author} {\bibfnamefont {M.}~\bibnamefont {Laine}}\ and\ \bibinfo {author} {\bibfnamefont {M.}~\bibnamefont {Losada}},\ }\bibfield  {title} {\bibinfo {title} {{Two loop dimensional reduction and effective potential without temperature expansions}},\ }\href {https://doi.org/10.1016/S0550-3213(00)00298-4} {\bibfield  {journal} {\bibinfo  {journal} {Nucl. Phys. B}\ }\textbf {\bibinfo {volume} {582}},\ \bibinfo {pages} {277} (\bibinfo {year} {2000})},\ \Eprint {https://arxiv.org/abs/hep-ph/0003111} {arXiv:hep-ph/0003111} \BibitemShut {NoStop}%
\bibitem [{\citenamefont {Athron}\ \emph {et~al.}(2024{\natexlab{b}})\citenamefont {Athron}, \citenamefont {Bal{\'a}zs}, \citenamefont {Fowlie}, \citenamefont {Morris},\ and\ \citenamefont {Wu}}]{Athron:2023xlk}%
  \BibitemOpen
  \bibfield  {author} {\bibinfo {author} {\bibfnamefont {P.}~\bibnamefont {Athron}}, \bibinfo {author} {\bibfnamefont {C.}~\bibnamefont {Bal{\'a}zs}}, \bibinfo {author} {\bibfnamefont {A.}~\bibnamefont {Fowlie}}, \bibinfo {author} {\bibfnamefont {L.}~\bibnamefont {Morris}},\ and\ \bibinfo {author} {\bibfnamefont {L.}~\bibnamefont {Wu}},\ }\bibfield  {title} {\bibinfo {title} {{Cosmological phase transitions: From perturbative particle physics to gravitational waves}},\ }\href {https://doi.org/10.1016/j.ppnp.2023.104094} {\bibfield  {journal} {\bibinfo  {journal} {Prog. Part. Nucl. Phys.}\ }\textbf {\bibinfo {volume} {135}},\ \bibinfo {pages} {104094} (\bibinfo {year} {2024}{\natexlab{b}})},\ \Eprint {https://arxiv.org/abs/2305.02357} {arXiv:2305.02357 [hep-ph]} \BibitemShut {NoStop}%
\end{thebibliography}%

\appendix
\section{Thermal sum-integrals}\label{app:sumintegrals}
We work in $D=4-2\epsilon$ Euclidean dimensions, with $d=3-2\epsilon$ spatial dimensions, and use the $\overline{\mathrm{MS}}$ measure
\begin{align}
    \int_p \equiv\left(\frac{e^\gamma \bar\mu^2}{4 \pi}\right)^\epsilon \int \frac{d^{3-2 \epsilon} p}{(2 \pi)^{3-2 \epsilon}}
\end{align}
For bosonic and fermionic sum-integrals we write, respectively, 
\begin{align}
    \sumint_{P\text{ or }\{P\}}\equiv T\sum_{n=-\infty}^\infty \int_p,
\end{align}
Thermal four-momenta are denoted by uppercase letters, $P=(p_0,p)$, where $p$ is
the spatial momentum. For bosonic sum-integrals, $p_0=2\pi T n$, while for
fermionic sum-integrals $p_0=(2n+1)\pi T$, with $n\in\mathbb{Z}$. We denote the
corresponding bosonic and fermionic momenta by $P$ and $\{P\}$, respectively.

Our conventions are mostly those of Refs.~\cite{Laine:2017hdk,Laine:2000kv}, translated to the normalization used in the main text.  In particular, the normalization of our logarithmic master integral below follows Ref.~\cite{Athron:2023xlk}, so does not contain the factor $1/2$ that is included in the definition of the corresponding function in Ref.~\cite{Laine:2017hdk}.

\subsection{One-loop logarithmic master integral}
For a bosonic one-loop determinant, 
\begin{align}
    J(m)&\equiv \sumint \log (P^2+m^2)\\
    &= \frac{J^{-1}}{\epsilon}+J^{0}+\epsilon J^1+\cdots
\end{align}
with
\begin{align}
J^{-1}(m)
&=-\frac{m^4}{2(4\pi)^2}
=-\frac{m^4}{32\pi^2},
\label{eq:app_Jminus1}\\
J^0(m)
&=\frac{m^4}{2(4\pi)^2}
\left[\ln\!\left(\frac{m^2}{\bar\mu^2}\right)-\frac32\right]
+\frac{T^4}{\pi^2}\,\mathcal J\left(\frac{m}{T}\right),
\label{eq:app_J0}
\end{align}
where
\begin{align}\label{eq:app_Jthermal}
\mathcal J(y)
&\equiv
\int_0^\infty dx\,x^2
\ln\!\left(1-e^{-\sqrt{x^2+y^2}}\right),\\
\Rightarrow\mathcal J(y\ll1)&=-\frac{\pi^4}{45}+\frac{\pi^2y^2}{12}+O(y^3).
\end{align}
In particular,
\begin{align}
J^0(0)=-\frac{\pi^2T^4}{45}.
\label{eq:app_Jzero}
\end{align}
The derivative identity
\begin{align}
\frac{\partial J(m)}{\partial m^2}=A(m)
\label{eq:app_Jderivative}
\end{align}
provides a useful check on all one-loop expressions below.

For fermions we define
\begin{align}
J_\text{F}(m)
&\equiv \sumint_{\{P\}}\ln(P^2+m^2)\\
 &=\frac{J_\text{F}^{-1}(m)}{\epsilon}+J_\text{F}^0(m)+O(\epsilon),
\label{eq:app_JFdef}
\end{align}
where the ultraviolet pole is the same as for the bosonic integral,
$J_\text{F}^{-1}(m)=J^{-1}(m)$, while
\begin{align}
J_\text{F}^0(m)
&=\frac{m^4}{2(4\pi)^2}
\left[\ln\!\left(\frac{m^2}{\bar\mu^2}\right)-\frac32\right]
+\frac{T^4}{\pi^2}\,\mathcal J_\text{F}(y),
\label{eq:app_JF0}\\
\mathcal J_\text{F}(y)
&\equiv
\int_0^\infty dx\,x^2
\ln\!\left(1+e^{-\sqrt{x^2+y^2}}\right).
\label{eq:app_JFthermal}
\end{align}
Thus a Dirac fermion with $n_\text{f}$ real fermionic degrees of freedom contributes $-n_\text{f}J_\text{F}^0(m)/2$ to the one-loop potential.
In the high-temperature limit, $y=m/T\ll1$, the thermal
functions have the expansions
\begin{align}
\mathcal{J}(y)
&=
-\frac{\pi^4}{45}
+\frac{\pi^2}{12}y^2
-\frac{\pi}{6}y^3
-\frac{y^4}{32}
\ln\left(\frac{y^2}{a_b}\right)
+O(y^6),
\\
\mathcal{J}_\text{F}(y)
&=
\frac{7\pi^4}{360}
-\frac{\pi^2}{24}y^2
-\frac{y^4}{32}
\ln\left(\frac{y^2}{a_f}\right)
+O(y^6),
\end{align}
where
\begin{equation}
a_b=16\pi^2 e^{3/2-2\gamma_E},
\qquad
a_f=\pi^2 e^{3/2-2\gamma_E}.
\end{equation}
N.B.\ at high temperatures, the logarithms combine, cancelling their mass dependence, and leaving only logarithms of $T/\bar{\mu}$.

\subsection{One-loop $A_n$ family}
The
additional
one-loop structures that appear in the potential computations are
\begin{align}
A_n(m)
&\equiv \sumint\frac{1}{(P^2+m^2)^n}\\
&=\frac{A^{-1}_n}{\epsilon}+A^0_n+\epsilon A^1_n+\cdots
\end{align}
and we abbreviate $A_1(m)\equiv A(m)$.  All members of this family follow from $A(m)$ through
\begin{align}\label{eq:app_An_general}
A_n(m)
&=\frac{(-1)^{n-1}}{(n-1)!}
\left(\frac{\partial}{\partial m^2}\right)^{n-1}A(m),
\qquad n\geq1,
\end{align}
with the same relation holding coefficient by coefficient in the
$\epsilon$ expansion, for $m^2>0$.

For $A(m)$ the coefficients needed in this work are
\begin{align}
A^{-1}(m)
&=-\frac{m^2}{(4\pi)^2},
\label{eq:app_Aminus1}\\
A^0(m)
&=\frac{m^2}{(4\pi)^2}
\left[\ln\!\left(\frac{m^2}{\bar\mu^2}\right)-1\right]
+\frac{T^2}{2\pi^2}I_1(y),
\label{eq:app_A0}\\
A^1(m)
&=-\frac{m^2}{(4\pi)^2}
\left[
\frac12\left[\ln\!\left(\frac{m^2}{\bar\mu^2}\right)\right]^2
\!\!-\ln\!\left(\frac{m^2}{\bar\mu^2}\right)
+1+\frac{\pi^2}{12}
\right]
\nonumber\\
&\quad
+\frac{T^2}{2\pi^2}
\left[
\left(\ln\!\frac{\bar\mu^2}{4T^2}+2\right)I_1\left(\frac{m}{T}\right)-I_2\left(\frac{m}{T}\right)
\right],
\label{eq:app_A1}
\end{align}
where
\begin{align}
I_1(y)
&\equiv
\int_0^\infty dx\,
\frac{x^2}{E_x}
n_\text{B}(E_x),
\label{eq:app_I1}\\
I_2(y)
&\equiv
\int_0^\infty dx\,
\frac{x^2 \ln x^2}{E_x}
n_\text{B}(E_x),
\label{eq:app_I2}
\end{align}
in terms of
\begin{align} \label{eq:nB}
E_x&\equiv\sqrt{x^2+y^2},
\quad
n_\text{B}(E)\equiv\frac{1}{e^E-1}.
\end{align}
With these definitions,
\begin{align}
I_1(0)&=\frac{\pi^2}{6},
&
I_2(0)&=\frac{\pi^2}{3}
\left[1-\gamma_E+\frac{\zeta'(2)}{\zeta(2)}\right],
\label{eq:app_Izero}
\end{align}
and therefore
\begin{align}\label{eq:app_Azero01}
A^{-1}(0)&=0,
 \qquad A^0(0)=\frac{T^2}{12},
\end{align}
\begin{align}\label{eq:app_Azero1}
A^1(0)
=\frac{T^2}{6}
\left[
\ln\!\left(\frac{\bar\mu e^{\gamma_E}}{2T}\right)
-\frac{\zeta'(2)}{\zeta(2)}
\right].
\end{align}
The relation
\begin{align}
\frac{d\mathcal J(y)}{d(y^2)}=\frac12 I_1(y)
\label{eq:app_JIrelation}
\end{align}
is another useful consistency check.

The derivative correction in Sec.~V requires
$A_3^0(m)$.  Keeping $y=m/T$ as the argument of $I_1$, Eq.~\eqref{eq:app_An_general} gives
\begin{align}
A_3^0(m)
&=\frac{1}{32\pi^2m^2}
+\frac{1}{16\pi^2T^2}
\left[
\frac{I_1''(y)}{y^2}-\frac{I_1'(y)}{y^3}
\right],
\label{eq:app_A3}
\end{align}
where primes denote derivatives with respect to $y$.  This expression is intended for the hard-scale field $m=m_\xi>0$; the massless $m\to 0_+$ limit of $A_3$ is infrared divergent.

\subsection{Two-loop sunset master integral}
The only two-loop master integral needed explicitly for the two-scalar model is
\begin{align}
H(m_1,m_2,m_3)
&\equiv
\sumint_{P,Q}
\frac{1}{(P^2+m_1^2)(Q^2+m_2^2)}
\nonumber\\[-0.4ex]
&\hspace{1.3cm}\times
\frac{1}{(P+Q)^2+m_3^2}
\nonumber\\
&=\frac{H^{-2}(m_1,m_2,m_3)}{\epsilon^2}
 +\frac{H^{-1}(m_1,m_2,m_3)}{\epsilon}
\nonumber\\[-0.4ex]
&\hspace{1.3cm}+H^0(m_1,m_2,m_3)+O(\epsilon).
\label{eq:app_Hdef}
\end{align}
For the mass assignment occurring in this paper, $H(m,m,0)$, the pole coefficients are
\begin{align}
H^{-2}(m,m,0)
&=-\frac{m^2}{(4\pi)^4},
\label{eq:app_Hminus2}\\
H^{-1}(m,m,0)
&=-\frac{m^2}{(4\pi)^4}
\left[3-2\ln\!\left(\frac{m^2}{\bar\mu^2}\right)\right]
\nonumber\\[-0.3ex]
&\quad+\frac{T^2}{(4\pi)^2}
\left[\frac1{12}+\frac{I_1(y)}{\pi^2}\right].
\label{eq:app_Hminus1}
\end{align}
Equivalently,
\begin{align}
H^{-1}(m,m,0)
=\frac{2A^0(m)}{(4\pi)^2}
+\frac{T^2}{12(4\pi)^2}
-\frac{m^2}{(4\pi)^4},
\label{eq:app_Hminus1_identity}
\end{align}
which is useful in the renormalisation-scale check below.

The finite part is 
\begin{widetext}
\begin{align}
H^0(m,m,0)
={}&-\frac{m^2}{(4\pi)^4}
\left[
7+\frac{\pi^2}{6}
-6\ln\!\left(\frac{m^2}{\bar\mu^2}\right)
+2\left[\ln\!\left(\frac{m^2}{\bar\mu^2}\right)\right]^2
\right]
\nonumber\\
&+\frac{T^2}{(4\pi)^2}
\Bigg\{
\left(\frac{1}{12}+\frac{I_1\left(\frac{m}{T}\right)}{\pi^2}\right)
\left[
\ln\!\left(\frac{\bar\mu^2}{T^2}\right)
+\ln\!\left(\frac{\bar\mu^2}{m^2}\right)
\right]
+(4-2\ln 2)\frac{I_1\left(\frac{m}{T}\right)}{\pi^2}
-\frac{I_2\left(\frac{m}{T}\right)}{\pi^2}
\nonumber\\
&\hspace{3.2cm}
+\frac{1}{6}
\left[
\gamma_E-\ln 2-\frac{\zeta'(2)}{\zeta(2)}
\right]
+\frac{\mathcal H\left(\frac{m}{T}\right)}{\pi^2}
\Bigg\}.
\label{eq:app_H0}
\end{align}
\end{widetext}
where the remaining finite thermal function is
\begin{equation}
\mathcal H(y)
\equiv
2\int_0^\infty dx\,x
\int_0^x dz\,z\,
\frac{n_\text{B}(E_x)}{E_x}
\frac{n_\text{B}(E_z)}{E_z}
\ln\!\left(\frac{x+z}{x-z}\right),
\label{eq:app_Hthermal}
\end{equation}
where $E_z\equiv\sqrt{z^2+y^2}$, and $E_x$ and $n_\text{B}$ are as in Eq.~\eqref{eq:nB}.
Equations~\eqref{eq:app_Hminus2}--\eqref{eq:app_H0} are the special $m_1=m_2=m$, $m_3=0$ case of the general thermal sunset formulas in Refs.~\cite{Laine:2017hdk,Laine:2000kv}.

\section{Explicit potential results}\label{app:2scalar}
\subsection{Two-scalar model}
We now show diagram by diagram the explicit contributions to the two-scalar potential used in Sec.~V. We use the same diagram conventions as in the main text: orange lines denote the massless hard-scale fluctuations of $\phi_\text{UV}$, while dashed lines denote the hard-scale $\xi$ field whose full background-dependent mass is kept in the propagator. Crosses are mass insertions and circled crosses are counterterm insertions.

Starting with LO, there are two one-loop determinants in addition to the tree-level potential. The massless $\phi_{\rm UV}$ determinant is

\begin{equation}
\begin{gathered}
\tikzset{every picture/.style={line width=0.75pt}} 
\begin{tikzpicture}[x=0.75pt,y=0.75pt,yscale=-1,xscale=1]

\draw  [color={rgb, 255:red, 245; green, 166; blue, 35 }  ,draw opacity=1 ] (80,30) .. controls (80,18.95) and (88.95,10) .. (100,10) .. controls (111.05,10) and (120,18.95) .. (120,30) .. controls (120,41.05) and (111.05,50) .. (100,50) .. controls (88.95,50) and (80,41.05) .. (80,30) -- cycle ;

\draw (190,29) node   { $\displaystyle=\frac{1}{2\epsilon } J^{-1}( 0) +\frac{1}{2} J^{0}( 0)$};
\end{tikzpicture}
\end{gathered}
\end{equation}
while the hard-scale $\xi$ determinant gives
\begin{equation}
\begin{gathered}
\tikzset{every picture/.style={line width=0.75pt}} 
\begin{tikzpicture}[x=0.75pt,y=0.75pt,yscale=-1,xscale=1]
\draw  [dash pattern={on 4.5pt off 4.5pt}] (80,30) .. controls (80,18.95) and (88.95,10) .. (100,10) .. controls (111.05,10) and (120,18.95) .. (120,30) .. controls (120,41.05) and (111.05,50) .. (100,50) .. controls (88.95,50) and (80,41.05) .. (80,30) -- cycle ;

\draw (200,29) node   {$\displaystyle=\frac{1}{2\epsilon } J^{-1}( m_{\xi }) +\frac{1}{2} J^{0}( m_{\xi })$};

\end{tikzpicture}
\end{gathered}
\end{equation}
Putting these two determinants together with the tree-level potential and leading-order counterterms gives
\begin{align}\label{eq:V_LO}
     V_{\rm LO}
    =&-
    \frac12 \mu_\phi ^2 \phi_{\rm IR}^2 +\frac{\lambda_\phi}{4!}\phi_{\rm IR}^4\nonumber \\
&
    -\frac{\pi^2 T^4}{90}+\frac{m^4_\xi}{64\pi^2}\left[ \ln\left(\frac{m_\xi^2}{\bar\mu^2}\right)-\frac{3}{2} \right] 
    +\frac{T^4}{2\pi^2} \mathcal{J}\left(\frac{m_\xi}{T}\right)
\end{align}
\begin{widetext}
At NLO, we have the one-loop mass insertion, the two double-bubble diagrams, the sunset, and the counterterm insertions , together with higher-order counterterms at tree-level. Some $O(\epsilon)$ terms are kept, which multiply $1/\epsilon$ poles.
Diagram-by-diagram, we have:

\begin{equation}
\begin{gathered}
\tikzset{every picture/.style={line width=0.75pt}} 
\begin{tikzpicture}[x=0.75pt,y=0.75pt,yscale=-1,xscale=1]

\draw  [color={rgb, 255:red, 245; green, 166; blue, 35 }  ,draw opacity=1 ] (80,30) .. controls (80,18.95) and (88.95,10) .. (100,10) .. controls (111.05,10) and (120,18.95) .. (120,30) .. controls (120,41.05) and (111.05,50) .. (100,50) .. controls (88.95,50) and (80,41.05) .. (80,30) -- cycle ;
\draw    (95,5) -- (105,15) ;
\draw    (95,15) -- (105,5) ;

\draw (124,30) node [anchor=west] [inner sep=0.75pt]  {$\displaystyle=m_{\phi }^2\left[\frac{1}{2\epsilon } A^{-1}( 0) +\frac{1}{2} A^{0}( 0)\right]$};
\end{tikzpicture}
\end{gathered}
\end{equation}
\begin{equation}
\begin{gathered}
\tikzset{every picture/.style={line width=0.75pt}} 
\begin{tikzpicture}[x=0.75pt,y=0.75pt,yscale=-1,xscale=1]

\draw  [color={rgb, 255:red, 245; green, 166; blue, 35 }  ,draw opacity=1 ] (80,30) .. controls (80,18.95) and (88.95,10) .. (100,10) .. controls (111.05,10) and (120,18.95) .. (120,30) .. controls (120,41.05) and (111.05,50) .. (100,50) .. controls (88.95,50) and (80,41.05) .. (80,30) -- cycle ;
\draw  [dash pattern={on 4.5pt off 4.5pt}] (80,70) .. controls (80,58.95) and (88.95,50) .. (100,50) .. controls (111.05,50) and (120,58.95) .. (120,70) .. controls (120,81.05) and (111.05,90) .. (100,90) .. controls (88.95,90) and (80,81.05) .. (80,70) -- cycle ;

\draw (122,32.4) node [anchor=north west][inner sep=0.75pt]   {$\displaystyle=\frac{g^{2}}{4}\left[\frac{A^{-1}( 0)}{\epsilon } +A^{0}( 0) +\epsilon A^{1}( 0)\right]\left[\frac{A^{-1}( m_{\xi })}{\epsilon } +A^{0}( m_{\xi }) +\epsilon A^{1}( m_{\xi })\right]$};

\end{tikzpicture}
\end{gathered}
\end{equation}
\begin{equation}
\begin{gathered}
\tikzset{every picture/.style={line width=0.75pt}} 
\begin{tikzpicture}[x=0.75pt,y=0.75pt,yscale=-1,xscale=1]

\draw  [dash pattern={on 4.5pt off 4.5pt}] (80,70) .. controls (80,58.95) and (88.95,50) .. (100,50) .. controls (111.05,50) and (120,58.95) .. (120,70) .. controls (120,81.05) and (111.05,90) .. (100,90) .. controls (88.95,90) and (80,81.05) .. (80,70) -- cycle ;
\draw  [dash pattern={on 4.5pt off 4.5pt}] (80,30) .. controls (80,18.95) and (88.95,10) .. (100,10) .. controls (111.05,10) and (120,18.95) .. (120,30) .. controls (120,41.05) and (111.05,50) .. (100,50) .. controls (88.95,50) and (80,41.05) .. (80,30) -- cycle ;

\draw (121,32.4) node [anchor=north west][inner sep=0.75pt]   {$\displaystyle=\frac{\lambda_\xi}{8}\left[\frac{A^{-1}( m_{\xi })}{\epsilon } +A^{0}( m_{\xi }) +\epsilon A^{1}( m_{\xi })\right]^{2}$};
\end{tikzpicture}
\end{gathered}
\end{equation}

\begin{equation}
\begin{gathered}
\tikzset{every picture/.style={line width=0.75pt}} 
\begin{tikzpicture}[x=0.75pt,y=0.75pt,yscale=-1,xscale=1]

\draw  [dash pattern={on 4.5pt off 4.5pt}] (80,30) .. controls (80,18.95) and (88.95,10) .. (100,10) .. controls (111.05,10) and (120,18.95) .. (120,30) .. controls (120,41.05) and (111.05,50) .. (100,50) .. controls (88.95,50) and (80,41.05) .. (80,30) -- cycle ;
\draw [color={rgb, 255:red, 245; green, 166; blue, 35 }  ,draw opacity=1 ]   (80,30) -- (120,30) ;

\draw (121,30) node [anchor=west] [inner sep=0.75pt]    {$\displaystyle=-\frac{g^{4} \phi _{\text{IR}}^{2}}{4}\left[\frac{1}{\epsilon ^{2}} H^{-2}( m_{\xi } ,m_{\xi } ,0) +\frac{1}{\epsilon } H^{-1}( m_{\xi } ,m_{\xi } ,0) +H^{0}( m_{\xi } ,m_{\xi } ,0)\right]$};

\end{tikzpicture}
\end{gathered}
\end{equation}

\begin{equation}
\begin{gathered}
\tikzset{every picture/.style={line width=0.75pt}} 
\begin{tikzpicture}[x=0.75pt,y=0.75pt,yscale=-1,xscale=1]

\draw  [color={rgb, 255:red, 245; green, 166; blue, 35 }  ,draw opacity=1 ] (80,30) .. controls (80,18.95) and (88.95,10) .. (100,10) .. controls (111.05,10) and (120,18.95) .. (120,30) .. controls (120,41.05) and (111.05,50) .. (100,50) .. controls (88.95,50) and (80,41.05) .. (80,30) -- cycle ;
\draw  [fill={rgb, 255:red, 255; green, 255; blue, 255 }  ,fill opacity=1 ] (96.72,6.23) .. controls (98.8,4.41) and (101.96,4.64) .. (103.77,6.72) .. controls (105.59,8.8) and (105.36,11.96) .. (103.28,13.77) .. controls (101.2,15.59) and (98.04,15.36) .. (96.23,13.28) .. controls (94.41,11.2) and (94.64,8.04) .. (96.72,6.23) -- cycle ;
\draw [fill={rgb, 255:red, 255; green, 255; blue, 255 }  ,fill opacity=1 ]   (96.46,6.46) -- (103.54,13.54) ;
\draw [fill={rgb, 255:red, 255; green, 255; blue, 255 }  ,fill opacity=1 ]   (103.54,6.46) -- (96.46,13.54) ;

\draw (121,13.4) node [anchor=north west][inner sep=0.75pt]   {$\displaystyle=\ \frac{1}{2} \delta m_{\phi }^{2}\left[\frac{A^{-1}( 0)}{\epsilon } +A^{0}( 0) +\epsilon A^{1}( 0)\right]$};
\end{tikzpicture}
\end{gathered}
\end{equation}

\begin{equation}
\begin{gathered}
\tikzset{every picture/.style={line width=0.75pt}} 
\begin{tikzpicture}[x=0.75pt,y=0.75pt,yscale=-1,xscale=1]

\draw  [dash pattern={on 4.5pt off 4.5pt}] (80,30) .. controls (80,18.95) and (88.95,10) .. (100,10) .. controls (111.05,10) and (120,18.95) .. (120,30) .. controls (120,41.05) and (111.05,50) .. (100,50) .. controls (88.95,50) and (80,41.05) .. (80,30) -- cycle ;
\draw  [fill={rgb, 255:red, 255; green, 255; blue, 255 }  ,fill opacity=1 ] (96.72,6.23) .. controls (98.8,4.41) and (101.96,4.64) .. (103.77,6.72) .. controls (105.59,8.8) and (105.36,11.96) .. (103.28,13.77) .. controls (101.2,15.59) and (98.04,15.36) .. (96.23,13.28) .. controls (94.41,11.2) and (94.64,8.04) .. (96.72,6.23) -- cycle ;
\draw [fill={rgb, 255:red, 255; green, 255; blue, 255 }  ,fill opacity=1 ]   (96.46,6.46) -- (103.54,13.54) ;
\draw [fill={rgb, 255:red, 255; green, 255; blue, 255 }  ,fill opacity=1 ]   (103.54,6.46) -- (96.46,13.54) ;

\draw (121,11.4) node [anchor=north west][inner sep=0.75pt]   {$\displaystyle=\ \frac{1}{2} \delta m_{\xi }^{2}\left[\frac{A^{-1}( m_{\xi })}{\epsilon } +A^{0}( m_{\xi }) +\epsilon A^{1}( m_{\xi })\right]$};
\end{tikzpicture}
\end{gathered}
\end{equation}
Here we have used the mass counterterms $\delta m^2_\phi$ and $\delta m^2_\xi$,  defined in the next section~\ref{app:renormalisation}. Substituting the master sum-integrals of Appendix A into these diagrams, the $1/\epsilon$ poles cancel and combining the finite pieces gives the complete NLO correction 
\begin{align}\label{eq:full2loopcompled}
    V_{\rm NLO}
    =&+
    \frac{m_\phi^2 T^2}{2\cdot12} 
    \nonumber \\ 
&+
\frac{T^2g^2}{12\cdot 4}\left(\frac{m^2_\xi}{(4\pi)^2}\left[-1+\ln\left(\frac{m^2_\xi}{\bar\mu^2}\right)\right] +\frac{T^2}{2 \pi^2}I_1\left(\frac{m_\xi}{T}\right)\right)\,\, \,
\nonumber \\ 
&+
\frac{\lambda_\xi}{8}\left(\frac{m^2_\xi}{(4\pi)^2}\left[-1+\ln\left(\frac{m^2_\xi}{\bar\mu^2}\right)\right] +\frac{T^2}{2 \pi^2}I_1\left(\frac{m_\xi}{T}\right)\right)^2
\,\,
\nonumber\\ 
&+
\left(-\frac{g^4 \phi_{\rm IR}^2}{4}\right)
\!\!\left(-\frac{m^2_\xi}{(4\pi)^4}\right)
    \!\!\left(
    7+\frac{\pi^2}{6}-6\ln\left(\frac{m^2_\xi}{\bar\mu^2}\right)+2\left[\ln\left(\frac{m^2_\xi}{\bar\mu^2}\right)\right]^2
    \right)
    \,\,
 \nonumber\\
 &+ \left(-\frac{g^4 \phi_{\rm IR}^2}{4}\right)\!\!\!
  \left( \frac{T^2}{(4\pi)^2}\right)\!\!\!
 \left[\!
 \Big(\frac{1}{12}+\frac{I_1\left(\frac{m_\xi}{T}\right)}{\pi^2}\Big)\!\! \Big(\ln\frac{\bar\mu^2}{T^2}+\ln\frac{\bar\mu^2}{m_\xi^2}\Big)
 +4
 \frac{ I_1\left(\frac{m_\xi}{T}\right)}{\pi^2} 
 -\frac{ I_2\left(\frac{m_\xi}{T}\right)}{\pi^2}  
 +\frac{ \mathcal{H}\left(\frac{m_\xi}{T}\right)}{\pi^2}\right]
 \nonumber\\
 &+ \left(-\frac{g^4 \phi_{\rm IR}^2}{4}\right)\!\!\!
  \left( \frac{T^2}{(4\pi)^2}\right)
  \left[
\frac16
\left(
\gamma_E
-\frac{\zeta'(2)}{\zeta(2)}
\right)
\right] \nonumber \\
  &+\frac{g^4\phi_{\rm IR}^2}{2(4\pi)^2}
  \left[
  \frac{m_\xi^2}{(4\pi)^2}
\left(
-1-\frac{\pi^2}{12}
+\ln\left(\frac{m_\xi^2}{\bar\mu^2}\right)
-\frac12
\ln^2\left(\frac{m_\xi^2}{\bar\mu^2}\right)
\right)
+
\frac{T^2}{2\pi^2}
\left(
\left(
2+\ln\frac{\bar\mu^2}{4T^2}
\right)I_1\left(\frac{m_\xi}{T}\right)
- I_{2}\left(\frac{m_\xi}{T}\right)
\right)
  \right]\nonumber \\
  &+\frac{g^4\phi_{\rm IR}^2T^2}{48(4\pi)^2}\left[
    2\gamma_E
    +\ln\left(\frac{\bar\mu^2}{4T^2}\right)
    -\frac{2\zeta'(2)}{\zeta(2)}
    \right]
\end{align}
\end{widetext}

\subsection{Leading-order potentials for the other models}
For completeness, we also give the explicit LO potentials for the three examples of Sec.~VI. The diagrammatic organisation is already shown in the main text, so here we only collect the final expressions.

\paragraph{Yukawa model.}
For the Yukawa model the hard fermionic determinant enters already at LO, together with the massless hard scalar contribution. This gives
\begin{align}\label{eq:yukawaLOapp}
V_{\text{eff}}^{\rm LO}={}&+\frac{1}{2} \mu_\phi^2 \phi_{\rm IR}^2+\frac{\lambda}{4!} \phi_{\rm IR}^4 +\frac{g}{3!}\phi_{\rm IR}^3+\sigma \phi_{\rm IR} \nonumber\\
&-\frac{\pi^2 T^4}{90}-\frac{n_\text{f}m_\psi^4}{64\pi^2}
\left[ \ln\left( \frac{m_\psi^2}{\bar\mu^2} \right) - \frac{3}{2} \right] \nonumber\\
&-\frac{n_\text{f} T^4}{2\pi^2} \mathcal{J}_\text{F} \left( \frac{m_\psi}{T} \right)
\end{align}
For a single Dirac fermion, $n_\text{f}=4$.

\paragraph{Abelian Higgs model.}
For the $U(1)$ model the full background-dependent gauge mass is kept in the hard-scale determinant. The LO potential is

\begin{align}\label{eq:abelianLOapp}
    V_{\text{eff}}^{\rm LO}=&
    -\frac{1}{2} \mu_\phi^2 \phi_{\rm IR}^2
    +\frac{\lambda}{4} \phi_{\rm IR}^4 - \frac{\pi^2 T^4}{90} \nonumber\\
& +\frac{3m_A^4}{64\pi^2}  \left[ \ln\left( \frac{m_A^2}{\bar\mu^2} \right) - \frac{5}{6} \right]
+\frac{3 T^4}{2\pi^2} \mathcal{J}\left( \frac{m_A}{T} \right)
\end{align}

\paragraph{Non-Abelian Higgs model.}
For the $SU(2)$ model the two gauge fields associated with the broken generators are hard, while the remaining massless modes give the field-independent thermal term. The LO potential is

\begin{align}\label{eq:SU(2)LOapp}
    V^{\rm LO}_\text{eff}=&-\frac{1}{2}\mu_\phi^2\phi_{\rm IR}^2+\frac{\lambda}{4} \phi_{\rm IR}^4
    - \frac{\pi^2 T^4}{30}
    \nonumber\\
    &+\frac{3m^4_A}{32\pi^2}
    \left[ \ln\left(\frac{m_A^2}{\bar\mu^2}\right)-\frac{5}{6}
    \right] +\frac{3T^4}{\pi^2} \mathcal{J}\left(\frac{m_A}{T}\right)
\end{align}

\section{Renormalisation of the two-scalar model}\label{app:renormalisation}

We define
\begin{align}
\beta_{\lambda_i}\equiv\frac{d\lambda_i}{d\ln\bar\mu},
\end{align}
and, when an absolute vacuum energy is retained, write
$V_{\rm tree}\supset+\Lambda$.  In this convention the one-loop
vacuum-energy beta function has a positive sign.

The beta functions 
at one and two-loop orders are
\begin{align}
(4\pi)^2\beta_\Lambda^{(1)}
&=\frac12\left(\mu_\phi^4+\mu_\xi^4\right),
\label{eq:app_betaLambda}\\
(4\pi)^4\beta_\Lambda^{(2)}
&=0,
\label{eq:app_betaLambda2}\\
(4\pi)^2\beta_{\lambda_\phi}^{(1)}
&=3(\lambda_\phi^2+g^4),
\label{eq:app_betalphi1}\\
(4\pi)^4\beta_{\lambda_\phi}^{(2)}
&=-\frac{17}{3}\lambda_\phi^3
-5\lambda_\phi g^4-12g^6,
\label{eq:app_betalphi2}\\
(4\pi)^2\beta_{\lambda_\xi}^{(1)}
&=3(\lambda_\xi^2+g^4),
\label{eq:app_betalxi1}\\
(4\pi)^4\beta_{\lambda_\xi}^{(2)}
&=-\frac{17}{3}\lambda_\xi^3
-5\lambda_\xi g^4-12g^6,
\label{eq:app_betalxi2}\\
(4\pi)^2\beta_{g^2}^{(1)}
&=g^2(\lambda_\phi+\lambda_\xi+4g^2),
\label{eq:app_betag1}\\
(4\pi)^4\beta_{g^2}^{(2)}
&=-9g^6-6g^4(\lambda_\phi+\lambda_\xi)
-\frac56g^2(\lambda_\phi^2+\lambda_\xi^2),
\label{eq:app_betag2}\\
(4\pi)^2\beta_{\mu_\phi^2}^{(1)}
&=\lambda_\phi\mu_\phi^2-g^2\mu_\xi^2,
\label{eq:app_betamphi1}\\
(4\pi)^4\beta_{\mu_\phi^2}^{(2)}
&=-\frac56\lambda_\phi^2\mu_\phi^2
-\frac12g^4\mu_\phi^2+2g^4\mu_\xi^2,
\label{eq:app_betamphi2}\\
(4\pi)^2\beta_{\mu_\xi^2}^{(1)}
&=\lambda_\xi\mu_\xi^2-g^2\mu_\phi^2,
\label{eq:app_betamxi1}\\
(4\pi)^4\beta_{\mu_\xi^2}^{(2)}
&=-\frac56\lambda_\xi^2\mu_\xi^2
-\frac12g^4\mu_\xi^2+2g^4\mu_\phi^2.
\label{eq:app_betamxi2}
\end{align}
Now let's shift our attention to the counterterms in the $\alpha$ expansion. The counterterm potential takes this form: 
\begin{align}
V^{\rm ct}=
\delta\Lambda-\frac12\delta\mu_\phi^2\phi_{\rm IR}^2
+\frac{\delta\lambda_\phi}{4!}\phi_{\rm IR}^4,
\end{align}
the LO pieces, at $O(\alpha^0 T^4)$ in the potential, are
\begin{align}
\delta\Lambda_{\rm LO}
&=\frac{\mu_\xi^4}{4(4\pi)^2\epsilon},
&
\delta\mu_{\phi,{\rm LO}}^2
&=\frac{g^2\mu_\xi^2}{2(4\pi)^2\epsilon},
\nonumber\\
\delta\lambda_{\phi,{\rm LO}}
&=\frac{3g^4}{2(4\pi)^2\epsilon}.
\label{eq:app_ctLOparameters}
\end{align}
At NLO ($\alpha^2$) the two-loop pieces that survive the strong-transition counting
are
\begin{align}
\delta\Lambda_{\rm NLO}
&=\frac{\lambda_\xi\mu_\xi^4}
{8(4\pi)^4\epsilon^2},
\label{eq:app_ctLambdaNLO}\\
\delta\mu_{\phi,{\rm NLO}}^2
&=\frac{-\mu_\xi^2}{4(4\pi)^4}
\left[-\frac{2g^4}{\epsilon}
+\frac{2g^4+g^2\lambda_\xi}{\epsilon^2}\right],
\label{eq:app_ctmphiNLO}\\
\delta\lambda_{\phi,{\rm NLO}}
&=\frac{1}{(4\pi)^4}
\left[-\frac{3g^6}{\epsilon}
+\frac{3(4g^6+g^4\lambda_\xi)}{4\epsilon^2}\right].
\label{eq:app_ctlphiNLO}
\end{align}
There are also the counterterm insertions in the NLO diagrams. For these, we define the two effective counterterm masses as
\begin{align}
\begin{cases}
\delta m_\phi^{2}(\phi_{\rm IR})
=-\delta\mu_{\phi,{\rm LO}}^{2}
+\tfrac12\delta\lambda_{\phi,{\rm LO}}\phi_{\rm IR}^{2},\\
\delta m_\xi^{2}(\phi_{\rm IR})
=+\delta\mu_{\xi,\text{LO}}^{2}
+\tfrac12\delta g_\text{LO}^2\phi_{\rm IR}^{2},
\end{cases}
\label{eq:app_effective_ct_masses}
\end{align}
and the additional one-loop counterterms needed at this order are
\begin{align}
\delta\mu_{\xi,\text{LO}}^2
&=\frac{\lambda_\xi\mu_\xi^2}{2(4\pi)^2\epsilon},
&
\delta g_\text{LO}^2
&=\frac{\lambda_\xi g^2+4g^4}{2(4\pi)^2\epsilon}.
\label{eq:app_ctheavy}
\end{align}

\end{document}